\documentclass[namedreferences]{SolarPhysics}
\usepackage[optionalrh]{spr-sola-addons} 
\usepackage{graphicx}        
\usepackage{color}           
\usepackage{url}             

\usepackage{lscape}
\usepackage{ccaption}
\usepackage{ulem}

\newcommand{\etal}{{\it et al.}}

\newcommand{\aap}{    {\it Astron. Astrophys.}}
\newcommand{\aaps}{   {\it Astron. Astrophys. Suppl.}}

\newcommand{\apj}{    {\it Astrophys. J.}}

\newcommand{\jgr}{    {\it J. Geophys. Res.}}

\newcommand{\pasj}{   {\it Pub. Astron. Soc. Japan}}

\newcommand{\solphys}{{\it Solar Phys.}}

\begin{document}

\begin{article}

\begin{opening}

\title{Coronal Temperature Diagnostic Capability of the \textit{Hinode}/X-Ray Telescope Based on Self-Consistent Calibration}

\author{N.~\surname{Narukage}$^{1}$\sep
        T.~\surname{Sakao}$^{2}$\sep
        R.~\surname{Kano}$^{1}$\sep
        H.~\surname{Hara}$^{1}$\sep
        M.~\surname{Shimojo}$^{3}$\sep
        T.~\surname{Bando}$^{1}$\sep
        F.~\surname{Urayama}$^{4}$\sep
        E.~\surname{DeLuca}$^{5}$\sep
        L.~\surname{Golub}$^{5}$\sep
        M.~\surname{Weber}$^{5}$\sep
        P.~\surname{Grigis}$^{5}$\sep
        J.~\surname{Cirtain}$^{6}$\sep
        S.~\surname{Tsuneta}$^{1}$
       }

\runningauthor{N. Narukage \etal}
\runningtitle{Coronal Temperature Diagnostic Capability of the \textit{Hinode}/X-Ray Telescope}
\institute{$^{1}$ National Astronomical Observatory of Japan (NAOJ),\\
                  2-21-1 Osawa, Mitaka, Tokyo, 181-8588, Japan\\
                  email: \url{noriyuki.narukage@nao.ac.jp}\\
           $^{2}$ Institute of Space and Astronautical Science/\\
                  Japan Aerospace Exploration Agency (ISAS/JAXA),\\
                  3-1-1 Yoshinodai, Sagamihara, Kanagawa, 229-8510, Japan\\
           $^{3}$ Nobeyama Solar Radio Observatory/\\
                  National Astronomical Observatory of Japan,\\
                  Minamimaki, Minamisaku, Nagano, 384-1305, Japan\\
           $^{4}$ Space Engineering Development Co., Ltd.,\\
                  1-12-2 Takezono, Tsukuba, Ibaraki, 305-0032, Japan\\
           $^{5}$ Smithsonian Astrophysical Observatory,\\
                  60 Garden Street, Cambridge, MA 02138, USA.\\
           $^{6}$ NASA/Marshall Space Flight Center,\\
                  AL 35812, USA
           }

\begin{abstract}
The \textit{X-Ray Telescope} (XRT) onboard the \textit{Hinode} satellite is an X-ray imager
that observes the solar corona with unprecedentedly high angular resolution (consistent with its 1$''$ pixel size).
XRT has nine X-ray analysis filters with different temperature responses.
One of the most significant scientific features of this telescope is its capability of diagnosing coronal temperatures
from less than 1~MK to more than 10~MK, which has never been accomplished before.
To make full use of this capability, accurate calibration of
the coronal temperature response of XRT is indispensable and is presented in this article.
The effect of on-orbit contamination is also taken into account in the calibration.
On the basis of our calibration results,
we review the coronal-temperature-diagnostic capability of XRT.
\end{abstract}
\end{opening}

\section{Introduction}
\label{introduction}

The outer atmosphere of the Sun, the solar corona, is most clearly discernible when seen in soft X-rays.
Since the early rocket experiments in the late 1960s, it has become widely recognized that soft X-ray
imagery of the Sun provides a powerful means to investigate physical conditions of hot plasmas (whose
temperature often exceeds 1~MK) that prevail in the corona.
Coronal imaging with the \textit{Soft X-ray Telescope} (SXT) onboard the \textit{Yohkoh} ({\it Solar-A\/}) satellite
\cite{oga91, tsu91} (operation period: 1991\,--\,2001)
covered a full solar activity cycle and
have revealed that magnetic reconnection plays an essential role in the energy release processes in the dynamic solar corona.


\begin{figure}
\centerline{\includegraphics[width=10.0cm,clip=]{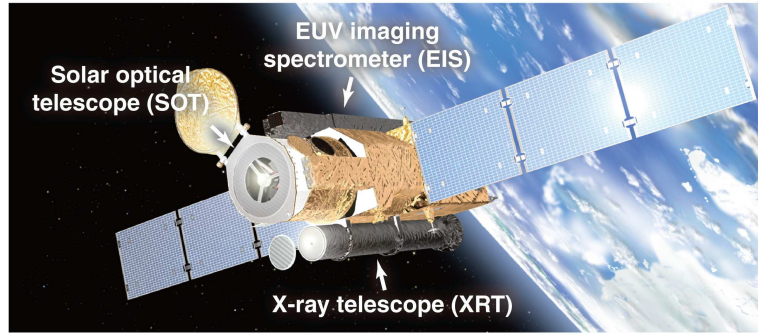}}
\caption{
The \textit{Hinode} satellite has three telescopes, namely \textit{Solar Optical Telescope} (SOT),
\textit{EUV Imaging Spectrometer} (EIS), and \textit{X-Ray Telescope} (XRT), to observe the Sun in different wavelengths.
The XRT is mounted at the bottom of this image, as indicated.
The size of \textit{Hinode} is approximately 2~m (from XRT to EIS) $\times$ 10~m (between both ends of solar paddles) $\times$ 4~m (between Sun-facing and rear ends).
}
\label{fig:hinode}
\end{figure}

The \textit{SOLAR-B} satellite was launched at 21:36 UT on 22 September 2006.
It was named \textit{``Hinode"}, which means sunrise in Japanese (\opencite{kos07}; Figure~\ref{fig:hinode}).
The \textit{X-Ray Telescope} (XRT) onboard \textit{Hinode} is a successor of the \textit{Yohkoh}/SXT.
It also employs grazing-incidence optics, but with
improved spatial resolution (consistent with 1$''$ CCD pixel size compared with $\approx$~2.5$''$ pixel size in
the case of SXT) while maintaining similar exposure cadence to that of SXT (only order of few milli-seconds for flare and order of few seconds for active region).
The XRT has the capability of imaging emission formed at much longer wavelengths than SXT. The combination of a backside thinned CCD and 
thin Al-mesh filter allow XRT to image emission significantly longward of 60~{\AA}.
This is a major difference from \textit{Yohkoh}/SXT in that the XRT can observe not
only high-temperature plasmas ($>$ 2~MK) seen with SXT, but also low-temperature ($<$ 2~MK, reaching
even below 1~MK) plasmas which comprise a significant amount of the corona.
With this enhanced temperature range for observing coronal plasmas, coupled with increased spatial
resolution, the XRT is able to perform detailed imaging observations of a wide variety of coronal
plasmas in a temperature range covering, continuously, from below 1~MK to well above 10~MK.
One of the most significant scientific features of the XRT is its coronal-temperature-diagnostic capability,
namely capability to make temperature maps, for such plasmas.

In order to have XRT perform coronal temperature diagnostics with its full capability,
we carefully calibrated the effective area of the XRT and its response to coronal temperatures
using not only ground-based test data but also on-orbit data observed in X-rays and visible light.
The effect of the on-orbit contamination, which manifested itself as decreasing intensity of the Sun's corona
with time as imaged by XRT, was also calibrated as accurately as possible.
On the basis of our calibration results,
we review the coronal-temperature-diagnostic capability of XRT with the filter-ratio method.

In Section~\ref{sec:Optical elements of XRT}, the optical elements of XRT are briefly mentioned.
In Section~\ref{sec:overview of calibration}, we show the overview of the calibration performed in this article.
In Section~\ref{sec:on-orbit calibration}, we summarize how to identify the contaminant and how to measure its accumulating thickness
on the focal-plane analysis filters (FPAFs) and CCD as a function of time.
This then identifies the effective area and temperature response of XRT including the contamination at every period.
In Section~\ref{sec:temperature}, we evaluate the coronal-temperature-diagnostic capability of XRT,
explain the filter-ratio method to derive the coronal temperature, and present suitable filter pairs for each coronal temperature.
An example of coronal temperature distribution with the XRT data is also shown.
Finally, we summarize the result of this article in Section~\ref{sec:summary}.

Additionally, in Appendix~\ref{sec:ground-based calibration}, the calibration of the nine X-ray FPAFs
with ground-based end-to-end test data is explained.
In 2002\,--\,2003, the X-ray transmission measurement of the FPAFs was performed
at the X-ray Astronomy Calibration and Testing (XACT) facility
of the Istituto Nazionale di Astrofisica / Osservatorio Astronomico di Palermo
``G.S. Vaiana" \cite{col94}.
However, some of the calibrated FPAFs were unfortunately damaged.
In this article, we characterize the complete flight set of FPAFs, namely both the un-damaged and
re-manufactured FPAFs, with another ground-based test performed in 2005.
Our calibrated thicknesses are consistent with the thicknesses available from XACT data.

An overview of the XRT scientific objectives,
design, and performance of the XRT telescope are summarized by \inlinecite{gol07},
while the X-ray camera is described by \inlinecite{kan08}.

\clearpage

\section{Optical Elements of XRT}
\label{sec:Optical elements of XRT}

\begin{figure}
\centerline{\includegraphics[width=11.0cm,clip=]{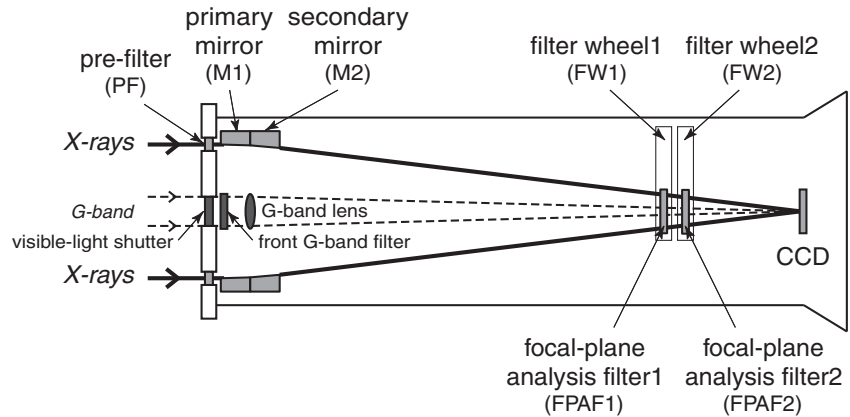}}
\vspace{3.0mm}
\caption{Optical elements of XRT.
The solid line shows the optical path in X-rays. The optical elements for X-rays are
the pre-filter (PF), primary mirror (M1), secondary mirror (M2), focal-plane analysis filters (FPAF1 and FPAF2) mounted on
filter wheels (FW1 and FW2, respectively), and CCD.
The optical path in visible light (G-band) is indicated by the dashed line.
The optical elements for visible light are front G-band filter, G-band lens, G-band filter mounted on filter wheel 2 (FW2), and CCD.}
\label{fig:optical elements}
\end{figure}

In order to describe in detail the calibrations performed on the XRT,
we first briefly describe the optical elements.
Figure~\ref{fig:optical elements} shows the optical elements and optical paths of XRT.
The XRT can take not only X-ray images but also visible light (G-band) images
for co-aligning images between XRT and other instruments mounted on \textit{Hinode}.
The solid and dashed lines in Figure~\ref{fig:optical elements} show the optical paths
in X-rays and visible light, respectively.

The XRT optical elements for X-rays consist of pre-filter (PF), two grazing-incidence mirrors (M1 and M2),
two focal-plane analysis filters (FPAF1 and FPAF2), and CCD.

\subsection{Optical Elements for X-rays}
\label{subsec:Optical elements for X-rays}

\begin{figure}
\centerline{\includegraphics[width=8.0cm,clip=]{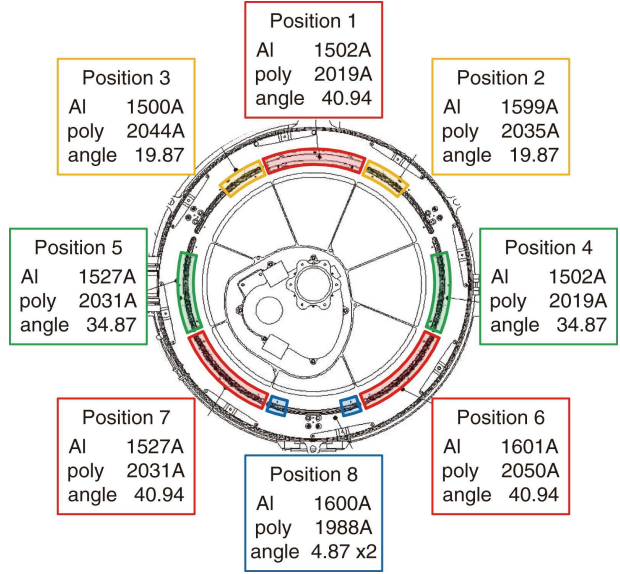}}
\vspace{3.0mm}
\caption{
Entrance aperture and pre-filter of XRT.
The pre-filter consists of eight fan-shaped annular segments made of Aluminum on a polyimide film.
The filter thicknesses measured by Luxel and the opening angles are summarized.
}
\label{fig:PF}
\end{figure}

\begin{figure}
\centerline{\includegraphics[width=10.0cm,clip=]{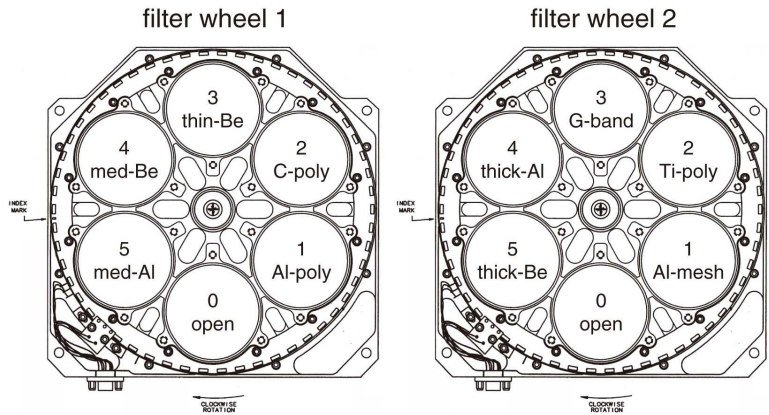}}
\vspace{3.0mm}
\caption{Filter wheels and focal-plane analysis filters.
XRT has two filter wheels: Filter Wheels 1 and 2 (FW1 and FW2).
Each filter wheel has six positions to mount the filters.
Five X-ray analysis filters are mounted on FW1, and FW2 has
four X-ray filters and one filter for visible-light observation (G-band).
The position number and the name of mounted filter are described
in each filter position.
}
\label{fig:FW}
\end{figure}

\begin{itemize}
\item[\textbf{\textit{i}) Pre-filter:}]
The pre-filter consists of eight fan-shaped annular filter segments each made of thin aluminum on a polyimide film (Figure~\ref{fig:PF}),
which were manufactured by Luxel Corporation.
The aperture shape was designed to obscure the locations of mirror-bonding pads to
avoid possible degradation in image quality caused by surface deformation of the mirror at the bonding pads.
In Figure~\ref{fig:PF}, the aperture areas are shown in color with their respective opening angles.
The average thickness of the entire pre-filter assembly is then given as $\sum (d \times \theta) / \sum \theta$, where $d$ and $\theta$ are
the thickness and opening angle of each filter segment, respectively.
The average pre-filter consists of 1538~{\AA} Al and 2030~{\AA} polyimide based on the measurement by manufacturer.

\item[\textbf{\textit{ii}) Mirrors:}]
The M1 and M2 mirrors are grazing-incidence annular mirror manufactured by Goodrich, each made of Zerodur.
On the basis of measurements by the manufacturer, the annular entrance aperture of the XRT primary mirror is located
between radii of $r_1$ = 17.042446~cm and $r_2$ = 17.074051~cm.
Considering the open angle of pre-filter (242.04$^{\circ}$, see Figure~\ref{fig:PF}), the geometric aperture area of XRT is
calculated to be $\pi \times (r_2^2 - r_1^2) \times (242.04 / 360)$ = 2.28~cm$^2$.
The grazing angle of X-rays at each XRT mirror is about 0.91$^{\circ}$ on average.

\item[\textbf{\textit{iii}) Focal-plane analysis filters:}]
XRT has two filter wheels: FW1 and FW2. Each filter wheel has six positions to mount focal-plane analysis filters (FPAFs).
As shown in Figure~\ref{fig:FW}, five X-ray analysis filters are mounted on FW1, and FW2 has
four X-ray filters and one filter for visible-light observation (G-band).
These nine X-ray filters are made of several kinds of metal and support, with different
thicknesses as summarized in Table~\ref{tbl:FPAF}.
The filters are designed to observe the corona in a temperature range from
less than 1~MK to more than 10~MK.
Focal-plane analysis filters on FW1 and FW2 (hereafter FPAF1 and FPAF2, respectively) can be selected independently,
even allowing combinations of filters both from FW1 and FW2 in series.

\item[\textbf{\textit{iv}) CCD:}]
X-rays are focused on a back-illuminated CCD \cite{kan08}.
\end{itemize}

The effective area [$A_\mathrm{eff}$] of XRT 
is defined by the product of the geometric aperture area [$A$] and the efficiency of all of the optical elements:
\begin{equation}
A_\mathrm{eff} = A \times \mathcal{T}_\mathrm{PF} \times R_\mathrm{M1} \times R_\mathrm{M2} \times \mathcal{T}_\mathrm{FPAF1} \times \mathcal{T}_\mathrm{FPAF2} \times QE_\mathrm{CCD} ,
\label{eq:A_eff}
\end{equation}
where $\mathcal{T}_\mathrm{PF}$ is the transmission of the pre-filter, $R_\mathrm{M1}$ and $R_\mathrm{M2}$ reflectivities
at the primary and secondary mirrors, $\mathcal{T}_\mathrm{FPAF1}$ and $\mathcal{T}_\mathrm{FPAF2}$ the transmissions of FPAF1 and FPAF2, and
$QE_\mathrm{CCD}$ the Quantum Efficiency of the CCD.
Because each efficiency is a function of wavelength,
$A_\mathrm{eff}$ is also a function of wavelength.
In this article, we calibrate $\mathcal{T}_\mathrm{PF}$, $\mathcal{T}_\mathrm{FPAF1}$, and $\mathcal{T}_\mathrm{FPAF2}$,
and then derive the effective area.

\subsection{Optical Elements for G-band}
\label{subsec:Optical elements for G-band}

When XRT takes visible-light images in G-band, the visible-light shutter is opened,
and FW1 and FW2 are set to the open and G-band filter position, respectively.
This G-band filter is a bandpass filter with a central wavelength of 4305.6~{\AA} and a bandwidth of 172.8~{\AA} (FWHM).
The visible light is focused onto the same CCD as X-rays (Figure~\ref{fig:optical elements}).
X-rays cannot reach the CCD through the G-band (dotted line) or the X-ray (solid line) paths,
because those X-rays are blocked by G-band filters employed at the front of the XRT and on FW2, respectively.

\begin{landscape}
\begin{table}
\caption{Focal-plane analysis filters and pre-filter.}
\label{tbl:FPAF}
\begin{tabular}{ccccccccc}
\hline
FW-pos$^{(a)}$ & filter name & \multicolumn{2}{c}{pre-delivery measurements at Luxel$^{(b)}$}               & & \multicolumn{4}{c}{calibrated values in this article} \\
\cline{3-4}\cline{6-9}
               &             & metal                               & support                                & & pure metal     & oxidized               & at fabrication$^{(c)}$ & support           \\
\hline
 1-0           & open        & --                                  & --                                     & & --             & --                     & --                     & --                \\
 1-1           & Al-poly     & Al$^{(d)}$ 1283{\AA} ($\pm$50{\AA}) & poly$^{(e)}$ 2656{\AA} ($\pm$100{\AA}) & & Al   1412{\AA} &  Al$_2$O$_3$  75{\AA}  &   1470{\AA}            & poly 2656{\AA}    \\
 1-2           & C-poly      & C$^{(f)}$  6038{\AA} ($\pm$50{\AA}) & poly 3478{\AA} ($\pm$100{\AA})         & & C    5190{\AA} &  --                    &   5190{\AA}            & poly 3478{\AA}    \\
 1-3           & thin-Be     & Be$^{(g)}$    9$\mu$ ($+$5$\mu$)    & --                                     & & Be  10.46$\mu$ &  BeO         150{\AA}  &  10.47$\mu$            & --                \\
 1-4           & med-Be      & Be   30$\mu$ ($+$5$\mu$/$-$2$\mu$)  & --                                     & & Be  26.89$\mu$ &  BeO         150{\AA}  &  26.90$\mu$            & --                \\
 1-5           & med-Al      & Al 12.5$\mu$ ($\pm$5$\%$)           & --                                     & & Al  12.25$\mu$ &  Al$_2$O$_3$ 150{\AA}  &  12.26$\mu$            & --                \\
 2-0           & open        & --                                  & --                                     & & --             & --                     & --                     & --                \\
 2-1           & Al-mesh     & Al 1605{\AA} ($\pm$50{\AA})         & mesh$^{(h)}$                           & & Al   1583{\AA} &  Al$_2$O$_3$ 150{\AA}  &   1700{\AA}            & 77\% trans. mesh  \\
 2-2           & Ti-poly     & Ti$^{(i)}$ 2345{\AA} ($\pm$50{\AA}) & poly 2522{\AA} ($\pm$100{\AA})         & & Ti   2338{\AA} &  TiO$_2$      75{\AA}  &   2380{\AA}            & poly 2522{\AA}    \\
 2-3           & G-band      & Glass                               & --                                     & & --             & --                     & --                     & --                \\
 2-4           & thick-Al    & Al   25$\mu$ ($\pm$10$\%$)          & --                                     & & Al  26.09$\mu$ &  Al$_2$O$_3$ 150{\AA}  &   26.1$\mu$            & --                \\
 2-5           & thick-Be    & Be  300$\mu$ ($\pm$30$\%$)          & --                                     & & Be 252.79$\mu$ &  BeO         150{\AA}  &  252.8$\mu$            & --                \\
               & pre-filter  & Al 1538{\AA} ($\pm$50{\AA})         & poly$^{(e)}$ 2030{\AA} ($\pm$100{\AA}) & & Al   1492{\AA} &  Al$_2$O$_3$  75{\AA}  &   1550{\AA}            & poly 2030{\AA}    \\
\hline
\end{tabular}
\begin{description}
\item[$^{(a)}$] ``FW" and ``pos" mean the filter wheel number and position on the filter wheel, respectively.
\item[$^{(b)}$] These values are described in the certification sheet by Luxel.
Note that the thicknesses measured by Luxel are only for reference.
\item[$^{(c)}$] The expected metal thickness when it was fabricated. At fabrication, the metals had not oxidized at all.
                This value is derived from the calibrated thicknesses of pure and oxidized metal using
                Equation~(\ref{eq:oxide}) in Appendix~\ref{sec:oxidization of metal}.
                This is shown for comparison with the value measured by Luxel.
\item[$^{(d)-(g), (i)}$] ``Al" means Aluminum, ``poly" polyimide (C$_{22}$H$_{10}$N$_{2}$O$_{5}$), ``C" Carbon, ``Be" Beryllium, and ``Ti" Titanium.
\item[$^{(h)}$] The mesh for Al-mesh filter is made of stainless steel.
\end{description}
\end{table}
\end{landscape}

\section{Overview of the Calibration}
\label{sec:overview of calibration}

\begin{figure}
\centerline{\includegraphics[width=12.0cm,clip=]{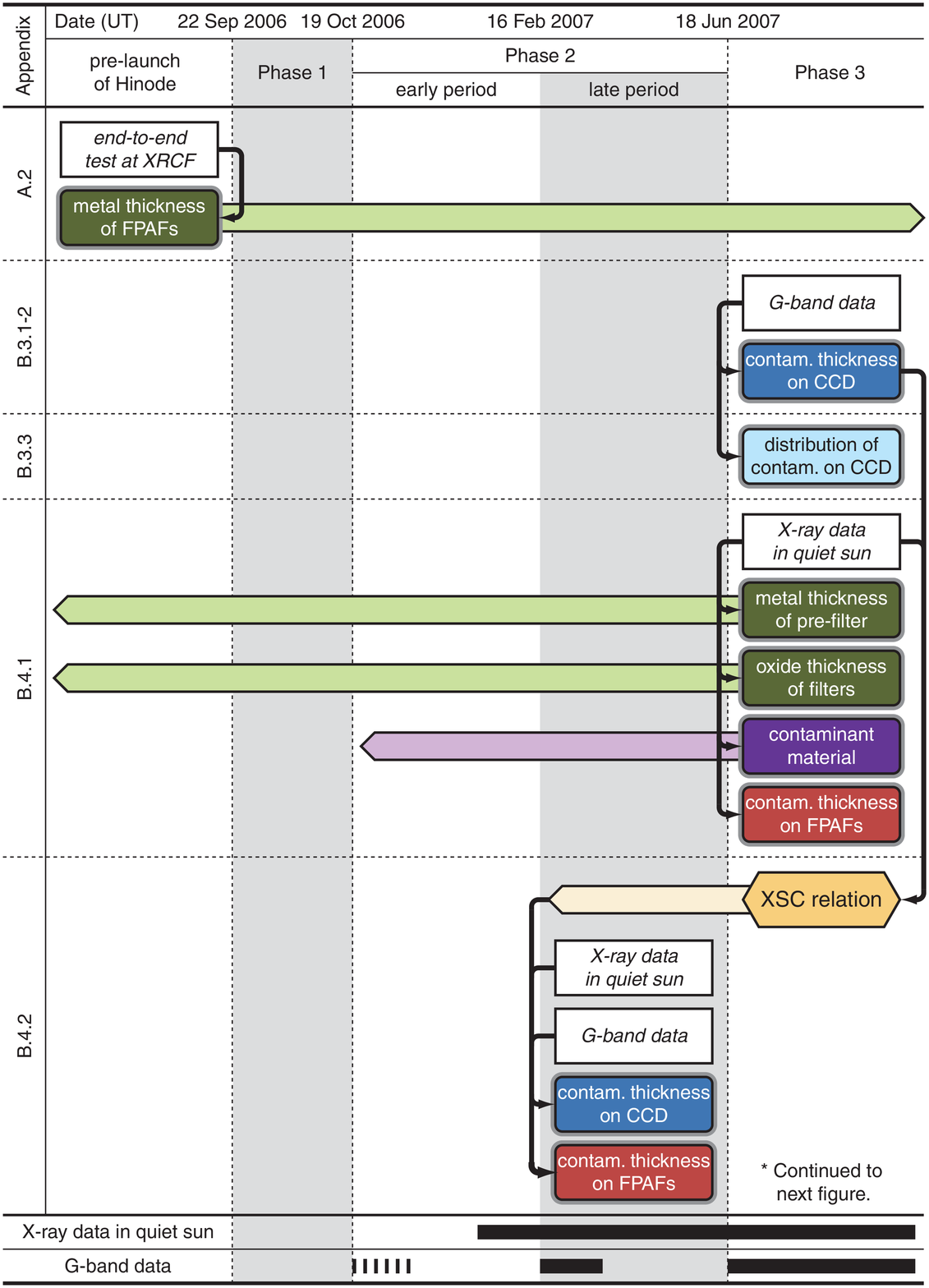}}
\vspace{3.0mm}
\caption{
Outline of calibration activity for XRT.
}
\label{fig:calibration}
\end{figure}

\begin{figure}
\centerline{\includegraphics[width=12.0cm,clip=]{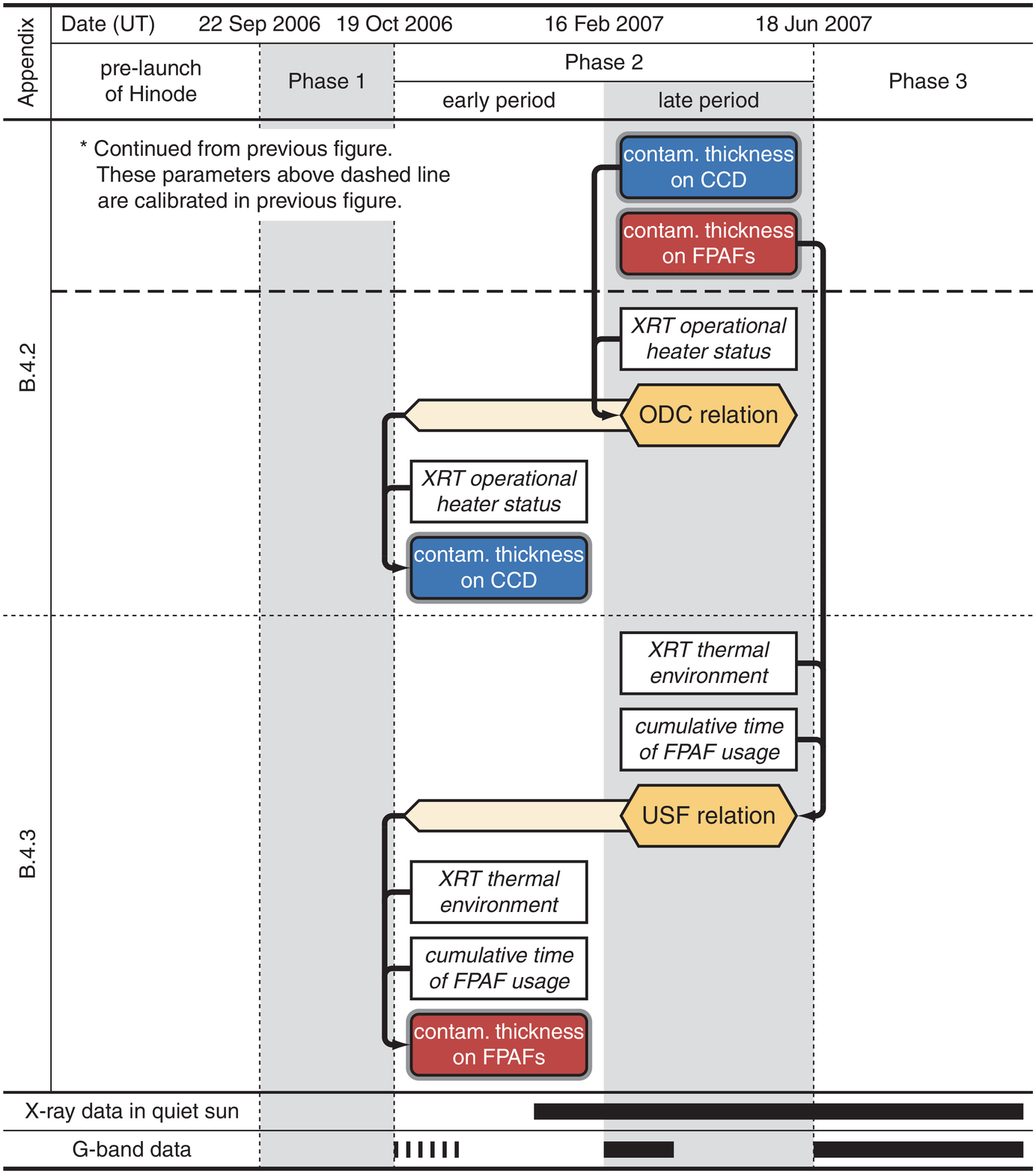}}
\vspace{3.0mm}
\contcaption{
Continued.
}
\end{figure}

In Section~\ref{sec:on-orbit calibration} and Appendices, we will carefully calibrate the spectral response of XRT
including the effect of the on-orbit contamination as accurately as possible.
The on-orbit contamination makes our calibrations complicated.
Therefore, before proceeding in detail,
we first give an outline of calibration activities in Figure~\ref{fig:calibration}.

In this figure, time passes from left to right.
After the launch of \textit{Hinode}, from the viewpoint of the XRT thermal environment,
there are three distinct intervals:
Phase~1 (22 September 2006 -- 19 October 2006), Phase~2 (19 October 2006 -- 18 June 2007), and Phase~3 (18 June 2007 -- present).
Phase~1 is the period from the launch to first light, when the CCD bakeout heater was kept on.
Phases~2 and 3 are the periods of normal operation of XRT without and with enabling (turning on) of the operational heater, respectively.
The details of each phase are described in Appendix~\ref{sec:thermal distribution inside XRT}.

At the bottom of Figure~\ref{fig:calibration}, the frequency of observations in X-rays and the G-band are shown by lines.
The solid lines indicate that many data sets were taken, dotted line a few data sets, and no line no signifies data sets.
From the viewpoint of this XRT observation, Phase~2 is divided into two periods, namely the early period (19 October 2006 -- 16 February 2007)
and the late period (16 February 2007 -- 18 June 2007).
In the early period,
there are no simultaneous observations in X-rays and G-band, while
in the late period, there are simultaneous observations.

The XRT calibrations were performed in order from top to bottom of the figure.
The white boxes indicate the data sets used for our calibration.
The black arrows connect the data sets or models to the calibrated results.
The calibrated XRT instrumental parameters are shown by green boxes.
The bars in light-green are extended from the green boxes to the times at which the calibrated results in the green boxes are valid.
Note that the results shown by green boxes and light-green bars are used for later calibration without indicating by black arrows.
The blue and red boxes are the calibration results of contaminant accumulated on the CCD and FPAFs, respectively.

In Phase~3,
since both X-ray and G-band data were taken sufficiently often (see bottom part of Figure~\ref{fig:calibration}),
the calibrations of both the XRT instrument and contamination are possible.
However, in Phase~2 the data in X-rays and/or G-band were lacking for the calibration.
In order to calibrate XRT for such episodes, we establish three models shown in yellow boxes.
For the late period of Phase~2 where G-band data are sparse, the model of the ``XSC (X-ray-suggested CCD contamination) relation" based on the result of Phase~3 is used
to calibrate the contaminant on both CCD and FPAFs as shown by a bar in light-yellow.
Meanwhile, for the early period of Phase~2 where the data are sparser than in the late period,
two models --- ``ODC (operational heater driven contamination)" and ``USF (FPAF-usage-suggested FPAF contamination)" relations ---
established in the late period are applied.
In Phase~1, we expect that the contaminant did not accumulate on both CCD and FPAFs,
because the CCD bake heater was kept on and the temperature around the FPAFs was warm.

\begin{figure}
\centerline{\includegraphics[width=10.0cm,clip=]{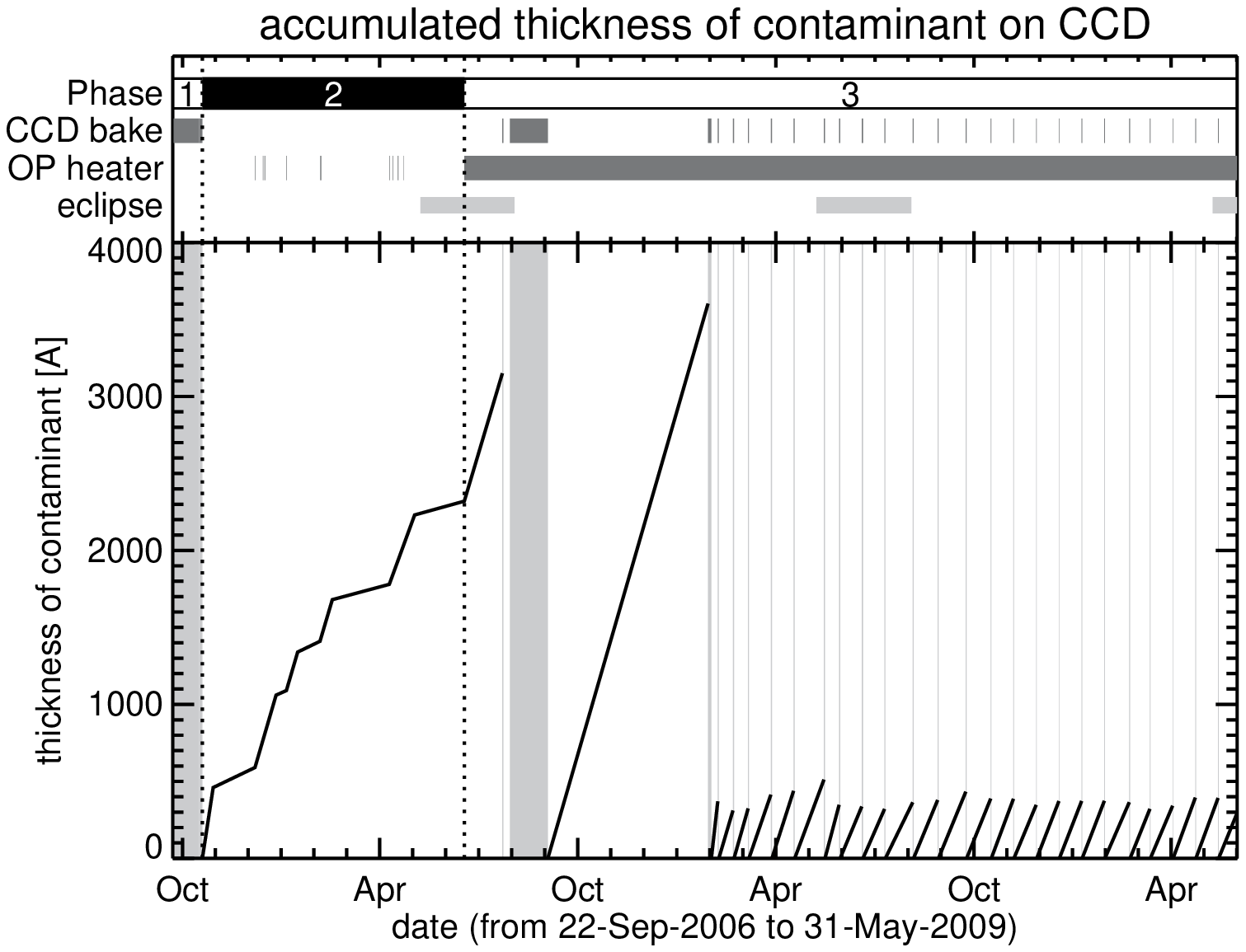}}
\caption{
Calibrated thickness of contaminant accumulated on the CCD, with bakeouts indicated by gray vertical lines.
}
\label{fig:contam on CCD}
\end{figure}

\begin{figure}
\centerline{\includegraphics[width=10.0cm,clip=]{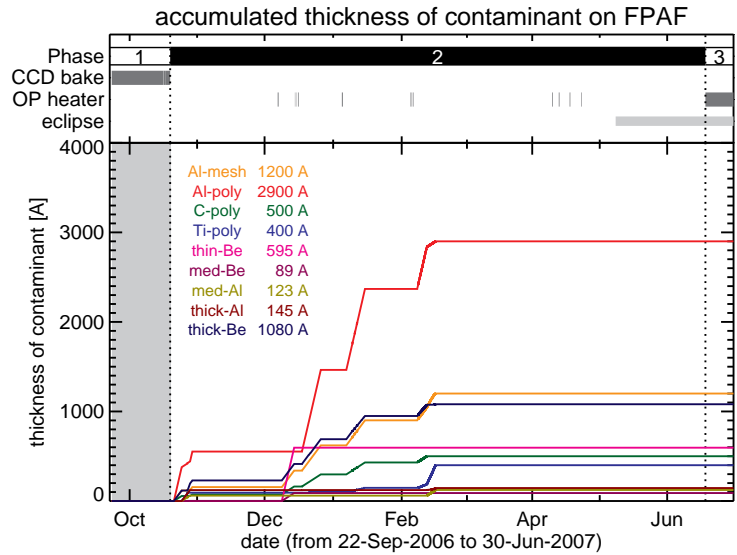}}
\caption{
Calibrated thicknesses of contaminant accumulated on the FPAFs.
The final thickness of contaminant on each FPAF is shown.
}
\label{fig:contam on FPAF}
\end{figure}

As shown in Figure~\ref{fig:calibration}, we calibrate the spectral response of XRT throughout its whole operation period.
Table~\ref{tbl:FPAF} summarizes the calibrated instrumental parameters of XRT with ground-based end-to-end test data (shown by green boxes in Figure~\ref{fig:calibration}).
The details of this ground-based calibration are described in Appendix~\ref{sec:ground-based calibration}.
Time-varying thicknesses of on-orbit contaminant accumulated on the CCD (blue boxes) and FPAFs (red boxes) are given by 
Figures~\ref{fig:contam on CCD} and \ref{fig:contam on FPAF}, respectively.
The contaminant material is identified to
have the chemical composition of a long-chain organic compound without silicon, refractive index of $\approx$~1.5, and density of $\approx$~1~g~cm$^{-3}$.
In Section~\ref{sec:on-orbit calibration} and Appendix~\ref{app:on-orbit calibration}, the on-orbit calibrations are described.

\section{On-orbit Calibration}
\label{sec:on-orbit calibration}

XRT took its first image on 23 October 2006.
After this first light, XRT has been taking several thousand images a day.
However, several months into the mission, we found that the X-ray intensity seen with XRT started to decrease continuously, especially, in the thinner filters.

\begin{figure}
\centerline{\includegraphics[width=10.0cm,clip=]{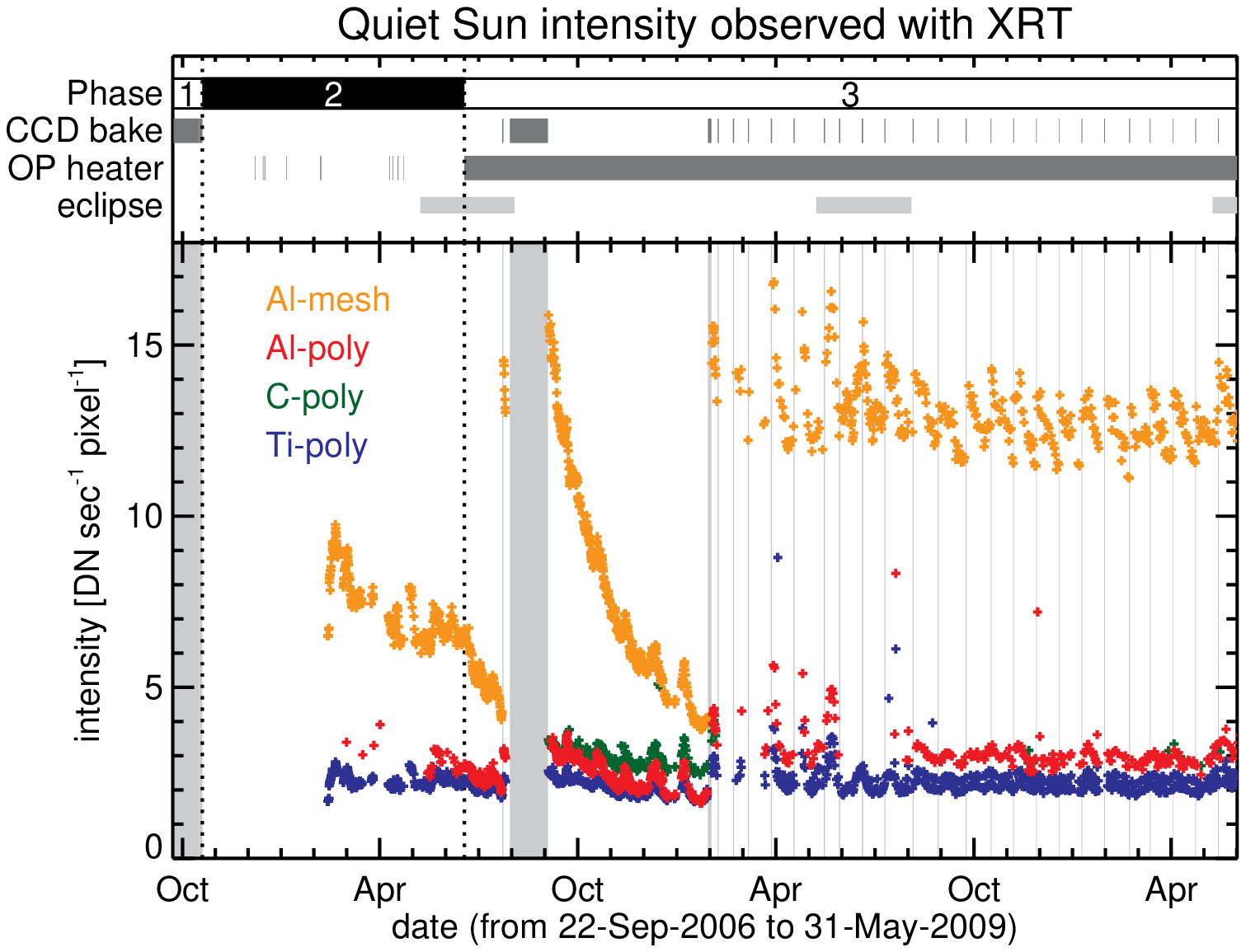}}
\vspace{3.0mm}
\centerline{\includegraphics[width=10.0cm,clip=]{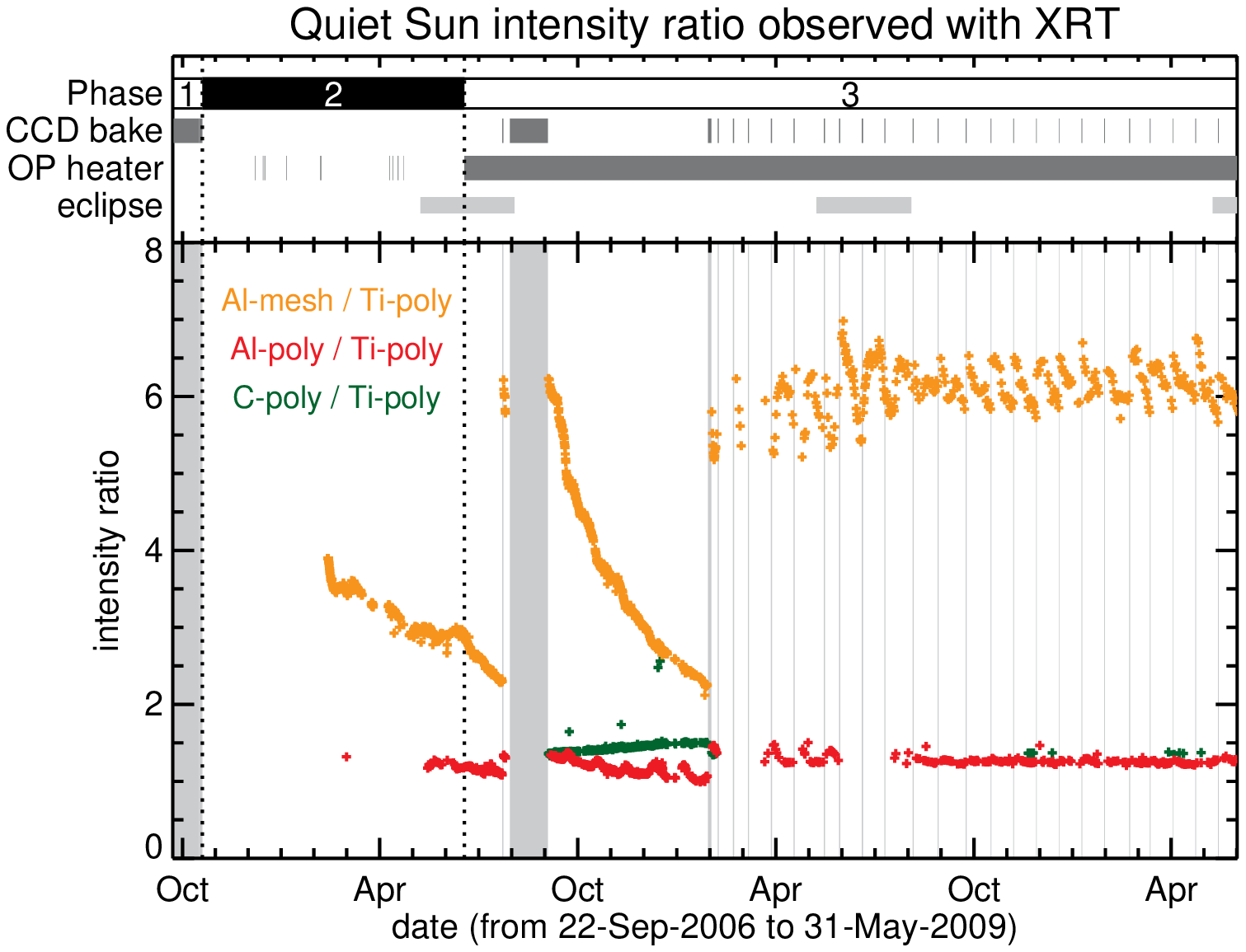}}
\caption{
Intensity of the quiet Sun observed with XRT.
top panel: Original intensity of the quiet Sun observed with four kinds of X-ray filters.
bottom panel: Intensity ratio of the quiet Sun normalized with the intensity observed with Ti-poly filter.
The gray area shows the duration of CCD bakeout.
\newline
\textbf{Note} -- The information about the phase, which is defined in Appendix~\ref{sec:thermal distribution inside XRT},
and status of bake out heater and operational heater (gray bars mean that these heaters are on) are shown in the top part of these plots.
\textit{Hinode} has been placed in a Sun-synchronous polar orbit that is above the day-night line on Earth,
where \textit{Hinode} can continuously observe the Sun for 24~hours a day for about nine months per year.
The remaining three months are called the ``eclipse season",
where the Sun is eclipsed by Earth for a maximum of 20~minutes in each 98-minute orbit.
This eclipse season is indicated by light-gray bars.
}
\label{fig:DN QS}
\end{figure}

The top panel of Figure~\ref{fig:DN QS} shows the intensity [DN sec$^{-1}$ pixel$^{-1}$] of the quiet Sun in the four X-ray filters of XRT.
In an intensity distribution (histogram) from a full-Sun corona, quiet-Sun areas manifest
themselves as a concentration in the distribution whose profile is well expressed by a Gaussian.
We took the center position of the Gaussian profile as representing the area-averaged quiet-Sun intensity.
The orange, red, green, and blue pluses indicate the quiet-Sun intensity
observed with the Al-mesh, Al-poly, C-poly, and Ti-poly filters, respectively.
The gray area shows episodes of CCD bakeout (see Table~\ref{tbl:bakeout} in Appendix~\ref{sec:CCD bakeout} for details).
The intensity decrease is seen in all filters, with that in the Al-mesh filter being most significant.
While intensity fluctuations in the short-term (several weeks)
is caused by solar activity such as the appearance of active regions,
the systematic intensity decrease can not be explained by solar activity.
For example, the intensity of the quiet Sun with the Al-mesh, Al-poly, C-poly, and Ti-poly filters decreases to 27\%, 59\%, 80\%, and 73\%, respectively,
during 142 days from 8 September 2007 to 28 January 2008.
We conclude that the decrease is caused by accumulation of materials, namely contaminants, obscuring the optical path of the XRT.

The X-ray intensity ratio is a function of the XRT filter response and coronal temperature (see Equation~(\ref{eq:R})),
and is not affected by the variation of the coronal density nor the distance between the Sun and \textit{Hinode} unlike the intensity.
Since such intensity ratio recovered to almost the same level
by each CCD bakeout as seen in Phase~3 in the bottom panel of Figure~\ref{fig:DN QS},
it is most likely that at least some fraction of the contaminants is accumulating on the CCD
and each bakeout reduces the contaminant to the same thickness.
The analysis of G-band intensity in Appendix~\ref{app:on-orbit calibration} gives the result that this ``same thickness" is actually zero.
Hence, we can say that each bakeout completely removes the contaminant from the CCD (except for the ``spots" discussed in Appendix~\ref{app:on-orbit calibration}).
Details of CCD bakeout are given in Appendix~\ref{sec:CCD bakeout}.

\begin{figure}
\centerline{\includegraphics[width=10.0cm,clip=]{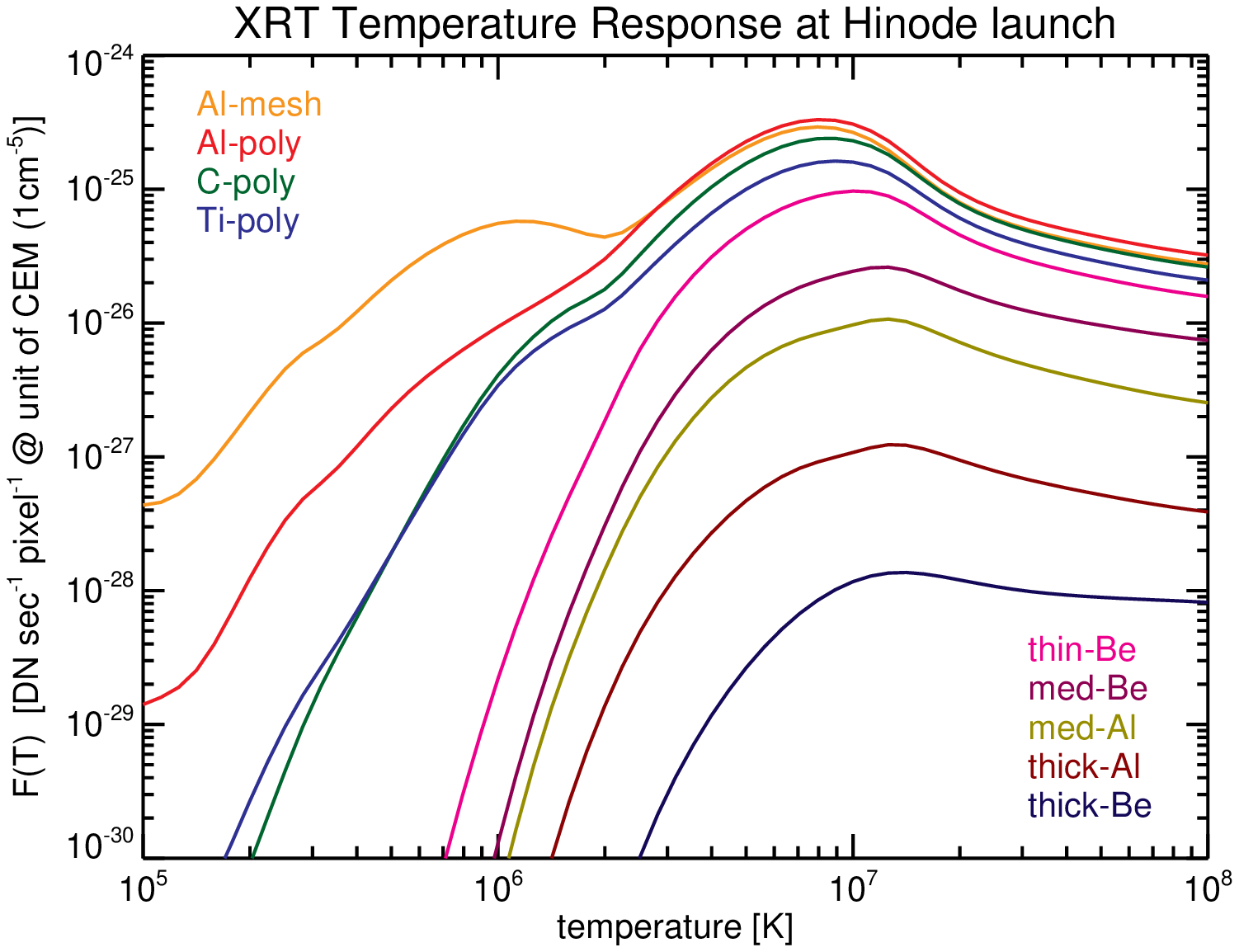}}
\vspace{3.0mm}
\centerline{\includegraphics[width=10.0cm,clip=]{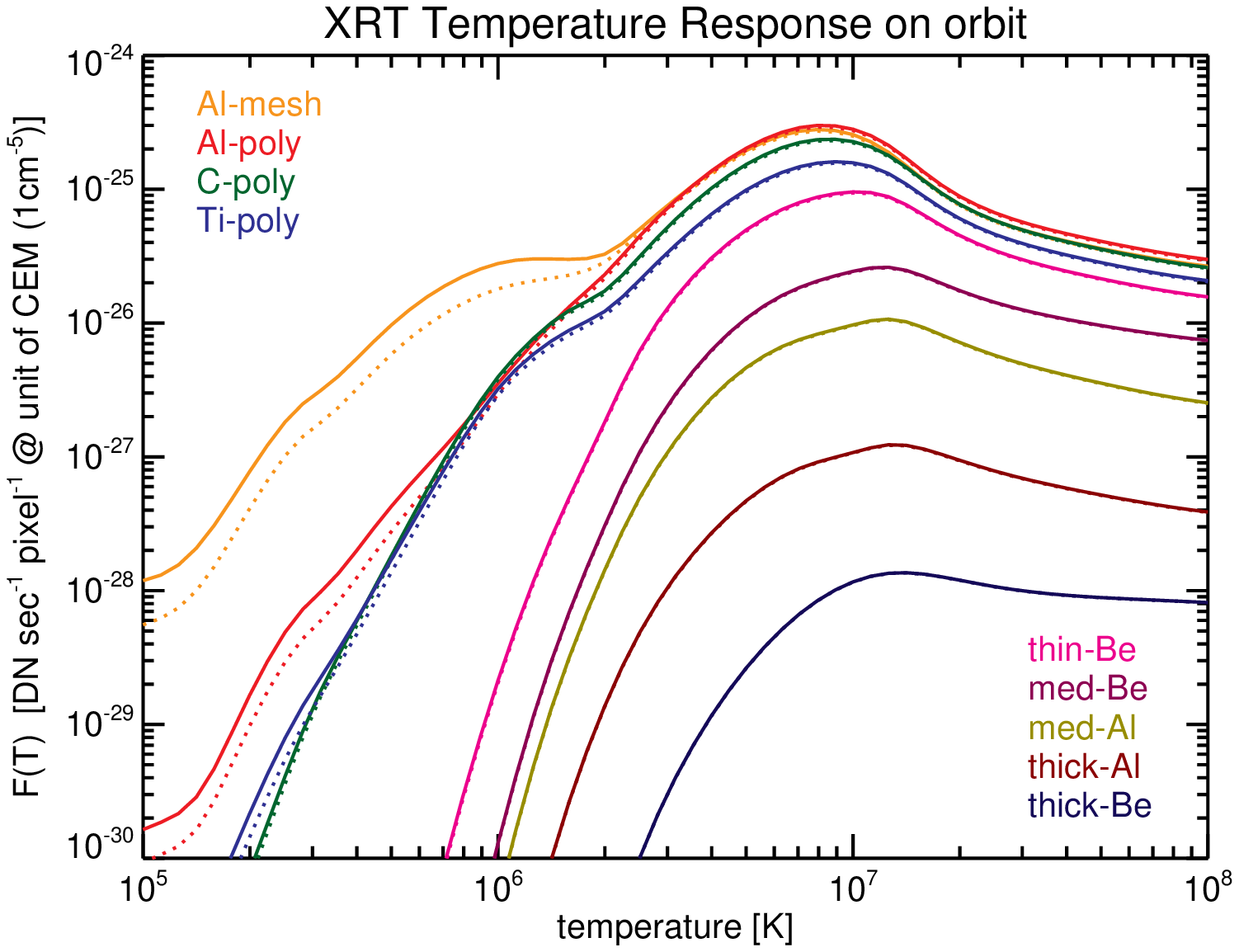}}
\caption{
top panel: Response of XRT to the coronal temperature at the launch of \textit{Hinode}, \textit{i.e.} when XRT had not been contaminated.
bottom panel: Response of XRT to coronal temperature on orbit considering the contaminant accumulated on the FPAFs and CCD.
The solid and dotted lines in the bottom panel show the temperature response just after CCD bakeout and one month after the bakeout in Phase~3, respectively.
We assume that 800~{\AA} of contaminant accumulated on the CCD one month after the bakeout.
}
\label{fig:response}
\end{figure}

We also find the intensity of the quiet Sun taken with Al-poly filter (red pluses) is weaker than with the C-poly filter (green pluses)
as clearly seen in the September 2007 to February 2008 time frame in the top panel of Figure~\ref{fig:DN QS}.
Strangely enough, however, the ground-based measurement at XRCF indicates that Al-poly filter should have a larger signal over the C-poly filter
throughout all coronal temperatures (see the top panel of Figure~\ref{fig:response}).
This discrepancy led us
to the notion that, in addition to the CCD, the FPAFs also suffered from the accumulation of contaminants
with different accumulated thickness for different filters.

\begin{table}
\caption{Behavior of contamination}
\label{tbl:contamination}
\begin{tabular}{cccccc}
\hline
period$^{(a)}$     & bake    & operational         & eclipse      & \multicolumn{2}{c}{contamination$^{(b)}$} \\
\cline{5-6}
                   & heater  & heater              & season       & on CCD                      & on FPAFs\\
\hline
Phase~1            & ON      & OFF                 & no           & none                        & none            \\
normal Phase~2     & OFF     & OFF                 & N/A          & $\nearrow$                  & $\longrightarrow$   \\
CCP in Phase~2     & OFF     & ON$\rightarrow$OFF  & no           & $\nearrow\nearrow\nearrow$  & $\nearrow\nearrow\nearrow$   \\
WCP in Phase~2     & OFF     & ON$\rightarrow$OFF  & close to     & $\nearrow\nearrow\nearrow$  & $\longrightarrow$   \\
Phase~3            & OFF     & ON                  & N/A          & $\nearrow\nearrow\nearrow$  & $\longrightarrow$   \\
CCD bakeout        & ON      & N/A                 & N/A          & completely removed          & $\longrightarrow$   \\
\hline
\end{tabular}
\begin{description}
\item[$^{(a)}$] Phases~1\,--\,3 are defined in Appendix~\ref{sec:thermal distribution inside XRT}.
                CCP and WCP represent ``cool contamination periods" and ``warm contamination period"
                defined in Appendix~\ref{subsubsec:contam on FPAF in Phase 2}, respectively.
\item[$^{(b)}$] ``$\nearrow\nearrow\nearrow$" and ``$\nearrow$" indicate
                the rapid and slow accumulation of contaminant, respectively.
                ``$\longrightarrow$" means that the thickness of contaminant is stable.
\end{description}
\end{table}

Here, we briefly summarize how we characterized the material of the contaminant and
the accumulation thicknesses on the CCD and FPAFs as a function of time.
First, we measure the thickness of contaminant accumulated on the CCD using G-band data and unique method
(see Appendix~\ref{subsec:contam analysis with G-band} in detail) as shown in Figure~\ref{fig:contam on CCD}.
Next, on the basis of the temporal evolution of the G-band and X-ray intensities, we identify material of contaminant
as the chemical composition of a long-chain organic compound without silicon, refractive index of $\approx$~1.5, and density of $\approx$~1~g~cm$^{-3}$.
Thickness of the pre-filter, oxidization thickness of the pre-filter and FPAFs, and thickness of contaminants accumulated on
each FPAF are also calibrated (see Appendix~\ref{subsec:contam analysis with X-ray} in detail)
as summarized in Table~\ref{tbl:FPAF} and Figure~\ref{fig:contam on FPAF}.
Table~\ref{tbl:contamination} shows the summary of the on-orbit contamination behavior.

\begin{figure}
\centerline{\includegraphics[width=10.0cm,clip=]{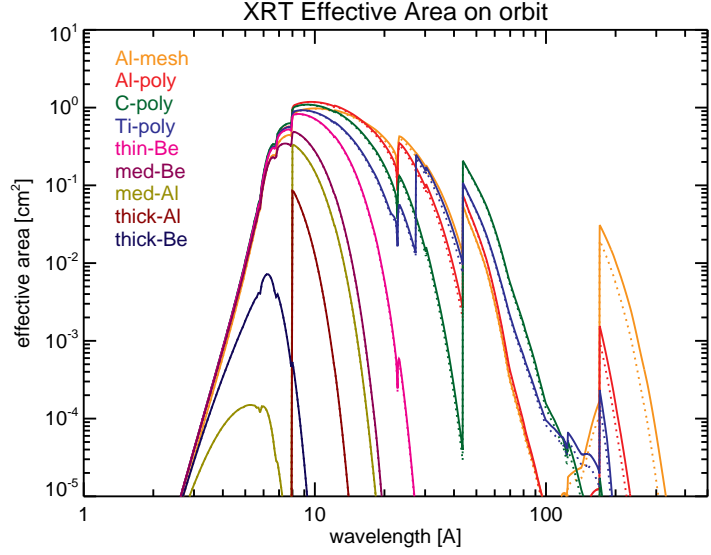}}
\caption{
Effective area of XRT on orbit considering the contaminant accumulated on the FPAFs and CCD.
The solid and dotted lines show the effective area just after CCD bakeout and one month after the bakeout in Phase~3, respectively.
We assume that 800~{\AA} of contaminant accumulated on the CCD one month after the bakeout.
}
\label{fig:eff area}
\end{figure}

Since we now have accurate knowledge of the response parameters (filter thicknesses, contaminant thicknesses, \textit{etc.})
at every epoch after launch of \textit{Hinode},
we can accurately derive the XRT effective area and the temperature response.
(The temperature response is explained in detail in Appendix~\ref{subsec:when XRT observes the solar spectra}.)
The calibrated effective area and temperature response are shown in Figures~\ref{fig:eff area} and \ref{fig:response}, respectively.
Let us next look into the temperature-diagnostics capability of XRT with the calibrated temperature response.

\section{Temperature Diagnostics with Filter-Ratio Method}
\label{sec:temperature}

The coronal-temperature-diagnostic capability is the most significant scientific attribute of XRT.
In Section~\ref{subsec:filter ratio method}, we explain the coronal temperature diagnostics with the filter-ratio method and the estimate
of statistical errors in the derived temperature due to photon noise for the XRT.
The filter-ratio method derives a mean temperature weighted by the filter responses and emission measure
from images observed with two different filters \cite{vai73}.
\inlinecite{har92} and \inlinecite{kan95} discussed coronal temperature diagnostics and
its statistical errors for the \textit{Soft X-ray Telescope} (SXT) onboard \textit{Yohkoh} (\textit{SOLAR-A}: \opencite{tsu91}), respectively.
In Section~\ref{subsec:suitable filter pair}, we summarize the suitable filter pairs of XRT
for coronal temperature diagnostics with the filter-ratio method.
In Section~\ref{subsec:filter ratio temperature}, we discuss the meaning of filter-ratio temperature derived with XRT.

\subsection{Filter-Ratio Method}
\label{subsec:filter ratio method}

\begin{figure}
\centerline{\includegraphics[width=12.0cm,clip=]{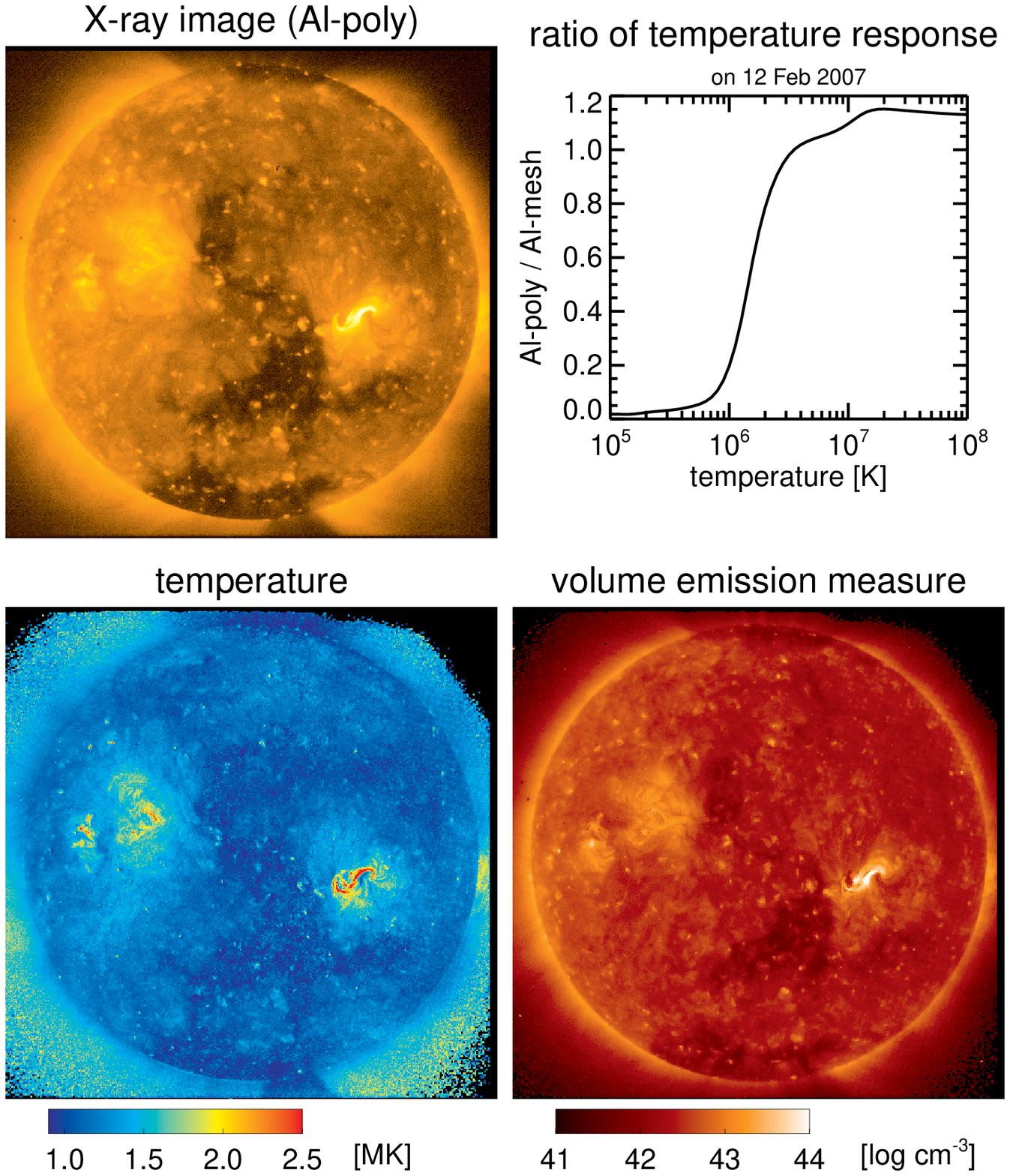}}
\caption{
Example of the XRT temperature diagnostics with the filter-ratio method applied to full-disk images taken with Al-mesh and Al-poly filters
on 12 February 2007.
Top-left panel: the original X-ray data taken with Al-poly filter.
Top-right panel: the ratio of temperature response between Al-poly and Al-mesh filters.
Bottom-left panel: the full-Sun temperature.
Bottom-right panel: the volume emission measure.
}
\label{fig:full sun temperature}
\end{figure}

Assuming that the corona is isothermal, the data number (DN) observed with XRT is
\begin{equation}
DN = \frac{F(T)}{S} \times VEM \times t
\label{eq:DN with VEM}
\end{equation}
for an isothermal corona at a temperature of $T$,
where $F(T)$, $VEM$, $t$, and $S$ are the XRT temperature responses, volume emission measure, exposure time, and
solar area detected in 1~pixel of CCD ($\approx$~560000~km$^2$
derived from plate scale of XRT $\approx$~1.03~arcsec~pixel$^{-1}$ \cite{shi07} and
the relation of 1~arcsec $\approx$~726~km arcsec$^{-1}$ on the solar surface), respectively.
Then, the ratio $R_{ij}(T)$ of normalized DNs per unit time
taken with two different filters is a function of only the coronal temperature $T$:
\begin{equation}
R_{ij}(T) \equiv \frac{DN_i / t_i}{DN_j / t_j} =  \frac{F_i(T) / S \times VEM}{F_j(T) / S \times VEM} = \frac{F_i(T)}{F_j(T)} .
\label{eq:R}
\end{equation}
The subscripts $i$ and $j$ specify the two different filters.

The coronal temperature can be estimated with the observed ratio of normalized DNs as
\begin{equation}
T = R_{ij}^{-1}\left( \frac{DN_i / t_i}{DN_j / t_j} \right)  ,
\label{eq:derived T}
\end{equation}
where $R_{ij}^{-1}$ is the inverse of $R_{ij}$.
Once the coronal temperature is obtained, we can derive the volume emission measure as
\begin{equation}
VEM \equiv \int n_\mathrm{e} n_\mathrm{H} ~\mathrm{d}l \mathrm{d}S = \frac{DN_i / t_i}{F_{i} / S}~\mathrm{or}~\frac{DN_j / t_j}{F_{j} / S} ,
\label{eq:derived VEM}
\end{equation}
where $n_\mathrm{e}$, $n_\mathrm{H}$, and $\mathrm{d}l$ are the electron number density [cm$^{-3}$], hydrogen number density [cm$^{-3}$],
and unit length along the line-of-sight [cm], respectively.
Figure~\ref{fig:full sun temperature} is an example of coronal temperature diagnostics with the filter-ratio method applied to
XRT data taken with Al-mesh and Al-poly filters on 12 February 2007.
The bottom-left and bottom-right panels are the full-Sun temperature and emission measure maps, respectively.

Next, we discuss the estimate of statistical errors in the derived temperature and emission measure due to Poisson noise
of incident photons into XRT.
Since Poisson noise of incident photons dominants over other noise sources (\textit{e.g.}, the camera system noise and dark current noise reported by \inlinecite{kan08}),
we consider only Poisson noise for the error estimate.
Note that \inlinecite{kan95} estimated those errors for \textit{Yohkoh}/SXT.
The differences between our estimate and that of \inlinecite{kan95} are
the adopted solar spectrum database and its units.
We adopt solar photon-number spectra in units of [photon cm$^{-2}$ sec$^{-1}$ sr$^{-1}$ {\AA}$^{-1}$]
from the CHIANTI atomic database version 6.0.1 (\opencite{der97}; \citeyear{der09}: which is the latest version, when we analyzed) with ionization equilibrium from \textsf{chianti.ioneq} \cite{der07} and
abundance from \textsf{sun\_coronal\_ext.abund} \cite{fel92, lan02},
while \inlinecite{kan95} used solar emissivity [erg cm$^{-2}$ sec$^{-1}$ sr$^{-1}$ {\AA}$^{-1}$] from Mewe's spectral data \cite{mew85, mew86}.

First, we derive two conversion factors as preparation for the error estimate.
When XRT observes an isothermal corona at a temperature of $T$ and a volume emission measure of $VEM$
with $p$ pixels of the CCD and $t$ [sec] exposure,
the incident photon number in a range of wavelength from $\lambda$ to $\lambda + d\lambda$ is
\begin{equation}
\mathrm{d}N = \left( \tilde{P}_\mathrm{iso\odot}\left(\lambda, T\right) \times s \times \frac{A_\mathrm{eff}\left(\lambda\right)}{f^2} \times \mathrm{d}\lambda \right) \times \frac{t}{S} \times VEM ,
\label{eq:dN}
\end{equation}
where $\tilde{P}_\mathrm{iso\odot}\left(\lambda, T\right)$ [cm$^{-2}$ sec$^{-1}$ sr$^{-1}$ {\AA}$^{-1}$] is
the photon-number spectrum emitted from an isothermal corona at a temperature of $T$ and an unit of column emission measure (CEM) of 1~cm$^{-5}$,
$s$ is the area of 1 CCD pixel ($13.5^2$~$\mu$m$^2$, see \inlinecite{kan08}),
$A_\mathrm{eff}\left(\lambda\right)$ is the effective area of XRT, and
$f$ is the focal length of XRT (2708~mm, see \inlinecite{gol07}).
The total photon number observed with XRT is, then,
\begin{equation}
N = \int \mathrm{d}N  .
\label{eq:N}
\end{equation}

\begin{figure}
\centerline{\includegraphics[width=10.0cm,clip=]{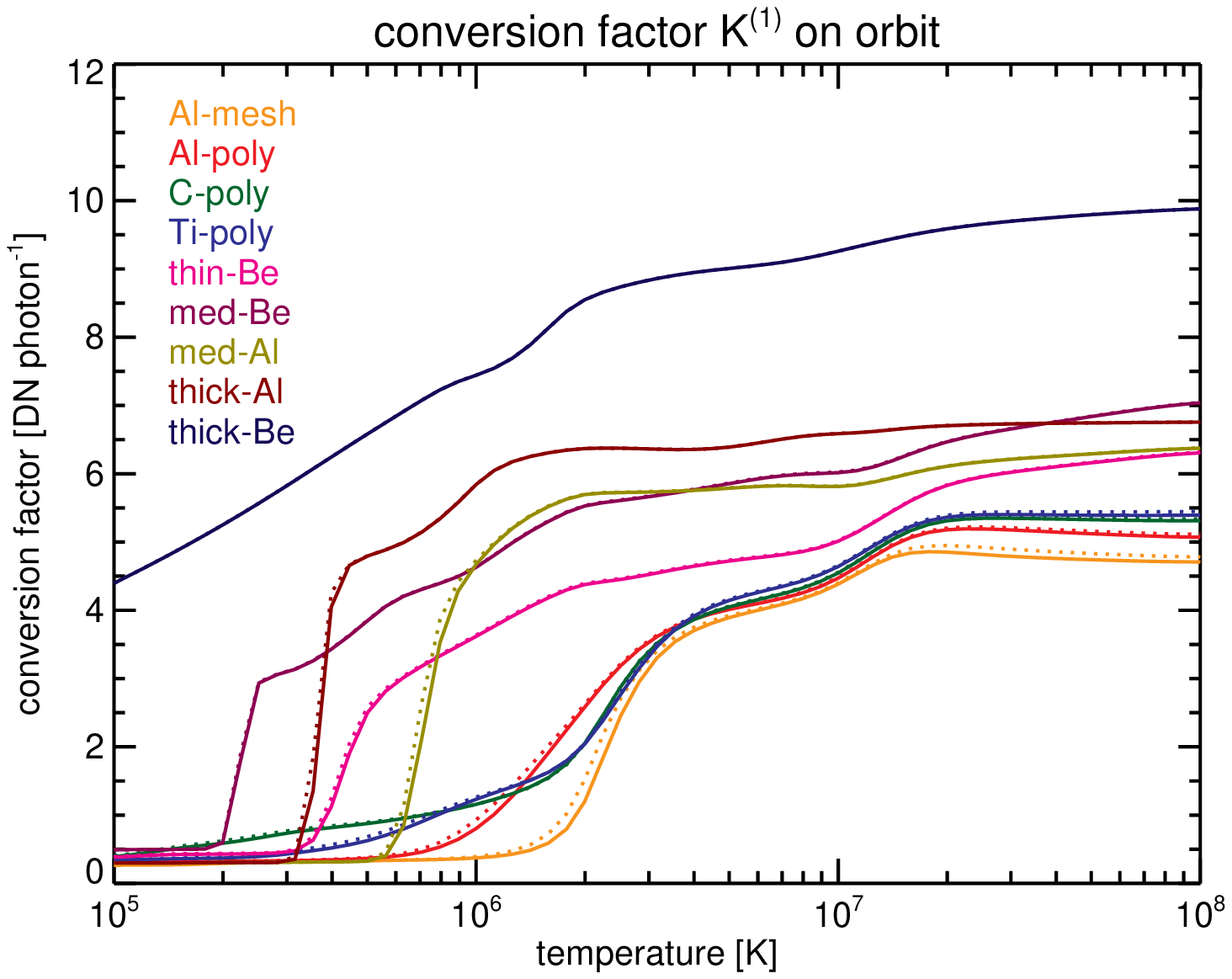}}
\vspace{3.0mm}
\centerline{\includegraphics[width=10.0cm,clip=]{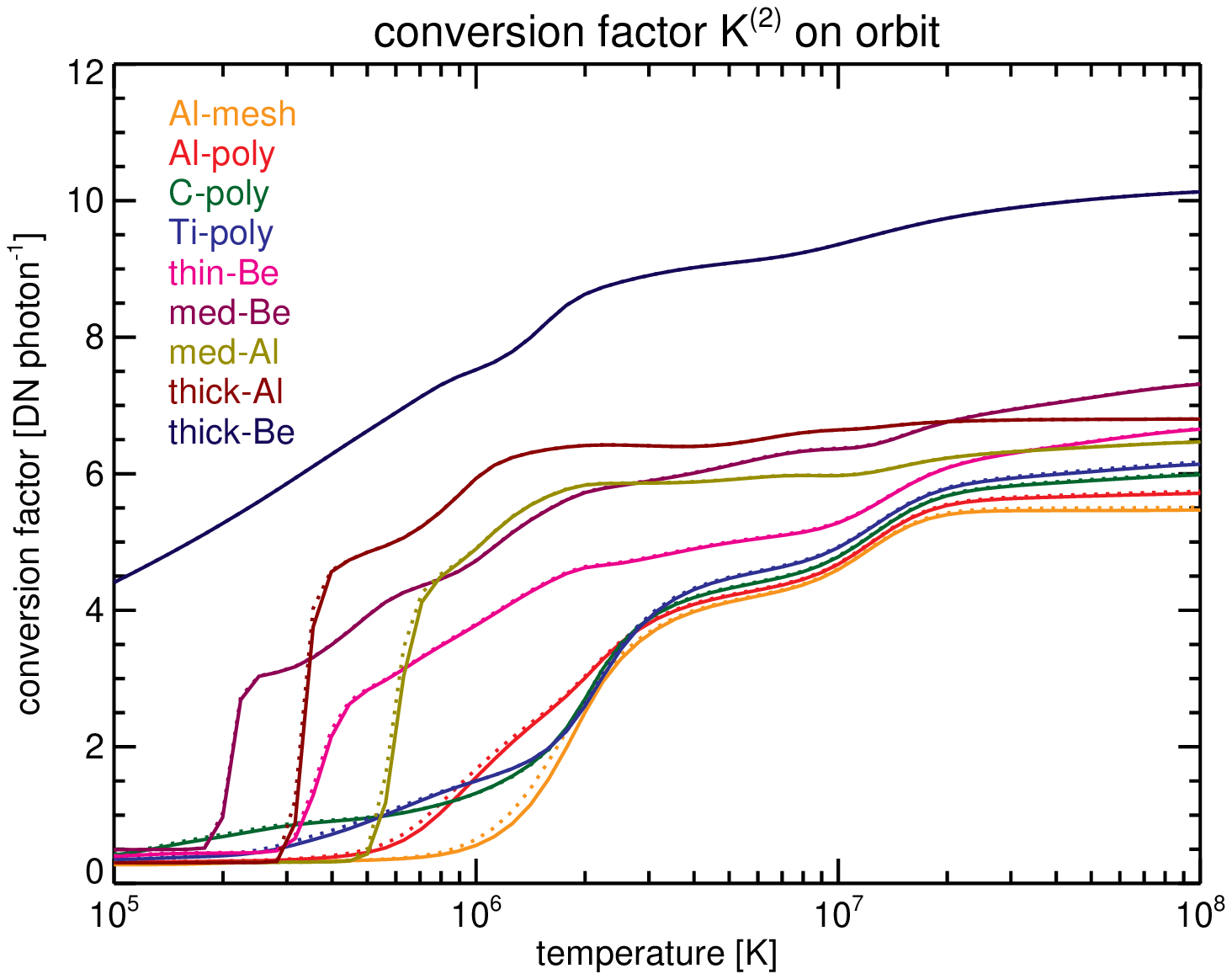}}
\caption{
top panel: Conversion factor [$K^{(1)}\left(T\right)$] of each FPAF from observed data number (DN) to incident photon number [$N$], namely
$N = DN / K^{(1)}\left(T\right)$.
bottom panel: Conversion factor [$K^{(2)}\left(T\right)$] of each FPAF from observed DN to the square of DN error [$\sigma_{DN}$],
namely $\sigma_{DN}^2 = K^{(2)}\left(T\right) \times DN$.
These figures show the conversion factors on orbit considering the contaminant accumulated on the FPAFs and CCD,
where the solid and dotted lines indicate the conversion factors just after CCD bakeout and one month after the bakeout in Phase~3, respectively.
We assume that 800~{\AA} of contaminant accumulated on the CCD one month after the bakeout.
}
\label{fig:K}
\end{figure}

The data number (DN) observed with XRT is derived as
\begin{eqnarray}   
DN\left(T\right) = \int \left( \tilde{P}_\mathrm{iso\odot}\left(\lambda, T\right) \times s \times \frac{A_\mathrm{eff}\left(\lambda\right)}{f^2} \times \frac{h c}{\lambda} \times \frac{1}{e \times 3.65 \times G} \right) \mathrm{d}\lambda  \nonumber \\
\times \frac{t}{S} \times VEM  
\label{eq:DN}
\end{eqnarray}
(from Equations~(\ref{eq:DN with VEM}) and (\ref{eq:F}) in Appendix~\ref{subsec:when XRT observes the solar spectra}),
where $h$, $c$, $e$, and $G$ are Planck's constant, speed of light, elementary electric charge, and the system gain of the CCD camera, which is 57.5 [electron DN$^{-1}$] in the XRT case \cite{kan08}, respectively.
When we define $K^{(1)}\left(T\right)$ as
\begin{equation}
K^{(1)}\left(T\right) \equiv \frac
{\int \left( \tilde{P}_\mathrm{iso\odot}\left(\lambda, T\right) \times A_\mathrm{eff}\left(\lambda\right) \times \frac{h c}{\lambda} \times \frac{1}{e \times 3.65 \times G} \right) \mathrm{d}\lambda}
{\int \left( \tilde{P}_\mathrm{iso\odot}\left(\lambda, T\right) \times A_\mathrm{eff}\left(\lambda\right) \right) \mathrm{d}\lambda} ,
\label{eq:K(1)}
\end{equation}
we can rewrite Equation~(\ref{eq:N}) as
\begin{equation}
N = \frac{DN}{K^{(1)}\left(T\right)} .
\label{eq:N-1}
\end{equation}
$K^{(1)}\left(T\right)$ is a conversion factor from observed DN to incident photon number.
The top panel of Figure~\ref{fig:K} indicates $K^{(1)}\left(T\right)$ for each filter.

Since $\mathrm{d}N$ follows a Poisson distribution, the standard deviation $\mathrm{d}\sigma$ of $\mathrm{d}N$ is expressed as
\begin{equation}
\mathrm{d}\sigma = \sqrt{\mathrm{d}N} .
\label{eq:sigma_dN}
\end{equation}
Since Equation~(\ref{eq:DN}) is rewritten as
\begin{equation}
DN = \int \left( \frac{h c}{\lambda} \times \frac{1}{e \times 3.65 \times G} \right) \mathrm{d}N ,
\label{eq:I-2}
\end{equation}
we can derive the error $\sigma_{DN}$ of $DN$ as
\begin{equation}
\sigma_{DN} = \sqrt{ \int \left( \frac{h c}{\lambda} \times \frac{1}{e \times 3.65 \times G} \right)^2 \left(\mathrm{d}\sigma\right)^2} = \sqrt{K^{(2)}\left(T\right) DN} ,
\label{eq:sigma_I}
\end{equation}
where
\begin{equation}
K^{(2)}\left(T\right) \equiv \frac
{\int \left[ \tilde{P}_\mathrm{iso\odot}\left(\lambda, T\right) \times A_\mathrm{eff}\left(\lambda\right) \times \left(\frac{h c}{\lambda} \times \frac{1}{e \times 3.65 \times G} \right)^{2} \right] \mathrm{d}\lambda}
{\int \left( \tilde{P}_\mathrm{iso\odot}\left(\lambda, T\right) \times A_\mathrm{eff}\left(\lambda\right) \times \frac{h c}{\lambda} \times \frac{1}{e \times 3.65 \times G} \right) \mathrm{d}\lambda} .
\label{eq:K(2)}
\end{equation}
$K^{(2)}\left(T\right)$ is a conversion factor from observed DN to the square of the DN error.
The bottom panel of Figure~\ref{fig:K} shows $K^{(2)}\left(T\right)$ for each filter.

We note that these two conversion factors $K^{(1)}$ and $K^{(2)}$ vary
in a part of detectable temperature range of XRT from $\approx$~1~MK to $\approx$~5~MK (see Figure~\ref{fig:K}),
while those factors for \textit{Yohkoh}/SXT was almost constant at three
throughout the whole detectable temperature range of $\approx$~2~MK to $>$~10~MK \cite{kan95}.

Following the Appendix of \inlinecite{kan95}, we can estimate the error of derived temperature and volume emission measure from
these conversion factors as
\begin{equation}
\frac{\sigma_T}{T} = \left| \frac{\mathrm{d}\ln{R_{ij}\left(T\right)}}{\mathrm{d}\ln{T}} \right|^{-1} \sqrt{ \frac{K_i^{(2)}\left(T\right)}{DN_i} + \frac{K_j^{(2)}\left(T\right)}{DN_j} }
\label{eq:sigma_T}
\end{equation}
and
\begin{equation}
\frac{\sigma_{VEM}}{VEM} = \left| \frac{\mathrm{d}\ln{R_{ij}\left(T\right)}}{\mathrm{d}\ln{T}} \right|^{-1} \sqrt{ \left[\frac{\mathrm{d}\ln{F_j\left(T\right)}}{\mathrm{d}\ln{T}}\right]^2 \frac{K_i^{(2)}\left(T\right)}{DN_i} + \left[\frac{\mathrm{d}\ln{F_i\left(T\right)}}{\mathrm{d}\ln{T}}\right]^2 \frac{K_j^{(2)}\left(T\right)}{DN_j} } .
\label{eq:sigma_VEM}
\end{equation}
These relations show how the errors in observed data number [$DN$] propagate to the errors in the derived temperature [$T$]
and volume emission measure [$VEM$].

We note that the temperature in the right-hand side of Equations~(\ref{eq:sigma_T}) and (\ref{eq:sigma_VEM}) should be the actual temperature of the coronal plasma.
However, we get the estimated coronal temperature with the filter-ratio method.
Hence, we used the derived temperature to estimate the errors instead of the actual temperature.
In the case of poor signal-to-noise ratio (S/N) data, the difference between actual and derived temperatures may be larger than in the case of good S/N data,
and the above treatment of $T$ in error estimate will cause the additional error.
We confirmed that 10\% S/N data makes this additional error sufficiently small.

On the basis of the error estimate discussed above, the full-Sun temperature map in Figure~\ref{fig:full sun temperature} is created.
In this map, the poor S/N pixels, namely pixels in the coronal hole and quiet Sun, are summed ($8 \times 8$ pixels binning maximum) to improve the S/N.
Using this good S/N data, the coronal temperature with the filter-ratio method can be derived
with less than 20\% errors across the Sun.
This diagnostic capability is the most significant scientific feature of XRT.

\subsection{Suitable Filter Pairs for Coronal Temperature Diagnostics with the Filter-Ratio Method}
\label{subsec:suitable filter pair}

\begin{figure}
\centerline{\includegraphics[width=12.0cm,clip=]{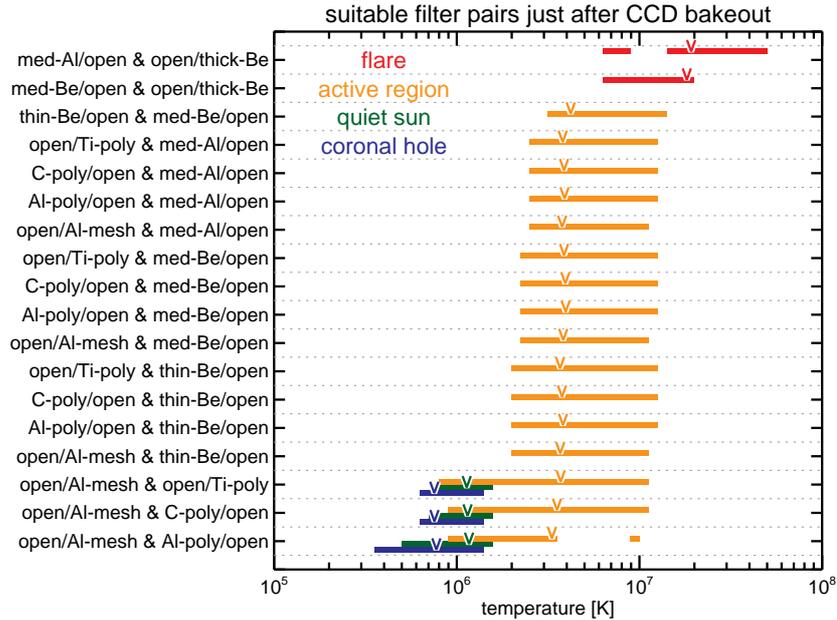}}
\caption{
Suitable filter pairs to derive the temperature with the filter-ratio method and their useful temperature ranges
just after the CCD bakeout in Phase~3, namely there is no contamination on the CCD but
the contaminant accumulated on the FPAFs.
The red, orange, green, and blue bars show the useful temperature ranges for flare, active region, quiet Sun, and coronal hole, respectively.
The ``\textsf{v}" marks indicate the filter-ratio temperatures derived from the DEMs of each region.
}
\label{fig:suitable filter pair 0}
\end{figure}

\begin{figure} 
\centerline{\includegraphics[width=12.0cm,clip=]{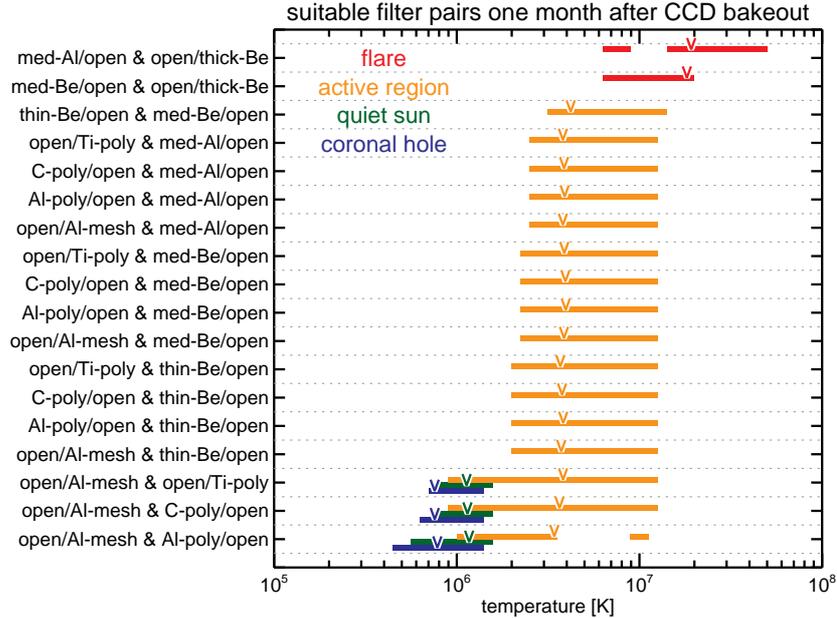}}
\caption{
Suitable filter pairs to derive the temperature with the filter-ratio method and their useful temperature ranges
one month after the CCD bakeout in Phase~3, namely there is contamination
not only on the FPAFs but also on the CCD.
We assume that 800~{\AA} of contaminant accumulated on the CCD.
The others are the same as in Figure~\ref{fig:suitable filter pair 0}.
}
\label{fig:suitable filter pair 800}
\end{figure}

\begin{table}
\caption{Typical values of coronal structures}
\label{tbl:coronal structure}
\begin{tabular}{cccccc}
\hline
coronal           & DEM                 & electron               & exposure & binning       \\
structure         & reference           & number density         & time     & size          \\
\hline
 flare            & \inlinecite{der79}  & $10^{11.0}$~cm$^{-3}$  & 1 sec    & $1 \times 1$  \\
 active region    & \inlinecite{war10}  &  $10^{9.4}$~cm$^{-3}$  & 10 sec   & $2 \times 2$  \\
 quiet Sun        & \inlinecite{bro06}  &  $10^{8.3}$~cm$^{-3}$  & 30 sec   & $4 \times 4$  \\
 coronal hole     & \inlinecite{ver78}  &  $10^{8.0}$~cm$^{-3}$  & 64 sec   & $8 \times 8$  \\
\hline
\end{tabular}
\end{table}

On the basis of the discussion in the previous section,
we present suitable filter pairs to derive temperatures with filter-ratio method for various coronal structures:
flares, active regions, quiet Sun, and coronal holes.
The typical values for each coronal structure are summarized in Table~\ref{tbl:coronal structure}.
The electron number density is calculated as the integral over temperature of DEM with the assumption that the line-of-sight depth is $10^5$~km.
The exposure time and binning size is determined from the time scale and spatial scale of each structure, respectively.
Figures~\ref{fig:suitable filter pair 0} and \ref{fig:suitable filter pair 800} show
the ``suitable filter pairs" and their ``useful temperature ranges"
for some representative cases of contamination accumulated on the CCD and the FPAFs.
The ``useful temperature range" means the range where a filter pair can estimate the temperature
(within 20\% accuracy) with the filter-ratio method for a given coronal structure
with the typical exposure time and binning size (see Table~\ref{tbl:coronal structure}).
The ``suitable filter pair" means that the pair has a wide ``useful temperature range" as shown in Figures~\ref{fig:suitable filter pair 0} and \ref{fig:suitable filter pair 800}.
For example, the Al-mesh and Ti-poly filter pair can have a wide temperature coverage for a wide variety of coronal structures,
coronal holes, quiet Sun, and active regions. Thus, this pair is suitable for full-disk synoptic observations.

\begin{figure}
\centerline{\includegraphics[width=12.0cm,clip=]{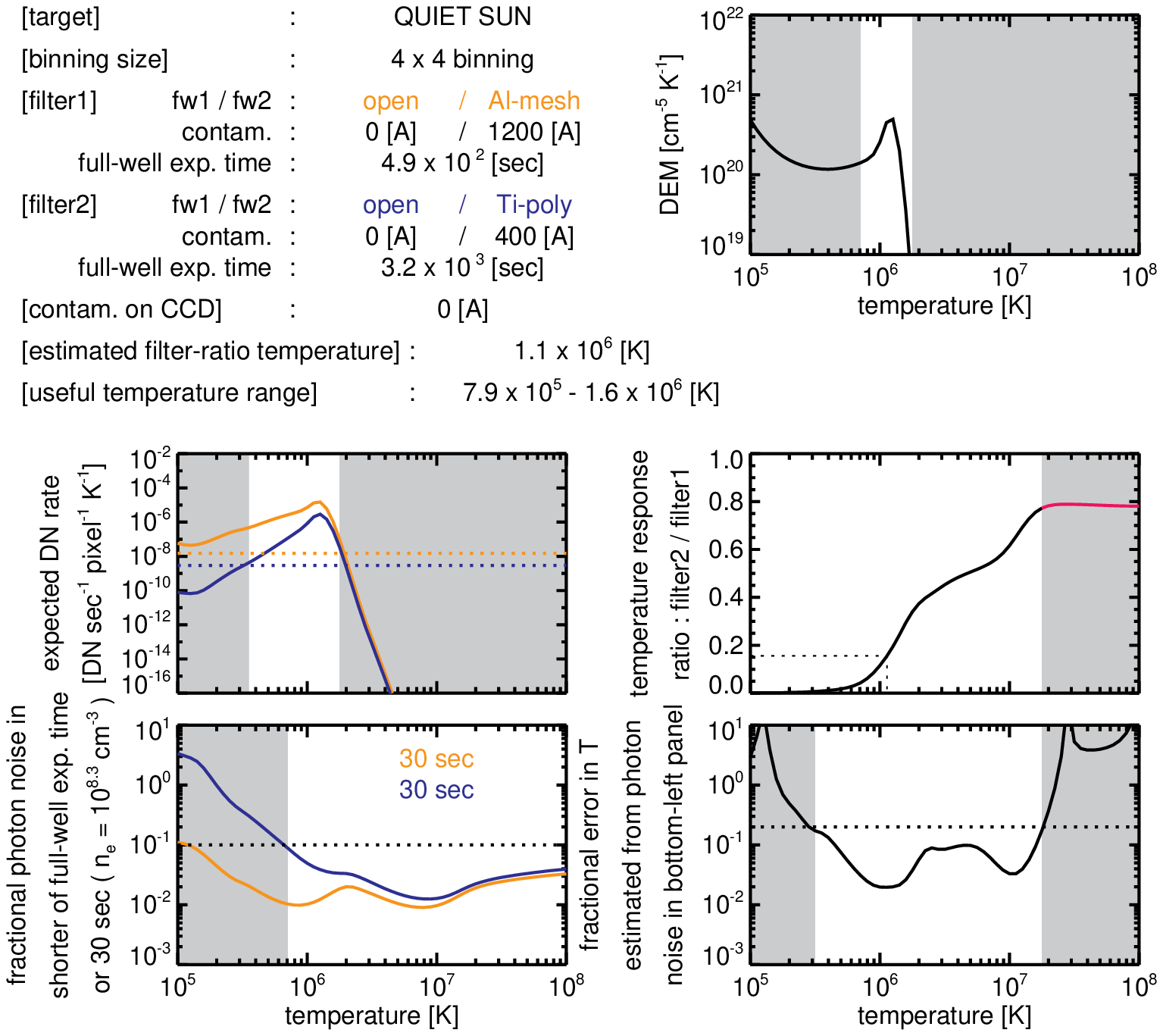}}
\vspace{3.0mm}
\caption{
Evaluation of the Al-mesh and Ti-poly filter pair for coronal temperature diagnostics with the filter-ratio method.
This is the case of the quiet Sun just after CCD bakeout in Phase~3.
This pair is suitable for temperature diagnostics in the quiet Sun.
}
\label{fig:filter pair1}
\end{figure}

\begin{figure}
\centerline{\includegraphics[width=12.0cm,clip=]{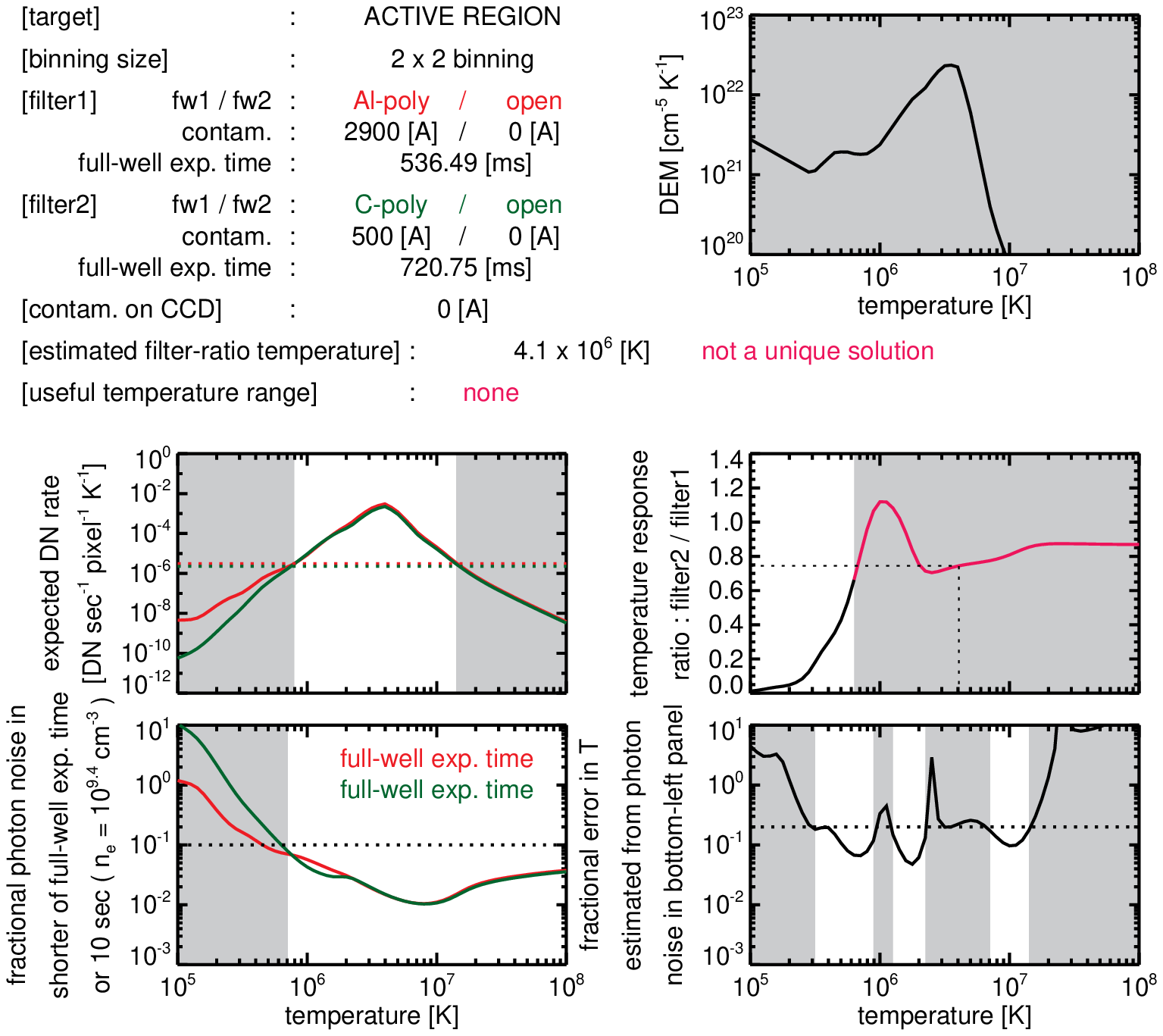}}
\vspace{3.0mm}
\caption{
Evaluation of the Al-poly and C-poly filter pair for coronal temperature diagnostics with the filter-ratio method.
This is the case of the active region just after CCD bakeout in Phase~3.
This pair is not suitable for temperature diagnostics in the active region.
}
\label{fig:filter pair2}
\end{figure}

Let us now examine how to derive the suitable filter pairs and their useful temperature ranges.
Figures~\ref{fig:filter pair1} and \ref{fig:filter pair2} are examples of
the filter-pair study for quiet Sun and active region, respectively.
A summary of the study is described in the top-left of these figures.
Here [target] indicates the coronal structure to be analyzed.
The typical differential emission measure (DEM) of such a target is plotted in the top-right side of these figures
(the references of typical DEMs are summarized in Table~\ref{tbl:coronal structure}).
The row with [binning size] is the recommended binning size to collect enough photons.
For a bright flare, [binning size] is $1 \times 1$ binning, that is no binning.
Meanwhile, for the dark coronal hole, [binning size] is $8 \times 8$ binning.
The binning size used in our calculation is shown in Table~\ref{tbl:coronal structure}.
The rows with [filter1] and [filter2] summarize the evaluated filter pair.
Labels ``fw1" and ``fw2" indicate used FPAFs on filter wheels~1 and 2, respectively.
The label ``contam." gives the the accumulated thickness of contaminant on each FPAF.
The label ``full-well exp. time" expresses the exposure time necessary to obtain the full-well intensity of the XRT CCD,
about 200000~electrons $\approx$~3500~DN, when XRT observes the coronal structure shown in [target]
with [filter1] or [filter2].
The full-well intensity means the maximum intensity to keep the linearity
between the incident photon energy and output intensity as shown by Equation~(\ref{eq:DN2}).
In the case of the intensity exceeding the full-well intensity, the actual output intensity becomes smaller than
the intensity expected from Equation~(\ref{eq:DN2}).
The row with [contam. on CCD] indicates the accumulated thickness of contaminant on the CCD.

Next, we explain six conditions that a filter pair shown in [filter1] and [filter2] is
suitable for the temperature diagnostics with the filter-ratio method.
The six conditions, (A)\,--\,(F), are given below.
The four panels at the bottom part of Figures~\ref{fig:filter pair1} and \ref{fig:filter pair2} show the temperature ranges (white areas)
which satisfy the following Conditions~(A)\,--\,(D).
\begin{itemize}
\item[\textbf{(A) Expected DN rate at each temperature:}] \mbox{}\par
The middle-left panel shows the expected DN rates (whose unit is DN sec$^{-1}$ pixel$^{-1}$ K$^{-1}$) at each temperature
when XRT with [filter1] and [filter2] observe the [target] which has the DEM shown in the top-right side of the figures.
Since the dynamic range of the XRT CCD is about 1000, we assume that XRT can detect temperature components
where DN rate exceeding 1/1000 of the peak DN rate in each expected DN rate plot.
Condition~(A) is that the expected DN rate exceeds the 1/1000 of the peak DN rate (shown by dotted lines) in both cases of [filter1] and [filter2].
\item[\textbf{(B) Low photon noise:}] \mbox{}\par
The bottom-left panel presents the photon noise, when XRT observes an isothermal plasma at each temperature given by the horizontal axis of the panel.
The electron number density at each temperature is set to be a constant value as the typical value described in Table~\ref{tbl:coronal structure} and vertical axis.
Hence, this plot shows how much photon noise is expected in the observed data when we obtain the filter-ratio temperature from such data.
The photon noise can be calculated with Equation~(\ref{eq:sigma_I}) as a function of temperature and DN.
The DN is derived from temperature, VEM, and exposure time as shown in Equation~(\ref{eq:DN with VEM}).
For the calculation of VEM, we use the above electron number density
and $10^5$~km as the line-of-sight depth. The effect of binning is also considered.
The exposure time adopted for this evaluation is shorter of the typical exposure time (see Table~\ref{tbl:coronal structure}) or
the exposure time to obtain the full-well intensity (see the label ``full-well exp. time").
The adopted exposure time for each filter are described in each filter color in the bottom-left panel.
The horizontal dotted line corresponds to 10\% photon noise (S/N~$= 10$).
We consider that the good S/N data (S/N~$> 10$) should be used for the temperature diagnostics,
because poor S/N data makes the error large as mentioned above.
Condition~(B) is that the photon noise should be lower than 10\%.
We note that Condition~(B) is not so strict, because we can sum many images to adequate enough photons.
When we collect enough photons with a single image, Condition~(B) should be satisfied.
\item[\textbf{(C) Profile of temperature response ratio:}] \mbox{}\par
The middle-right panel indicates the temperature response ratio of [filter1] and [filter2] shown by Equation~(\ref{eq:R}).
Using this ratio, we can estimate the coronal temperature from the observed intensity ratio.
The horizontal dashed lines in Figures~\ref{fig:filter pair1} and \ref{fig:filter pair2}
show the expected intensity ratio, when XRT observes [target] with [filter1] and [filter2].
In the case of Figure~\ref{fig:filter pair1}, the temperature is uniquely derived as shown by the vertical dashed line.
The derived temperature is described by [estimated filter-ratio temperature] in the top-left panel of the figure.
However, in the case of Figure~\ref{fig:filter pair2}, the temperature cannot be uniquely determined.
In this case, [estimated filter-ratio temperature] shows the closest temperature to the peak temperature of DEM in the top-right panel as a reference.
The temperature range where the intensity ratio gives the alternative temperatures is indicated by the magenta line and gray area. 
Condition~(C) is that the temperature is derived uniquely from the filter-ratio method.
\item[\textbf{(D) Error in estimated temperature:}] \mbox{}\par
The bottom-right panel shows the fractional error in estimated temperature caused by 10\% photon noise.
The error in estimated temperature is calculated with Equation~(\ref{eq:sigma_T}) as a function of temperature and DN.
The DN is derived with the method described in Condition~(B).
We take an estimate error of 20\%, which is twice the value of the photon noise, as the threshold for a reasonable temperature estimate.
This is Condition~(D).
The horizontal dotted line is located at the 20\% error in estimated temperature.
\item[\textbf{(E) No saturation:}] \mbox{}\par
The shortest exposure time for XRT is 1~ms.
Hence, the ``full-well exp. time" should be longer than 1~ms.
For bright coronal structures, namely active regions and flares, the suitable filter pairs are limited by this condition.
\item[\textbf{(F) Filter-ratio temperature:}] \mbox{}\par
The useful temperature range with the filter-ratio method is derived from satisfying the Conditions~(A)\,--\,(E)
as shown by [useful temperature range] and in white area in the top-right DEM panel.
We call the filter pair suitable
if the [estimated filter-ratio temperature] falls within this [useful temperature range].
This is Condition~(F).
\end{itemize}
The suitable filter pairs for coronal temperature diagnostics with filter-ratio method should satisfy all of the Conditions~(A)\,--\,(F).
These suitable filter pairs depend on the target, filter pair, and accumulation of contaminant.
Examples are summarized in Figures~\ref{fig:suitable filter pair 0} and \ref{fig:suitable filter pair 800}
for some representative cases of contamination accumulated on the CCD and the FPAFs
with their [useful temperature range] and [estimated filter-ratio temperature] shown by bars and ``\textsf{v}" marks, respectively.
These suitable filter pairs and their useful temperature range
are useful for understanding the temperature diagnostic capability of XRT
and making the XRT observation plan for the temperature analysis.
In the analysis with actual data taken with XRT, it is also possible to estimate the temperature
with filter pairs other than those we described here as suitable filter pairs,
since the suitable filter pairs are derived with the typical values of coronal structures (DEMs, density and binning size) summarized in Table~\ref{tbl:coronal structure}.

We note that since the actual solar corona may have a multi-temperature structures,
the photon noise for Condition~(B) should be estimated with a multi-temperature (DEM).
However, we consider that the treatment in Condition~(B) is valid, since we confirmed that the photon noise derived with the filter-ratio temperature well represents
one derived with DEM, at least, at the estimated filter-ratio temperature as will be discussed in Appendix~\ref{sec:meaning of filter-ratio noise}.
In Condition~(D), we evaluated the error in estimated filter-ratio temperature. Although the error is derived under the assumption
that the corona is the single filter-ratio temperature, the derived error is useful as the accuracy-index of estimated filter-ratio temperature
which depends on the photon noise and the shape of temperature responses in each filter.

\subsection{Meaning of Filter-Ratio Temperature Derived with XRT}
\label{subsec:filter ratio temperature}

\begin{figure}
\centerline{\includegraphics[width=8.0cm,clip=]{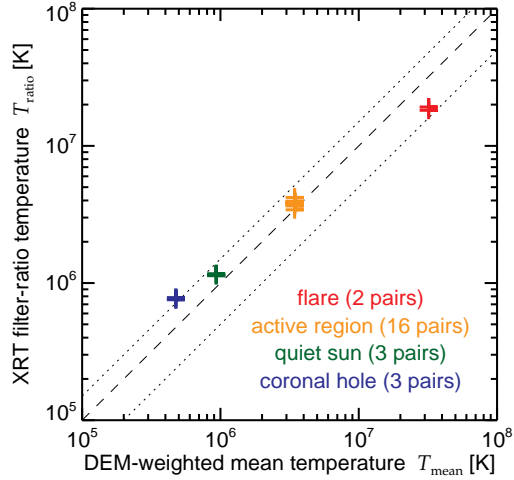}}
\caption{
Correlation between DEM-weighted mean temperature [$T_\mathrm{mean}$] and XRT filter-ratio temperature [$T_\mathrm{ratio}$]
for four DEMs shown in Table~\ref{tbl:coronal structure}.
This plot is the case when 1 month passed after the CCD bakeout, \textit{i.e.} 800~{\AA} of contaminant on the CCD.
The dashed and dotted lines indicate the positions where $T_\mathrm{ratio}$ is equal to $T_\mathrm{mean}$,
and where the difference between $T_\mathrm{mean}$ and $T_\mathrm{ratio}$ is 50\%, respectively.
The number of analyzed filter pairs is shown in parenthesis.
}
\label{fig:filter ratio temperature}
\end{figure}

The actual solar corona may have a multi-temperature structures, and
the temperature derived with the filter-ratio method is a mean temperature weighted by the filter responses.
Unlike filter-ratio temperatures from narrow-band instruments (see discussion in \inlinecite{mar02}),
filter-ratio temperatures from broad-band instruments (\textit{e.g.}, \textit{Hinode}/XRT and \textit{Yohkoh}/SXT) are expected to provide
a temperature which is close to the mean temperature [$T_\mathrm{mean} \equiv \frac{\int T \times DEM ~\mathrm{d}T}{\int DEM ~\mathrm{d}T}$]
weighted by differential emission measure (DEM) of the coronal observation target as demonstrated by \inlinecite{act99} for \textit{Yohkoh}/SXT.
In the case of \textit{Hinode}/XRT, Figure~\ref{fig:filter ratio temperature} shows the correlation between
the DEM-weighted mean temperature [$T_\mathrm{mean}$] and filter-ratio temperature [$T_\mathrm{ratio}$]
for four DEMs (regions) summarized in Table~\ref{tbl:coronal structure}.
The $T_\mathrm{ratio}$ is calculated for all suitable filter pairs
which are investigated in Section~\ref{subsec:suitable filter pair} and summarized in Figure~\ref{fig:suitable filter pair 800}.
The values of $T_\mathrm{ratio}$ derived with different filter pairs are different, \textit{i.e.}
the thinner and thicker filter pairs give the lower and higher $T_\mathrm{ratio}$, respectively
(see the ``\textsf{v}" marks in Figures~\ref{fig:suitable filter pair 0} and \ref{fig:suitable filter pair 800}).
Additionally, $T_\mathrm{ratio}$ is higher than $T_\mathrm{mean}$ in the lower temperature region ($T < 1$~MK: coronal hole),
and $T_\mathrm{ratio}$ is lower than $T_\mathrm{mean}$ in the higher temperature region ($T > 10$~MK: flare).
This is caused by the bias of XRT temperature response (Figure~\ref{fig:response}) which has a sensitivity to mainly 1\,--\,10~MK plasma.
However, the difference between $T_\mathrm{mean}$ and $T_\mathrm{ratio}$ is less than 50\% (within two dotted lines in Figure~\ref{fig:filter ratio temperature}) for all filter pairs.
Hence, we claim that the filter-ratio temperature with XRT (the broad-band instruments) is useful for the quantitative analysis, \textit{e.g.}, coronal energetics.
Additionally, the filter-ratio method has the following advantages:
\begin{itemize}
\item[\textit{i})] the filter-ratio method applied to XRT data can estimate a coronal temperature over a wide field of view at once, in contrast to a spectrometer.
\item[\textit{ii})] the filter-ratio method can investigate rapid temporal evolution of coronal temperature with a time scale well below one minute,
since XRT can quickly take a data set with a pair of filters.
\end{itemize}
Hence, the temperature diagnostics obtained with the filter-ratio method and the advanced DEM analysis
with data sets taken with multiple filters of the XRT or \textit{EUV Imaging Spectrometer} (EIS) onboard \textit{Hinode} are complementary to each other.

\section{Summary}
\label{sec:summary}

We calibrated the effective area and coronal temperature response of the XRT
using ground-based test data and also on-orbit data in X-rays and visible light,
with the effect of contamination taken into account to the best of our knowledge.
The calibrated thicknesses of pre-filter and FPAFs are summarized in Table~\ref{tbl:FPAF}.
The time-varying thicknesses of the contaminant accumulated on the CCD and FPAFs are shown in
Figures~\ref{fig:contam on CCD} and \ref{fig:contam on FPAF}, respectively.

\begin{figure}
\centerline{\includegraphics[width=12.0cm,clip=]{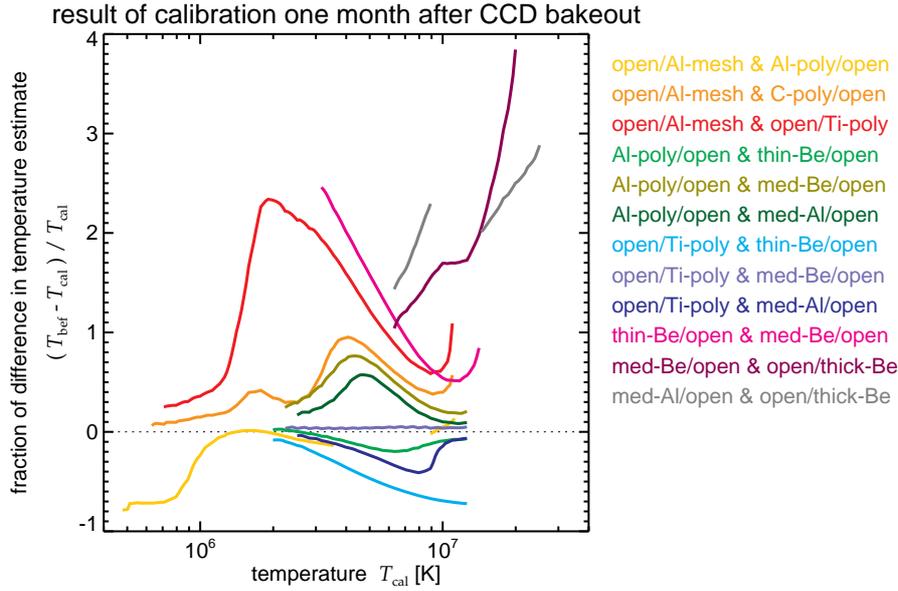}}
\caption{
Comparison between filter-ratio temperatures with the calibrated results in this article [$T_\mathrm{cal}$] and
with the instrumental parameters before calibration [$T_\mathrm{bef}$].
The horizontal and vertical axes of this plot show
$T_\mathrm{cal}$ and
the fraction of the difference in the temperature estimate defined as $(T_\mathrm{bef} - T_\mathrm{cal}) / T_\mathrm{cal}$, respectively.
The dotted line indicates the position where these temperatures are the same.
}
\label{fig:result 800}
\end{figure}

Here, we examine how much our calibration results improve the coronal-temperature-diagnostic capability of XRT.
Figure~\ref{fig:result 800} shows a comparison between filter-ratio temperatures
with our calibrated results and with the instrumental parameters before calibration \cite{gol07}
for some suitable filter pairs (see Figure~\ref{fig:suitable filter pair 800}).
This plot is created as following:
We derive the expected intensity ratios with our calibrated parameters
at each temperature [$T_\mathrm{cal}$] in the useful temperature range of each filter pair (see Figure~\ref{fig:suitable filter pair 800}).
Next, we derive the temperature [$T_\mathrm{bef}$] with the above-derived ratios and instrumental parameters before calibration.
The vertical axis shows the fraction of the difference in the temperature estimate defined as $(T_\mathrm{bef} - T_\mathrm{cal}) / T_\mathrm{cal}$.
Note that this plot varies with the accumulation of contaminants on the CCD and FPAFs as a function of time.
This plot shows the case when one month passed after the CCD bakeout, \textit{i.e.} 800~{\AA} of contaminant was deposited on the CCD.
While some filters happen to provide consistent temperature estimates in some temperature ranges,
for most filters there are large deviations as illustrated in Figure~\ref{fig:result 800}.
Hence, we claim that our calibration result is indispensable for accurately estimating coronal temperatures with XRT.

On the basis of our calibration results,
we reviewed the coronal-temperature-diagnostic capability of XRT with the filter-ratio method.
XRT has a wide variety of suitable filter pairs to estimate coronal temperatures with the filter-ratio method.
For reference, we briefly mention the multi-temperature diagnostics.
Using simultaneous data from more than two filters allows one to obtain more information on
the temperature distribution, and to resolve ambiguity in the results of the filter-ratio method
(\textit{i.e.}, it increases the ability to pass Condition~(C) in Section~\ref{subsec:suitable filter pair}).
With many XRT filters, at least more than four filters \cite{gol07}, one can try to derive the differential emission measure (DEM).
In this situation, the calculation of DEM with XRT data including the error estimate itself is the topic of interest (\opencite{sch09a}; \citeyear{sch09b}; \citeyear{sch10}).
The XRT is a powerful instrument not only to observe the morphology of coronal structures in detail
but also to derive coronal physical quantities.

\subsection{Softwares for the coronal temperature diagnostics with XRT}
\label{sec:SSW}

Lastly, we itemize the software used for the XRT calibrations and the coronal temperature diagnostics described in this article.
These routines are distributed in the \textsf{hinode/xrt} directory tree of \textsf{Solar Software} \cite{fre98}.

\begin{itemize}
\item[\textsf{xrt\_time2contam.pro}] \mbox{}\par gives the total thickness of contaminant on the FPAFs and the CCD as a function of time (Figures~\ref{fig:contam on CCD} and \ref{fig:contam on FPAF}).
\item[\textsf{xrt\_eff\_area.pro}]   \mbox{}\par gives the XRT effective area (Figure~\ref{fig:eff area}).
\item[\textsf{xrt\_flux.pro}]        \mbox{}\par gives the XRT temperature response (Figure~\ref{fig:response}).
\item[\textsf{xrt\_cvfact.pro}]      \mbox{}\par gives the conversion factor [$K^{(1)}\left(T\right)$] from observed DN to incident photon number (Equation~(\ref{eq:N-1}) and the top panel of Figures~\ref{fig:K}).
                                        With the \textsf{/error} keyword, the conversion factor [$K^{(2)}\left(T\right)$] from observed DN to the square of DN error is given (Equation~(\ref{eq:sigma_I}) and the bottom panel of Figures~\ref{fig:K}).
\item[\textsf{xrt\_teem.pro}]        \mbox{}\par returns the estimated coronal temperature and emission measure with the filter-ratio method (Figure~\ref{fig:full sun temperature}).
\end{itemize}

The XRT team have provided more software than the above. For example,
\begin{itemize}
\item[\textsf{make\_xrt\_wave\_resp.pro}] \mbox{}\par gives the XRT effective areas, and provides option for user-specified spectral emission models.
\item[\textsf{make\_xrt\_temp\_resp.pro}] \mbox{}\par gives the XRT temperature responses, and provides option for user-specified spectral emission models.
\end{itemize}
The above two softwares are useful to examine the effect of uncertainties in the atomic physics, \textit{i.e.} difference in spectral emission models, on the temperature analysis.
For the DEM analysis,
\begin{itemize}
\item[\textsf{xrt\_dem\_iterative2.pro}]  \mbox{}\par gives DEM solutions for multi-filter datasets, and provides option for Monte Carlo-style analysis of uncertainties.
\end{itemize}
The details how to analyze the XRT data with those softwares are described in
the XRT analysis guide (\url{http://xrt.cfa.harvard.edu/resources/documents/XAG/XAG.pdf}).

\appendix

\section{Ground-Based Calibration}
\label{sec:ground-based calibration}

\begin{figure}
\centerline{\includegraphics[width=10.0cm,clip=]{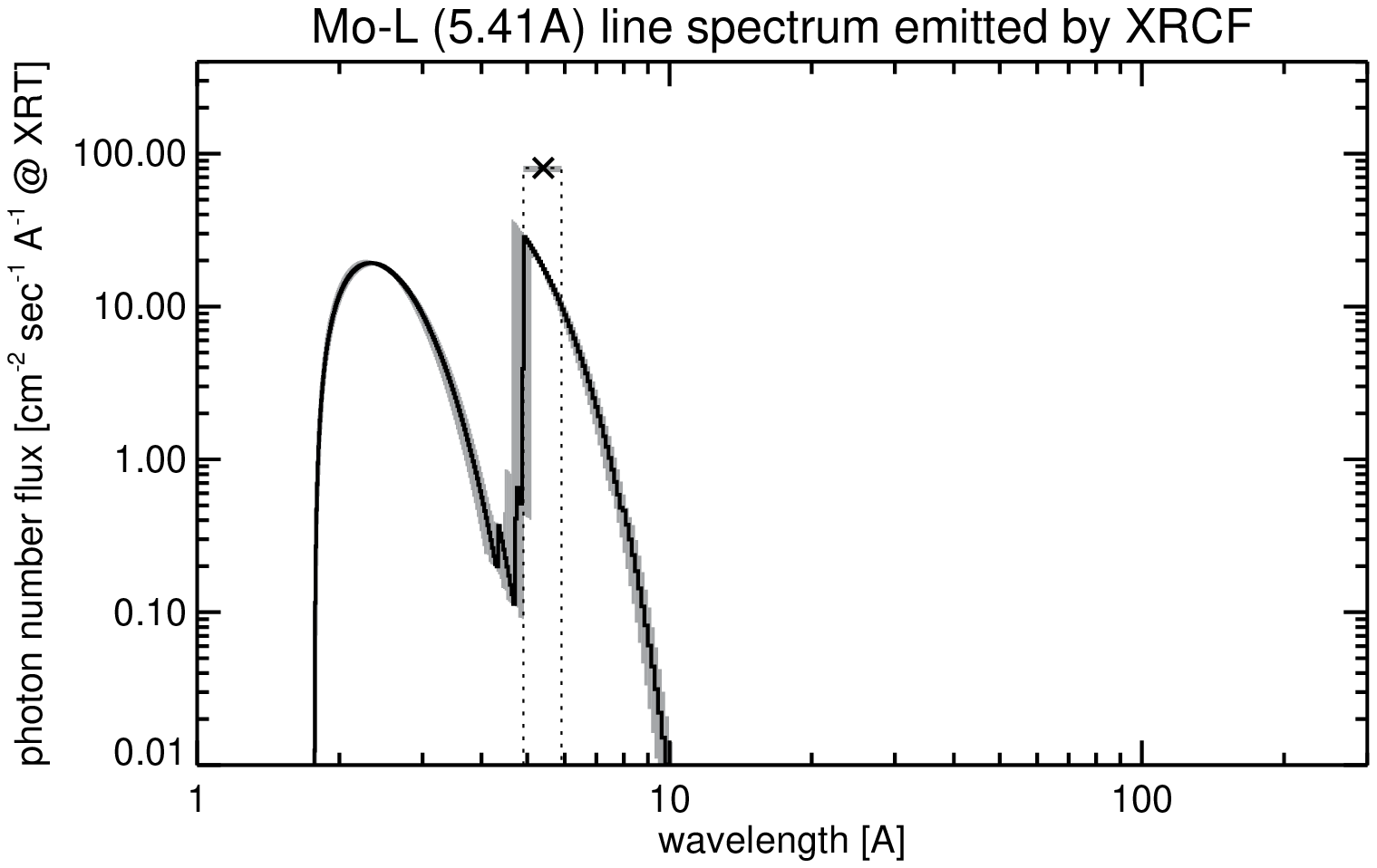}}
\centerline{\includegraphics[width=6.0cm,clip=]{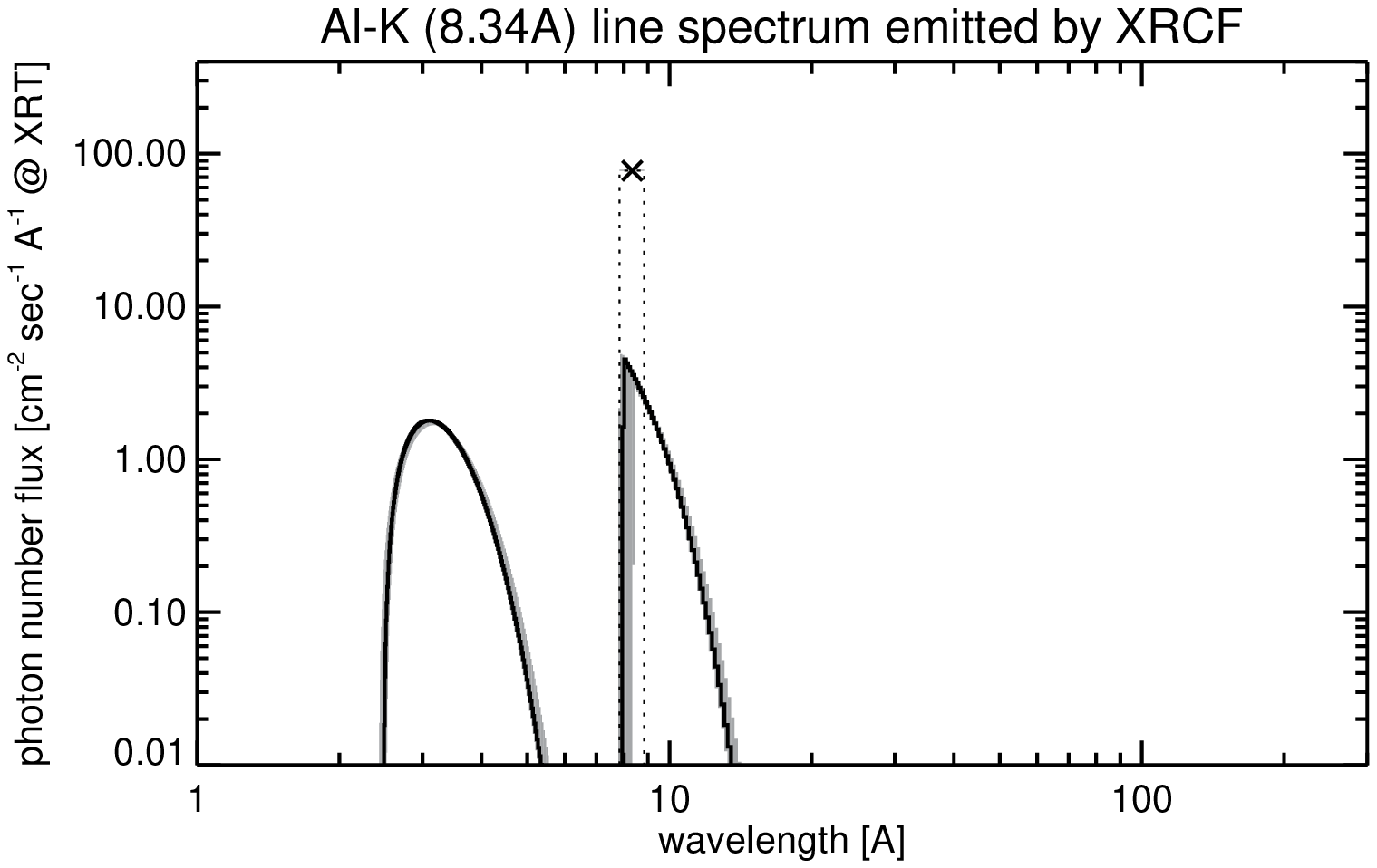}
            \hspace*{0.0cm}
            \includegraphics[width=6.0cm,clip=]{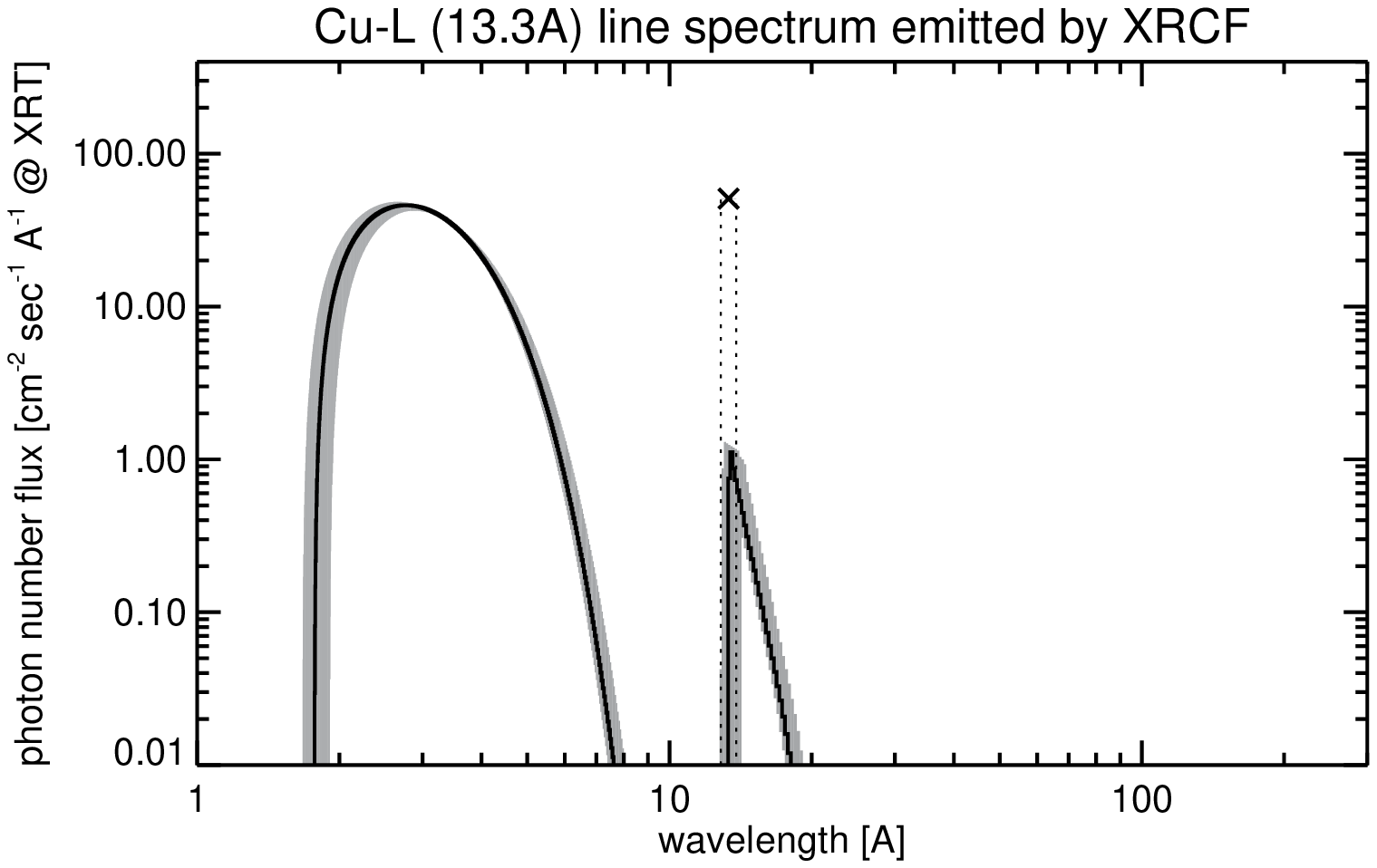}
           }
\vspace{0.5cm}
\centerline{\includegraphics[width=6.0cm,clip=]{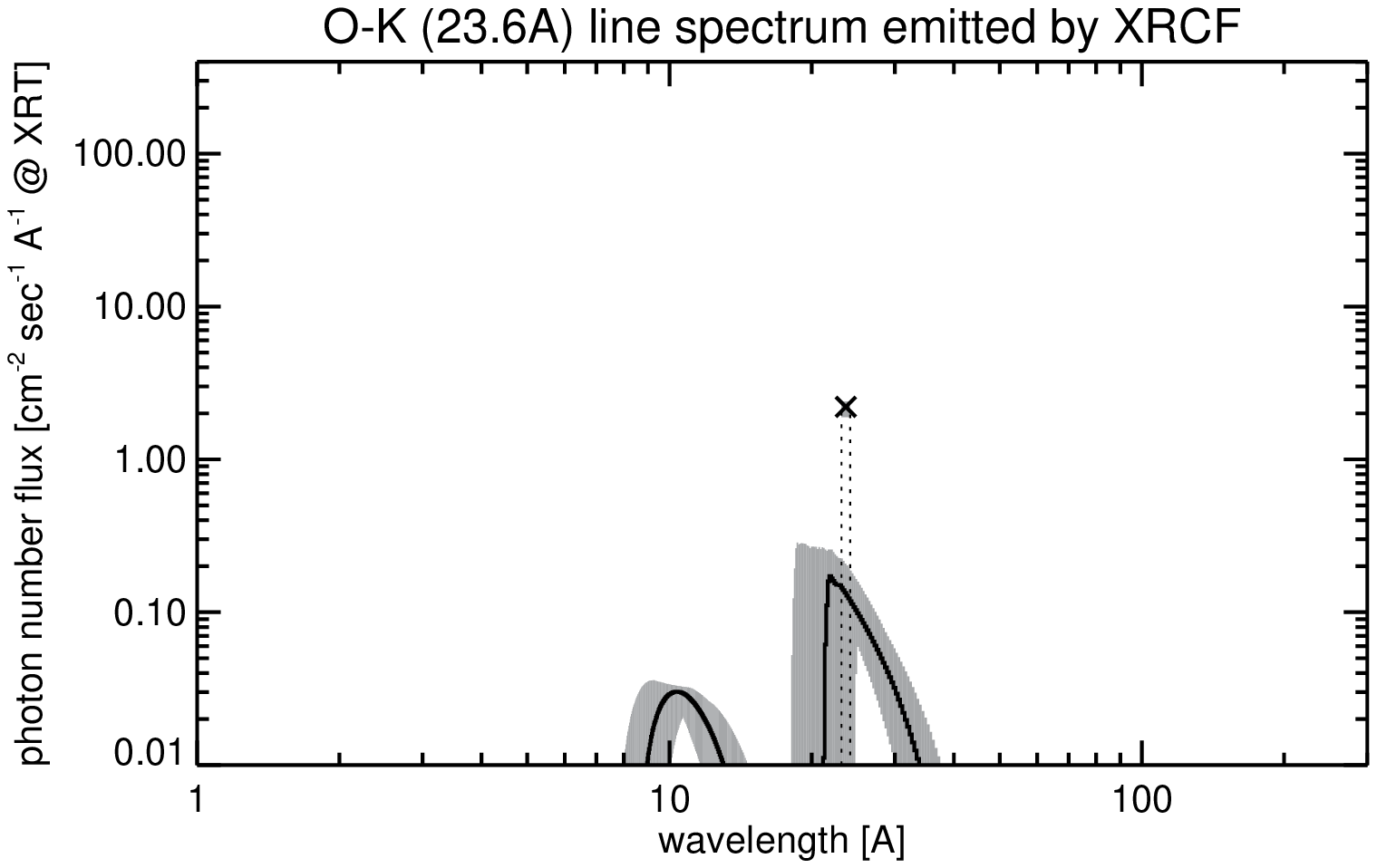}
            \hspace*{0.0cm}
            \includegraphics[width=6.0cm,clip=]{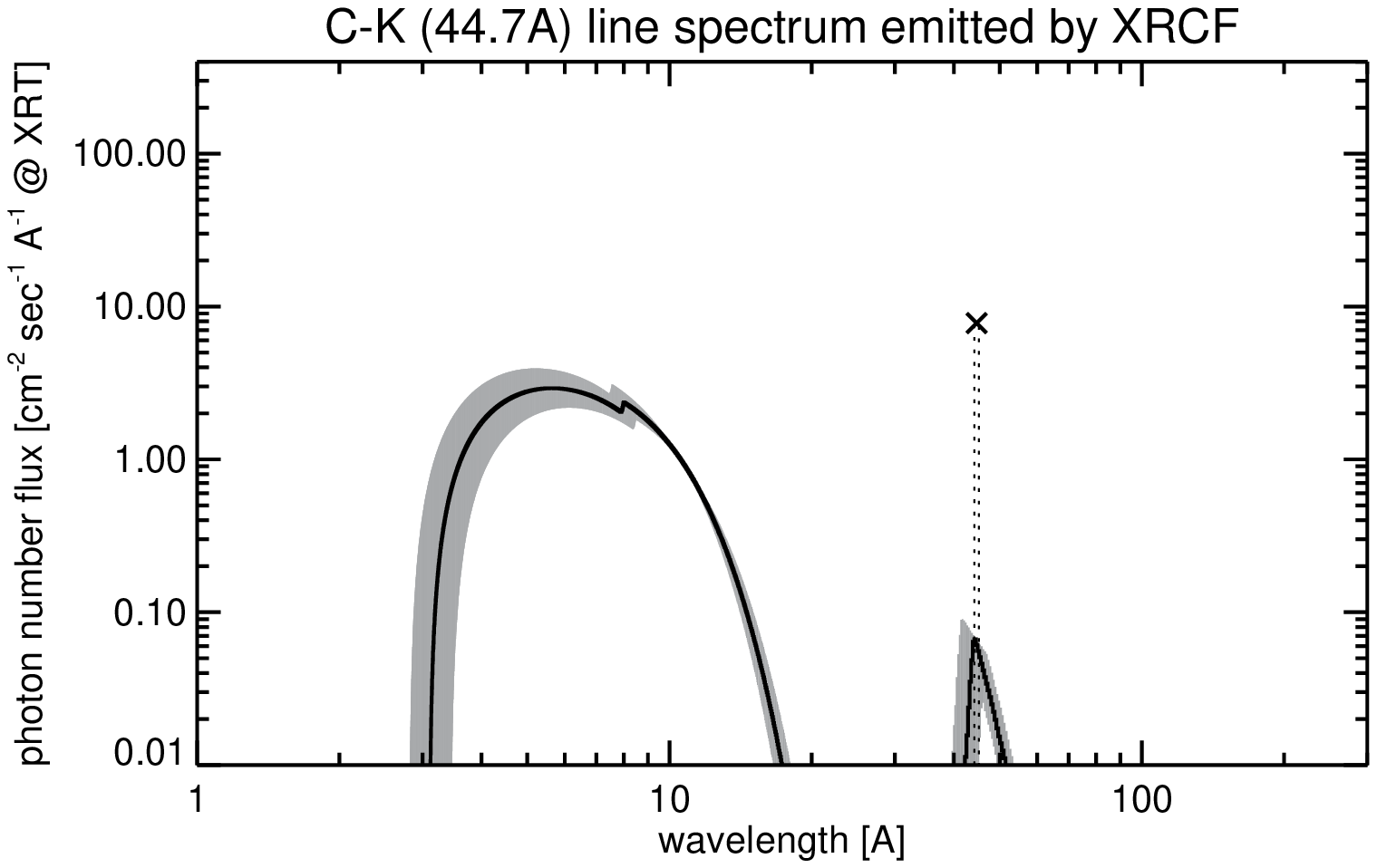}
           }
\vspace{3.0mm}
\caption{Spectra of characteristic X-ray lines and continua emitted by XRCF. These are the incident spectra into XRT.
The solid lines are the spectra of continuum bremsstrahlung.
The crosses ($\times$) show the wavelength (horizontal axis) and photon-number flux in the unit of [cm$^{-2}$ sec$^{-1}$]
(vertical axis) emitted by the characteristic X-ray lines.
If we indicate the characteristic X-ray lines as spectra in units of [cm$^{-2}$ sec$^{-1}$ {\AA}$^{-1}$],
the spectra can be indicated with the dotted lines, whose width and hight are 1~{\AA} and the same value as
photon-number flux in the unit of [cm$^{-2}$ sec$^{-1}$], respectively. The gray areas show the error bar of the estimated spectra.
}
\label{fig:XRCF spectrum}
\end{figure}

Measurements of the X-ray transmission of focal-plane analysis filters (FPAFs) were made with five characteristic X-ray lines
as part of the end-to-end throughput test with the flight XRT telescope (except for the flight pre-filters and sun-shield plate
in front of the telescope) performed at the X-ray Calibration Facility (XRCF) at NASA/Marshall Space Flight Center in June 2005.
Spectra of the X-ray lines used for measurement are shown in Figure~\ref{fig:XRCF spectrum}
(see also Appendix~\ref{sec:XRCF spectrum} for details).

\subsection{Check of XRCF X-Ray Spectra}
\label{subsec:check of XRCF spectra}

In the end-to-end test, XRT pre-filters were not installed.
Therefore, the effective area as measured at the XRCF can be written as
\begin{equation}
A_\mathrm{eff@XRCF} = A \times R_\mathrm{M1} \times R_\mathrm{M2} \times \mathcal{T}_\mathrm{FPAF1} \times \mathcal{T}_\mathrm{FPAF2} \times QE_\mathrm{CCD} .
\label{eq:A_eff at XRCF}
\end{equation}
With FW1 and FW2 set to the open positions, this can be simplified as
\begin{equation}
A_\mathrm{eff@XRCF}^\mathrm{open} = A \times R_\mathrm{M1} \times R_\mathrm{M2} \times QE_\mathrm{CCD} .
\label{eq:A_eff at XRCF open}
\end{equation}
With the manufacturer-supplied information on the geometric aperture area [$A$], reflectivities of primary and secondary mirrors [$R_\mathrm{M1}$ and $R_\mathrm{M2}$], and
the Quantum Efficiency [$QE_\mathrm{CCD}$] of the CCD,
we can check whether the XRCF X-ray spectra (Figure~\ref{fig:XRCF spectrum}) were well estimated, by comparing
signal from the XRT CCD (in terms of data number (DN) which represents the number of electrons generated) at the open/open position
and that expected from energy spectra of the X-ray beams at XRCF measured with a flow proportional counter (FPC)
(see Appendix~\ref{subsec:when XRT observes the XRCF spectra} for details).

\begin{figure}
\centerline{\includegraphics[width=5.5cm,clip=]{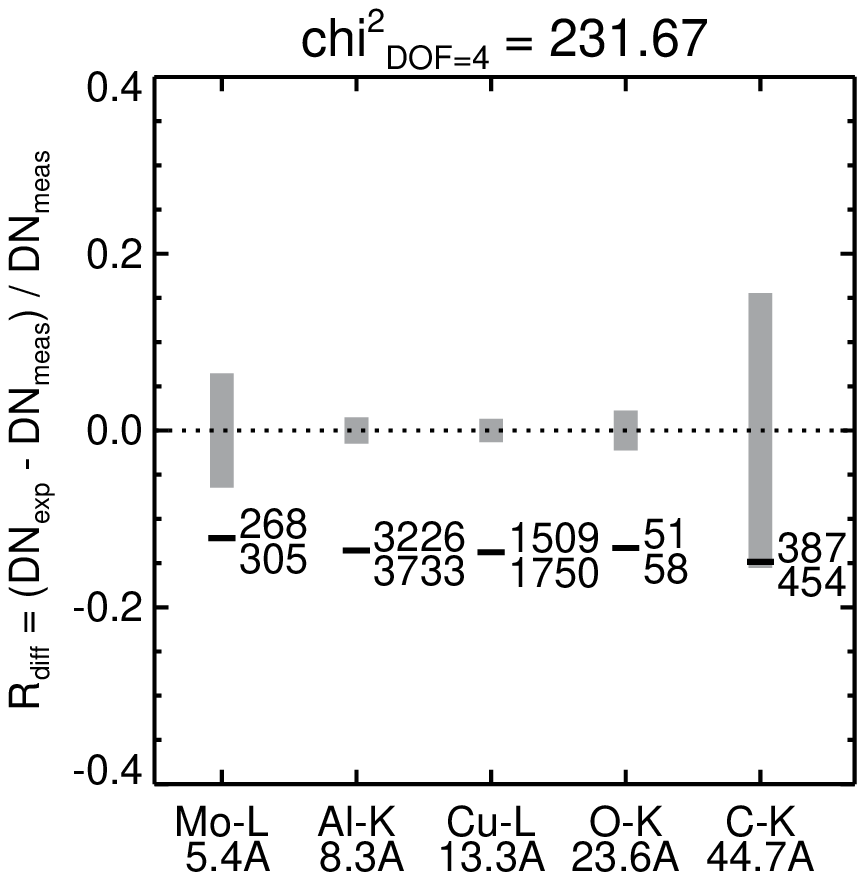}
            \hspace*{0.0cm}
            \includegraphics[width=5.5cm,clip=]{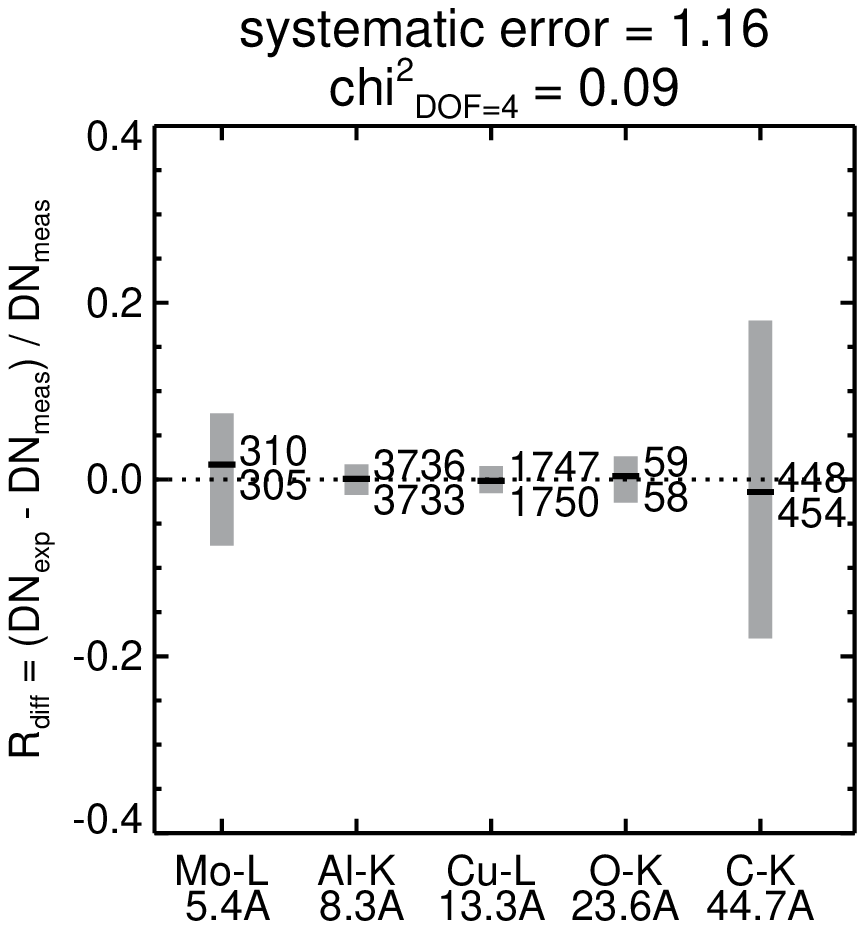}
           }
\vspace{3.0mm}
\caption{
Left panel shows the residual between the DN measured with XRT CCD ($DN_\mathrm{meas}$) and DN expected with XRCF spectra
($DN_\mathrm{exp}$) with one second exposure in the case where FW1 and FW2 are set at the open position.
Black marks ($-$) indicate the differences defined as
$R_\mathrm{diff} \equiv (DN_\mathrm{exp}-DN_\mathrm{meas})/DN_\mathrm{meas}$.
The top and bottom values near the black marks ($-$) shows the value of
$DN_\mathrm{exp}$ and $DN_\mathrm{meas}$, respectively.
The gray bar is the $1\sigma$ error bar.
Right panel is the best correlative case, where the systematic error of 1.16 is taken into account.
}
\label{fig:filter open}
\end{figure}

The left panel of Figure~\ref{fig:filter open} shows the result of the comparison.
The vertical axis shows fractional difference $R_\mathrm{diff} \equiv \left(DN_\mathrm{exp} - DN_\mathrm{meas}\right) / DN_\mathrm{meas}$,
where $DN_\mathrm{meas}$ and $DN_\mathrm{exp}$ are the measured DN with the XRT CCD and the expected DN from XRCF spectra.
If $DN_\mathrm{meas}$ and $DN_\mathrm{exp}$ are consistent, $R_\mathrm{diff}~\approx~0$.
The $1\sigma$ error is shown by gray bars with observation error of CCD
(\textit{e.g.}, photon noise and fluctuation of dark level) and the estimate error of XRCF spectrum
taken into account as the source of errors.
We found that $R_\mathrm{diff}$ of each X-ray line is almost constant at $-0.14$ as shown in left panel of Figure~\ref{fig:filter open}.
This suggests that the spectral shape of XRCF beam is well estimated, but that the intensity is underestimated.
We found that ``$DN_\mathrm{exp} \times 1.16$" gives the best fit to the data with $\chi^2_\mathrm{DOF=4} = 0.09$
as shown in the right panel of Figure~\ref{fig:filter open}, where DOF stands for the degree of freedom.
We consider that this coefficient of 1.16 is the systematic error
caused by the ambiguous aperture area of FPC and slight difference in the locations of XRT and FPC (see Figure~\ref{fig:XRCF}),
because we do not know the exact values of them and we apply the approximate values provided from XRCF.
Hereafter, we derive $DN_\mathrm{exp}$ correcting this systematic error.
The fact that $DN_\mathrm{meas}$ and $DN_\mathrm{exp}$ show good match across the entire wavelength range in the right panel of Figure~\ref{fig:filter open}
gives support to the reliability of the calibration result on the thicknesses of FPAFs with the XRCF X-ray spectra.

\subsection{Calibration of Focal-Plane Analysis Filters}
\label{subsec:calibration of FPAF}

Among the materials used for FPAFs, aluminum, titanium and beryllium can form thin oxidization layers on their surfaces.
Hence, the effect of oxidization should correctly be taken into account when calibrating FPAFs.
Oxidization of filters decreases transmission of FPAFs, especially at longer wavelengths,
because the added oxygen absorbs X-rays.
Considering oxidization,
the transmission of FPAFs ($\mathcal{T}_\mathrm{FPAF1}$ and $\mathcal{T}_\mathrm{FPAF2}$ for FW1 and FW2, respectively) are expressed as
\begin{equation}
\mathcal{T}_\mathrm{FPAF} = \mathcal{T}_\mathrm{pure} \times \mathcal{T}_\mathrm{ox} \times \mathcal{T}_\mathrm{supp} ,
\label{eq:T_FPAF}
\end{equation}
where $\mathcal{T}_\mathrm{pure}$, $\mathcal{T}_\mathrm{ox}$, and $\mathcal{T}_\mathrm{supp}$ are
the transmission of pure metal (unoxidized metal), oxidized metal, and support material, respectively.
The transmission [$\mathcal{T}$] of the material is determined by the material and its thickness [$d$] as
\begin{equation}
\mathcal{T} = \exp \left( - \frac{d}{l_\mathrm{att}\left(\lambda\right)} \right) ,
\label{eq:T}
\end{equation}
where $l_\mathrm{att}\left(\lambda\right)$ is the attenuation length for the material at the X-ray wavelength $\lambda$.
The attenuation length of respective atoms ($Z$ = 1\,--\,92, where $Z$ is the atomic number) at an energy range of
50\,--\,30000~eV is derived by \inlinecite{hen93} and can be obtained from the ``X-Ray Interactions With Matter"
(\url{http://henke.lbl.gov/optical_constants/}).
Using Equations~(\ref{eq:A_eff at XRCF}), (\ref{eq:A_eff at XRCF open}), (\ref{eq:T_FPAF}), (\ref{eq:T}) and also Equation~(\ref{eq:DN1})
in Appendix~\ref{subsec:when XRT observes the XRCF spectra},
$DN_\mathrm{exp}$ at an exposure time of $t = 1$~sec is written as
\begin{eqnarray}
DN_\mathrm{exp} = \int \left( P\left(\lambda\right) \times A_\mathrm{eff@XRCF}^\mathrm{open}\left(\lambda\right) \times \exp \left( - \frac{d_\mathrm{pure}}{l_\mathrm{att}\left(\lambda\right)} \right) \times \mathcal{T}_\mathrm{ox}\left(\lambda\right) \right. \nonumber \\
\left. \times \mathcal{T}_\mathrm{supp}\left(\lambda\right) \times \frac{C}{\lambda} \right) \mathrm{d}\lambda ,
\label{eq:DN estimated}
\end{eqnarray}   
where $d_\mathrm{pure}$ is the pure metal thickness of FPAFs,
and $P$ is the photon-number spectrum of the X-ray beam at XRCF (see Figure~\ref{fig:XRCF spectrum}).
$C$ gives the conversion from number of photons at wavelength $\lambda$ to signal DN generated on the CCD,
and is expressed as $C = t \times h c / e / 3.65 / G$, where $G$ is the system gain of the CCD camera, which is 57.5 [e DN$^{-1}$] for XRT \cite{kan08}.
In performing the fitting, we set the support thickness $d_\mathrm{supp}$, namely $\mathcal{T}_\mathrm{supp}$, as a fixed parameter.
The reason for this is described below.
For $\mathcal{T}_\mathrm{ox}$, we adopt the oxidized-metal thickness derived with the on-orbit data,
because after the end-to-end test at the XRCF and until the launch of \textit{Hinode}, FPAFs were continuously purged with dry nitrogen
to prevent additional oxidization.

On the basis of the analysis in Appendix~\ref{subsubsec:X-ray analysis in Phase 3},
the oxidized-metal thickness for the polyimide side can be regarded as 0~{\AA}
while that for the metal side (exposed to the ambient atmosphere) can be consistently set to 75~{\AA},
irrespective of the metal material.
We note that our 75~{\AA} oxidized-metal thickness is consistent with the measurement of a well-oxidized aluminum filter by \inlinecite{pow90}.
Hence, in this calibration, the fitting parameter is the metal thickness [$d_\mathrm{metal}$] of each FPAF only.

\begin{figure}
\centerline{\includegraphics[width=5.5cm,clip=]{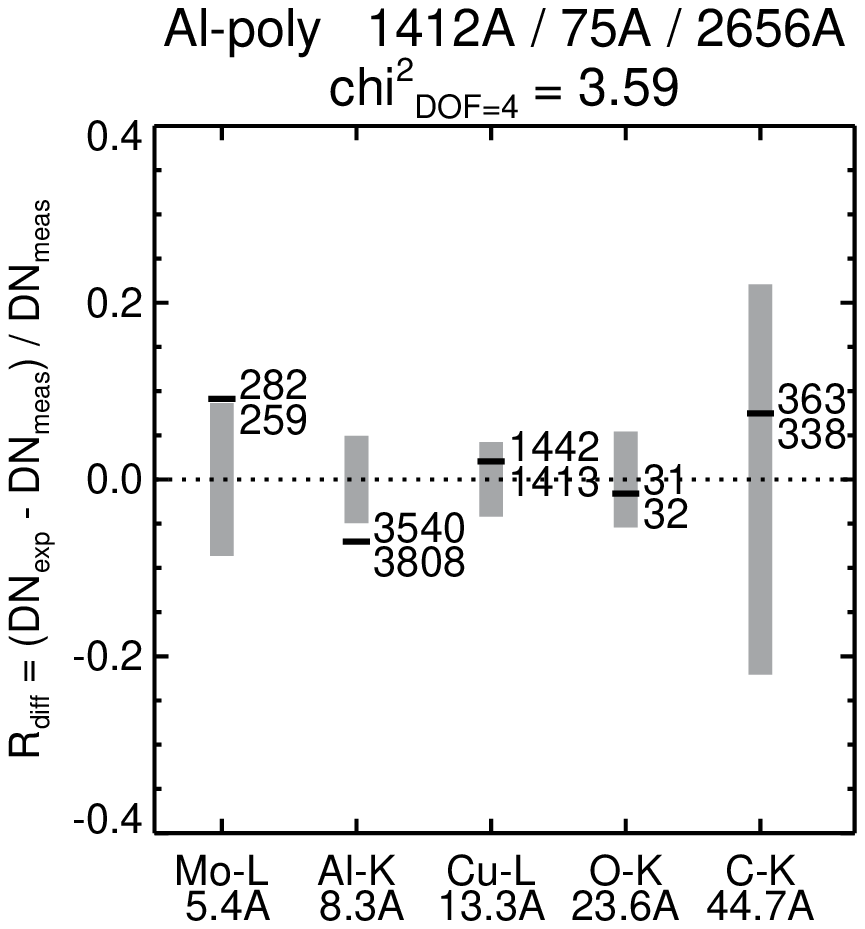}
            \hspace*{0.0cm}
            \includegraphics[width=5.5cm,clip=]{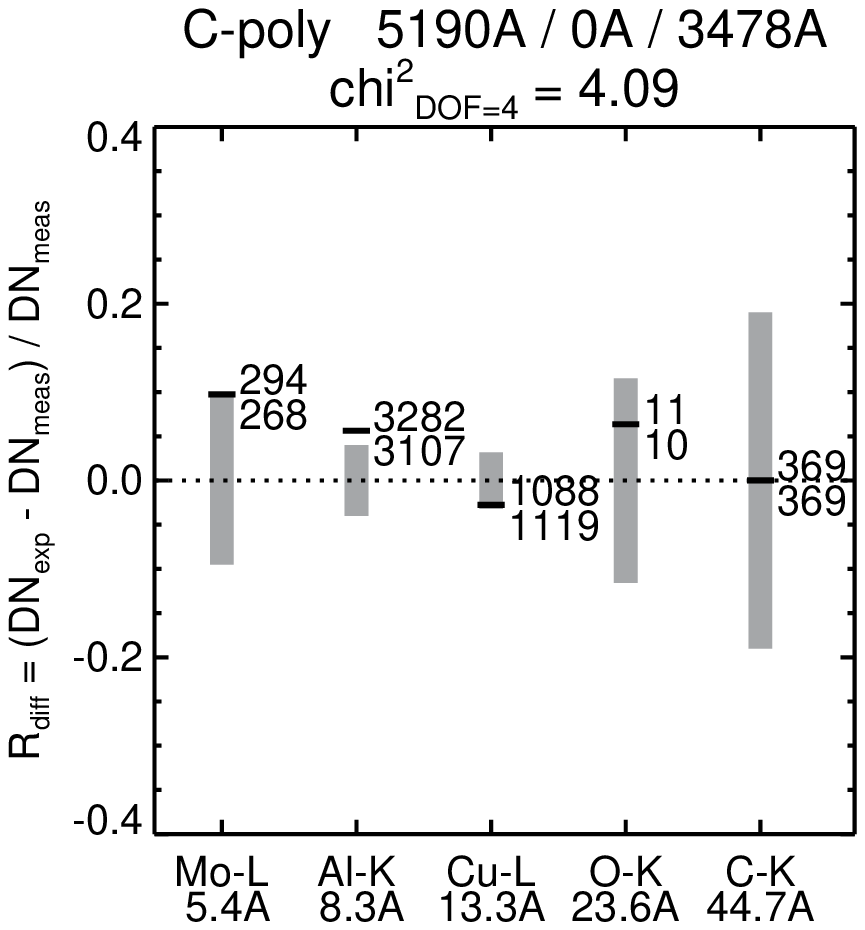}
           }
\centerline{\includegraphics[width=5.5cm,clip=]{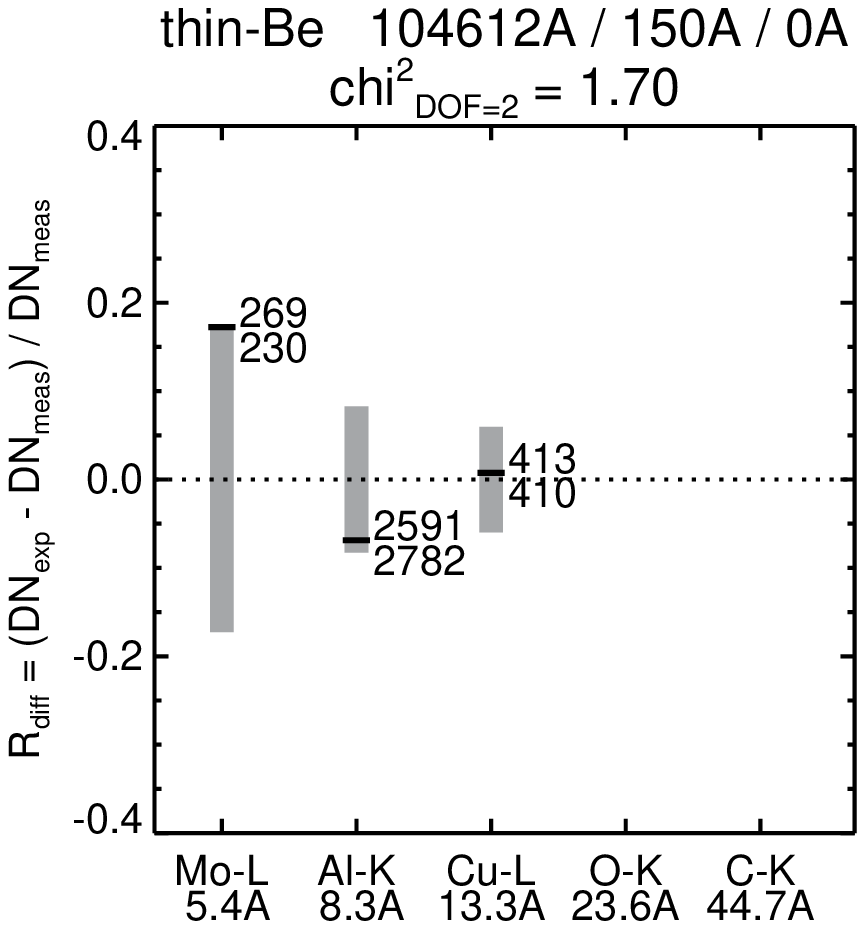}
            \hspace*{0.0cm}
            \includegraphics[width=5.5cm,clip=]{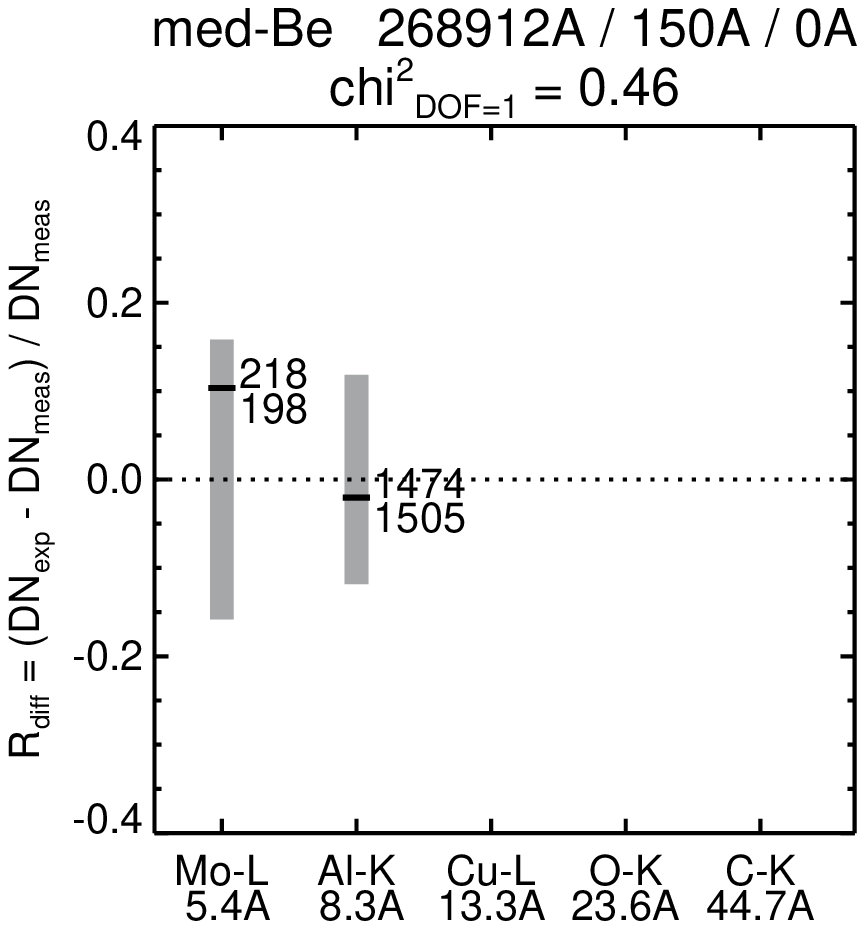}
           }
\centerline{\includegraphics[width=5.5cm,clip=]{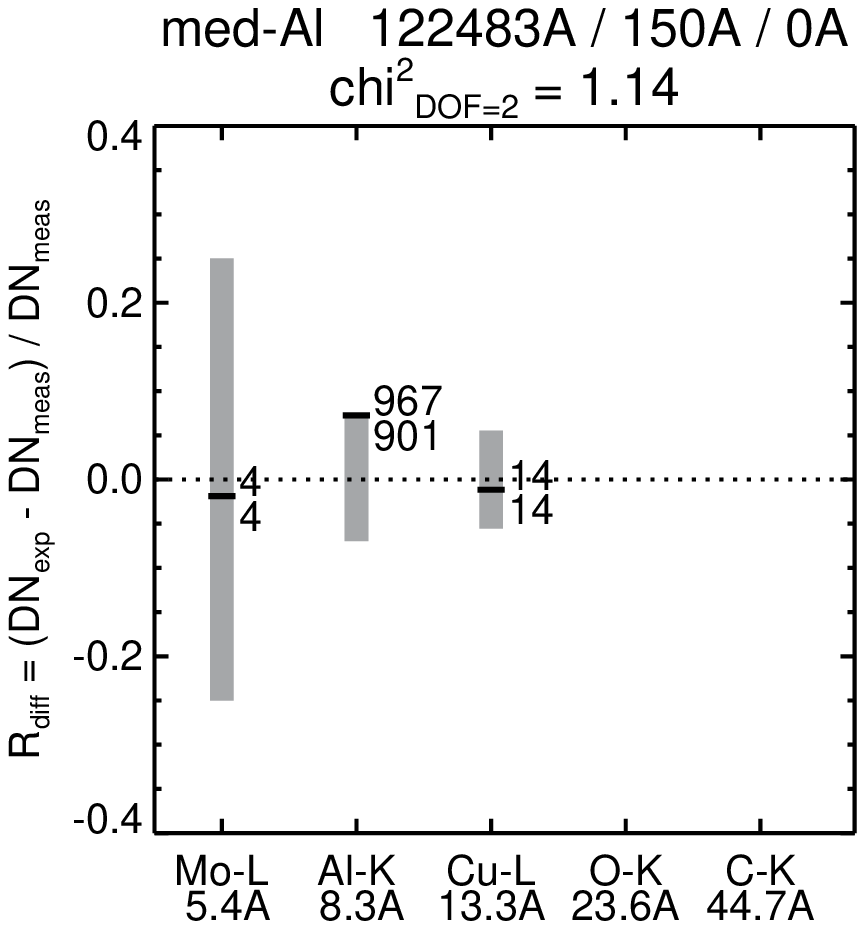}
            \hspace*{0.0cm}
            \hspace*{5.5cm}
           }
\vspace{3.0mm}
\caption{
Calibrated thicknesses of focal-plane analysis filters mounted on filter wheel~1.
The thicknesses described in the left, middle, and right side  of ``/" show the
estimated pure metal, oxidized metal, and support thicknesses, respectively.
Black marks ($-$) indicate the differences defined as
$R_\mathrm{diff} \equiv (DN_\mathrm{exp}-DN_\mathrm{meas})/DN_\mathrm{meas}$,
where $DN_\mathrm{meas}$ is the DN measured with XRT CCD and 
$DN_\mathrm{exp}$ is the DN expected with XRCF spectra.
The top and bottom values near the black marks ($-$) show the value of
$DN_\mathrm{exp}$ and $DN_\mathrm{meas}$ with one second exposure, respectively.
The gray bar is the $1\sigma$ error bar.
}
\label{fig:FPAF1}
\end{figure}

\begin{figure}
\centerline{\includegraphics[width=5.5cm,clip=]{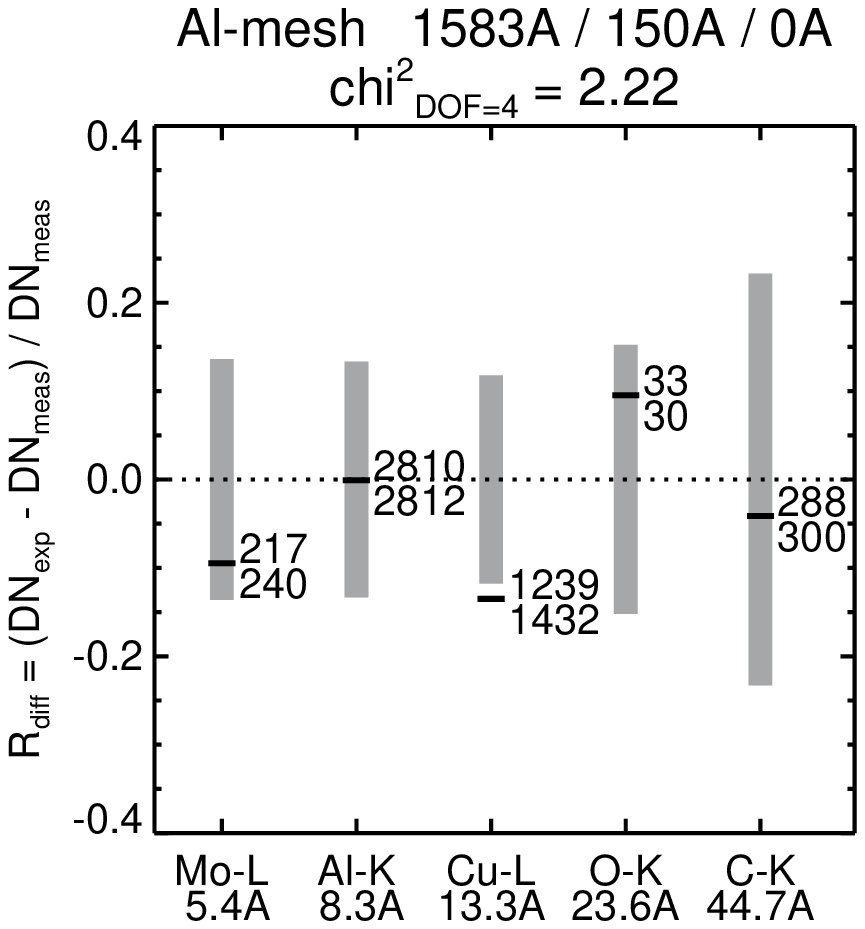}
            \hspace*{0.0cm}
            \includegraphics[width=5.5cm,clip=]{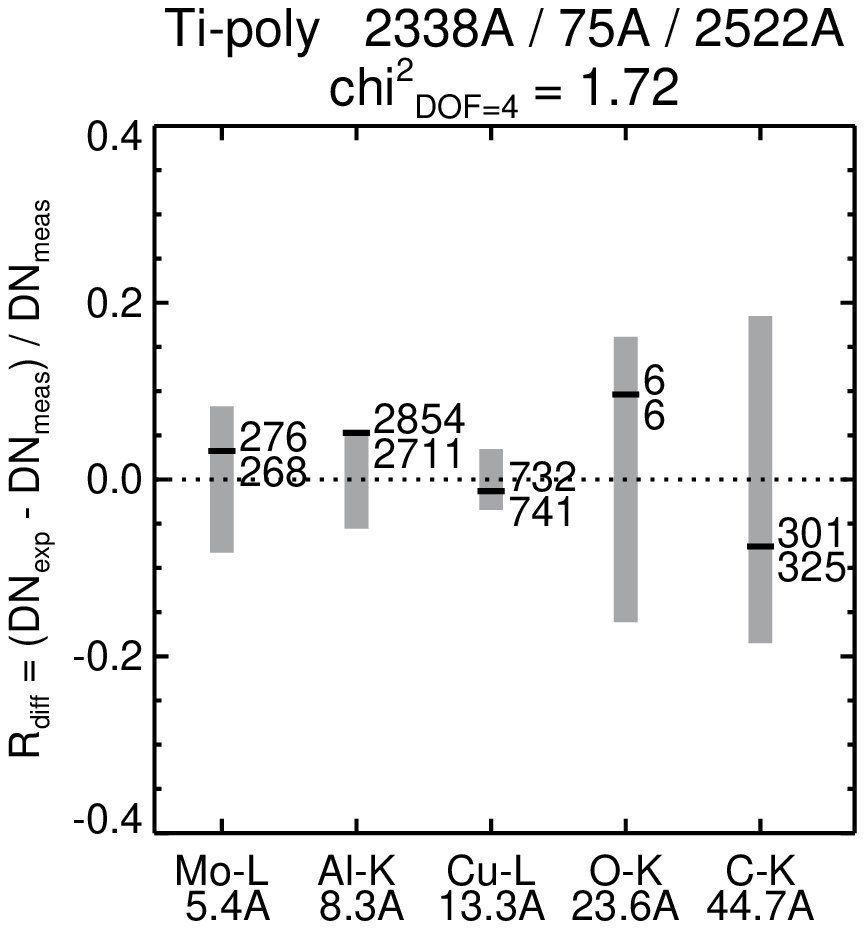}
           }
\centerline{\includegraphics[width=5.5cm,clip=]{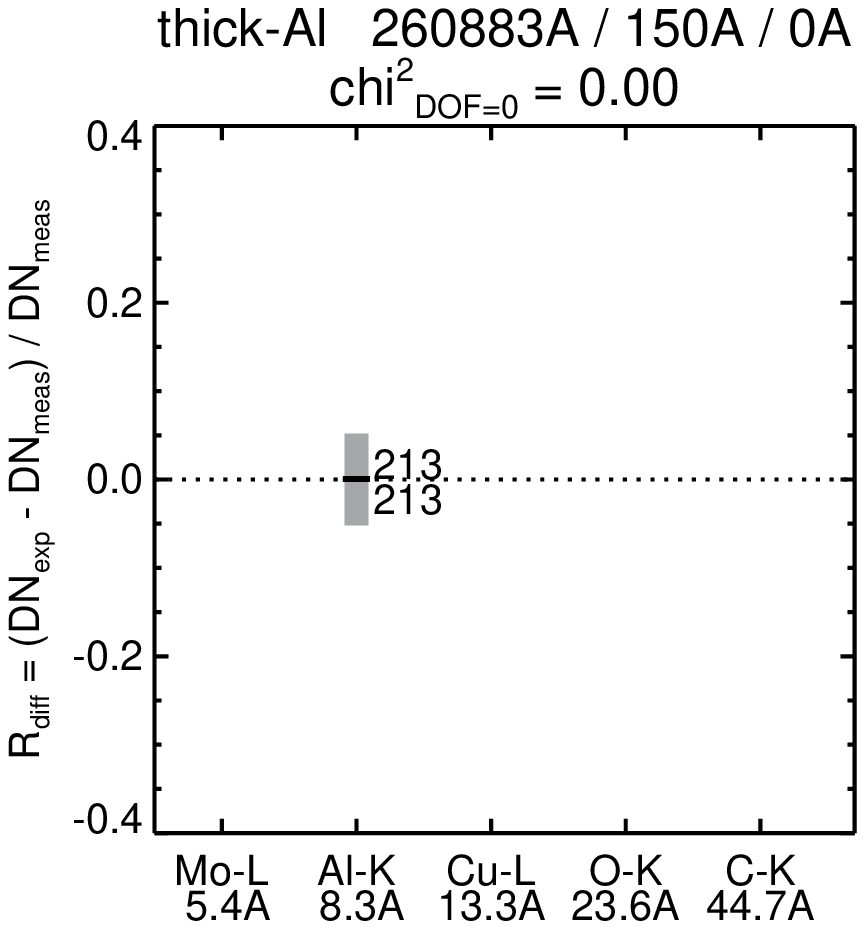}
            \hspace*{0.0cm}
            \includegraphics[width=5.5cm,clip=]{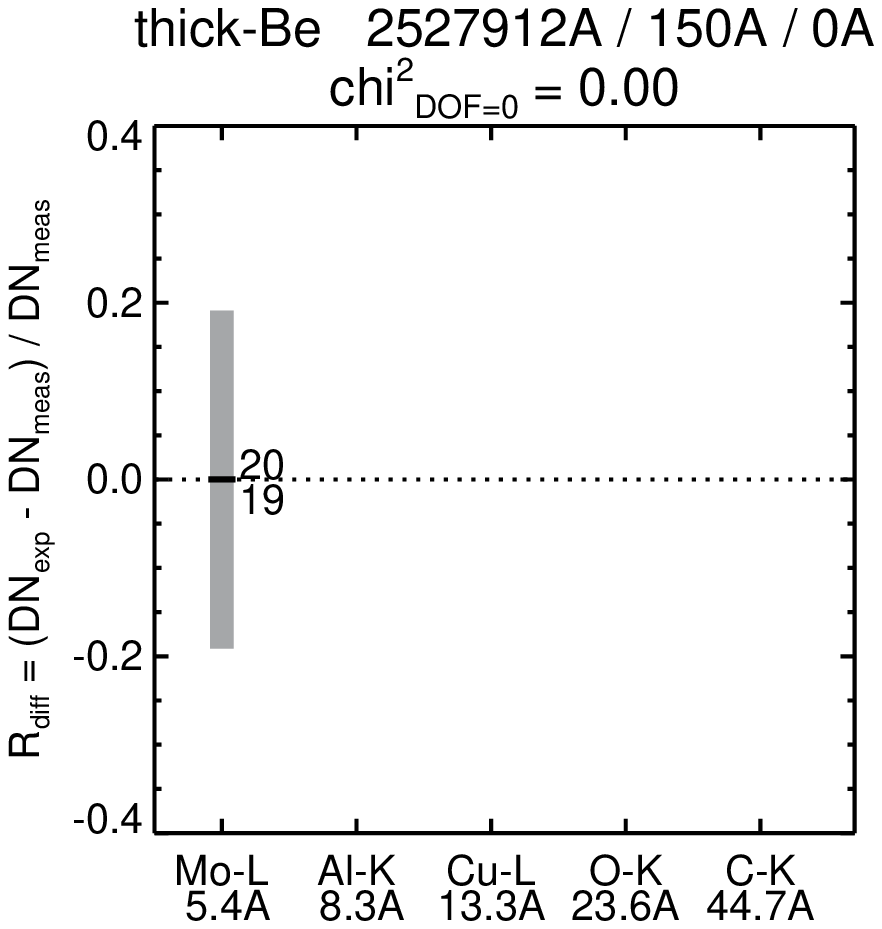}
           }
\vspace{3.0mm}
\caption{
Calibrated thicknesses of focal-plane analysis filters mounted on filter wheel~2.
The meaning of this figure is same as Figure~\ref{fig:FPAF1}.
}
\label{fig:FPAF2}
\end{figure}

The results of fitting for each FPAF are shown in Figure~\ref{fig:FPAF1} and Figure~\ref{fig:FPAF2}.
For all filters, the fitted value [$R_\mathrm{diff}$] is determined within the $1\sigma$ error.
The calibrated thicknesses of the FPAFs are summarized in Table~\ref{tbl:FPAF}.

In this fitting analysis, there are three notes:
\begin{itemize}
\item[\textit{i})]
The thickness of polyimide film, which is used in Al-poly, C-poly, and Ti-poly filters as supports of these thin metal filters.
According to the filter manufacturer (Luxel), the thickness of polyimide was well controlled.
Therefore we fix the polyimide thickness to the values provided by the manufacturer (see Table~\ref{tbl:FPAF}).
\item[\textit{ii})]
The fraction of the open area of the support mesh for the Al-mesh filter.
On the basis of the microscope measurement performed by the manufacturer, we identified that the geometrical open area of mesh is 77\%.
In the calibration of Al-mesh filter thickness, we adopted the value of 77\% with scattering from the mesh pattern
and the X-ray shape on it taken account.
The detail of annulus transmission is described in Appendix~\ref{variation in annulus transmission}.
\item[\textit{iii})]
The last note is about the calibration of thick filters, namely med-Be, thick-Al, and thick-Be filters.
Since low-energy X-rays do not have enough transmission through the thick filters,
we were able to use one or two X-ray lines to calibrate these thick filters.
Thus we consider that the calibration of the thick filters is less certain.
However, according to Equation~(\ref{eq:T}),
even if the calibrated thicknesses of the thick filters have some error, 
the effect on their transmission is small.
Hence we accept the thicknesses derived with XRCF data for now.
To obtain further knowledge of the thicknesses of the thick filters, we need to take enough data sets
with intense X-ray sources, \textit{i.e.} active regions and flares, with thick filters.
Calibration with on-orbit observation data is a project for future work.
\end{itemize}

\begin{table}
\caption{Materials used in our calibration}
\label{tbl:material}
\begin{tabular}{ccc}
\hline
material                                   & molecular weight & density             \\
\hline
Al                                         &  26.98            &  2.699~g~cm$^{-3}$  \\
C                                          &  12.01            &  2.2~g~cm$^{-3}$    \\
Be                                         &   9.01            &  1.848~g~cm$^{-3}$  \\
Ti                                         &  47.87            &  4.54~g~cm$^{-3}$   \\
Al$_2$O$_3$                                & 101.96            &  3.97~g~cm$^{-3}$   \\
TiO$_2$                                    &  79.87            &  4.26~g~cm$^{-3}$   \\
BeO                                        &  25.01            &  3.01~g~cm$^{-3}$   \\
polyimide (C$_{22}$H$_{10}$N$_{2}$O$_{5}$) &  --               &  1.43~g~cm$^{-3}$   \\
\hline
\end{tabular}
\end{table}

In Table~\ref{tbl:FPAF}, we see that, for each metal filter, the difference between our thickness
and that from the manufacturer is almost within the uncertainties
reported by the manufacturer except for the Al-poly (difference is 200~{\AA}) and C-poly (800~{\AA}).
In the case of the Al-poly filter, the manufacturer found a discrepancy between their measured thickness and
that inferred from their own transmission measurement in visible light,
and noted that the Al-poly filter should be thicker than their reported value.
This information is consistent with our result.
In the case of the C-poly filter, the difference is significantly large.
This could have been caused by the different method of measurement.
As we mentioned, we derived the thickness with the transmission in X-rays and attenuation length.
The attenuation length depends on the density of material.
In our calibration, we used the typical density of material summarized in Table~\ref{tbl:material}
with the typical density of graphite adopted for our analysis on C-poly filter.
However, it often happens that the assumption of a typical density is different from the actual density for the filter,
especially in the case of C-poly filter, because carbon is not a metal.
According to the manufacturer of filters, they have estimated that the density of their carbon film is about 90\% of typical value of graphite, namely about 1.9~g~cm$^{-3}$,
because the carbon is evaporated using an electron beam in the process of making thin carbon filter.
Considering this difference in density, the X-ray transmission of the C-poly filter estimated with our calibrated thickness and typical density
is consistent with that calculated with manufacturer's thicknesses and $\approx$~90\% of typical density,
because our calibrated thickness (5190~{\AA}, see Table~\ref{tbl:FPAF}) is about 90\% of manufacturer's thicknesses (6038~{\AA}).

\section{On-Orbit Calibration}
\label{app:on-orbit calibration}

\subsection{Thermal Distribution inside XRT}
\label{sec:thermal distribution inside XRT}

\begin{figure}
\centerline{\includegraphics[width=10.0cm,clip=]{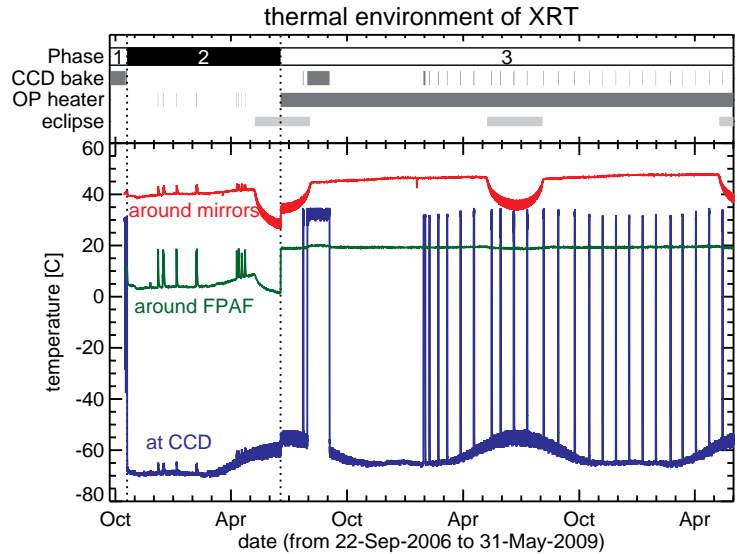}}
\caption{
Thermal environment of XRT.
The red, green, and blue lines indicate the temperature at the mirror support plate
(\textsf{XRTD\_TEMP\_11}), at the focal-plane shutter motor (\textsf{XRTD\_TEMP\_22}),
and at the CCD (\textsf{XRTE\_CCD\_TEMP}), respectively, where \textsf{XRTD\_TEMP\_11}, \textsf{XRTD\_TEMP\_22}, and \textsf{XRTE\_CCD\_TEMP} are
names in the \textit{Hinode} telemetry.
}
\label{fig:XRT status}
\end{figure}

Figure~\ref{fig:XRT status} shows the temporal evolution of the XRT thermal environment.
The red, green, and blue lines indicate the temperature at the mirror support plate,
at the focal-plane shutter motor, and at the CCD, respectively.
From thermal viewpoint, there are three distinct episodes in XRT operation which are shown as
Phases~1, 2, and 3 in Figure~\ref{fig:XRT status}.
Phase~1 is the period from the launch of \textit{Hinode} to first light,
where the CCD bakeout heater was kept on to avoid contaminants accumulating on the CCD.
When the CCD bakeout heater is on, we can say that no contaminants accumulate on the CCD,
since the temperature of CCD becomes about $30^{\circ}$C, which is higher than its surroundings.
Phases~2 and 3 are the periods of normal operation of XRT without and with enabling (turning on) the operational heater
for the telescope tube, respectively. With the operational heater enabled,
the rear end of the XRT telescope tube (in front of the CCD camera when viewed from the Sun)
is warmed up to about $+20^{\circ}$C.
In these periods, the CCD was cooled down to a temperature of about $-60^{\circ}$C to reduce the dark noise.
Since the CCD is the coolest component in XRT, it is possible that the contaminant selectively
accumulates on the CCD.
The operational heater was originally kept off in Phases~1 and 2 due to
technical reasons, and was then enabled from 18 June 2007.
The period after this date is labeled as Phase~3.

Let us now examine the relationship between the thermal environment of XRT and observed decrease in quiet-Sun X-ray intensity.
The bottom panel of Figure~\ref{fig:DN QS} shows temporal evolution of the X-ray intensity ratio normalized with the intensity observed
with the Ti-poly filter to remove possible variation in emission measure for the quiet Sun
(see Equations~(\ref{eq:I iso}) or (\ref{eq:I multi}) in Appendix~\ref{subsec:when XRT observes the solar spectra}).
In the subsequent analysis, we made an assumption that the profile of the differential emission measure in the quiet Sun is almost stable,
and the variation in the observed quiet-Sun intensity ratio can be attributed solely to the accumulated contaminant.
During Phases~2 and 3, the CCD temperature was almost constant at around $-60^{\circ}$C.
The decrease in the intensity ratio was continuously observed in these phases,
indicating that the contaminant kept accumulating on the CCD in both Phases~2 and 3 (except the bakeout periods).
The X-ray intensity ratio is recovered to almost the same level
by each CCD bakeout as seen in Phase~3 in the bottom panel of Figure~\ref{fig:DN QS}.
This implies two things:
\begin{itemize}
\item[\textit{i})] Each bakeout reduces the contaminant on the CCD to the same thickness (except for the ``spots" discussed below).
\item[\textit{ii})] Thickness of contaminant accumulated on each FPAF remains unchanged during Phase~3.
\end{itemize}
The result~\textit{i}) is supported by the behavior of G-band intensity where
the G-band intensity returned (reduced) to almost the same level by the CCD bakeouts (see Figure~\ref{fig:G-band}).
As will be mentioned in Appendix~\ref{subsubsec:contam on FPAF in Phase 2},
we found that on/off operations of the operational heater when the temperature around the FPAFs is cooler than
$5^{\circ}$C cause accumulation of contaminant on the FPAFs.
Hence, we suspect that the accumulation of contaminant on the FPAFs took place in Phase~2 while not in Phases~1 and 3.

Meanwhile, the first and third bakeouts created lumps of contaminant which are distributed over the whole CCD
as spot-like patterns (see Figure~\ref{fig:spot map} in Appendix~\ref{sec:Spot of contamination on CCD}).
These spots of contaminant cannot be removed by the subsequent bakeouts.
We consider that excess amount of contaminant accumulated on CCD before the first and third bakeouts caused these spots.
The detail of the spots is described in Appendix~\ref{sec:Spot of contamination on CCD}.
In this regard, we consider that the contamination on the CCD comprises two seemingly different components:
One is laminar contaminant that accumulates across the CCD as time goes on and can be well removed by a CCD bakeout
when the accumulation thickness is small enough.
The other is ``condensed" contaminant forming spot-like patterns over the CCD.
This condensation of contaminants occurs in association with a CCD bakeout when the accumulation thickness is
large (threshold thickness is not known) and can not be removed by any further CCD bakeouts.
In this article, we characterize only the laminar contaminant for contamination on the CCD.
Detailed characterization of the spot-shaped contamination is a subject for future study.

We consider that no contamination accumulated on the pre-filter,
because the temperature of the front end of the telescope (around mirrors) has been the warmest portion inside the telescope
with the temperature at around or even exceeding $40^{\circ}$C in all phases.

\subsection{Visible Light Analysis -- Contamination on the CCD}
\label{subsec:contam analysis with G-band}

\subsubsection{Method}
\label{subsubsec:contam monitor method}

\begin{figure}
\centerline{\includegraphics[width=9.0cm,clip=]{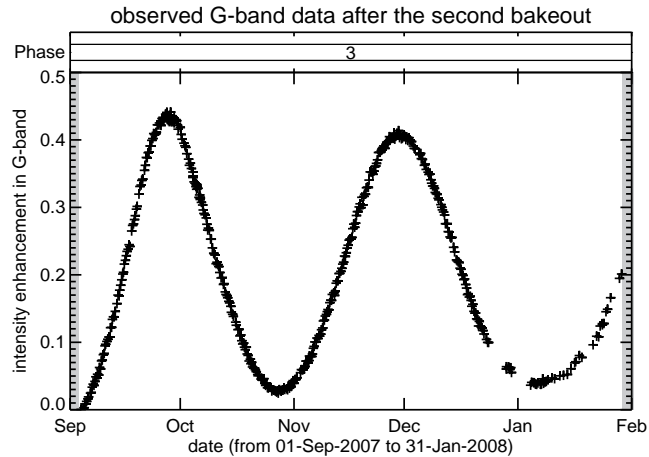}}
\caption{
G-band intensity enhancement observed after the second CCD bakeout from 512~$\times$~512 pixels at CCD center.
The first G-band data just after the second bakeout has been taken as the reference for intensity enhancement.
The gray areas in the left and right sides of this plot are the second and third CCD bakeout periods, respectively.
}
\label{fig:G-band 2nd bake}
\end{figure}

We explain our method to measure the thickness of contaminant
accumulated on the CCD using visible light (G-band) data.
Figure~\ref{fig:G-band 2nd bake} shows temporal evolution of G-band intensity after the second CCD bakeout on 3 September 2007.
This plot is made from average G-band intensity over 512~$\times$~512 pixels area at the center of the CCD
for full-Sun (synoptic) images where the Sun center is located at the same position of the CCD (near the center of the CCD).
Notable features in this plot are \textit{i}) the intensity shows enhancement (non-negative increase) from
the first G-band data just after the second bakeout, and \textit{ii}) the intensity shows periodic oscillation.
The period of oscillation is about 53.3~days.
The maximum intensity enhancement is about 40\% of the reference intensity.
This amount of enhancement cannot be explained by the fluctuation of solar intensity
which is less than 1\% in visible light during the 11~year solar cycle.

\begin{figure}
\centerline{\includegraphics[width=7.0cm,clip=]{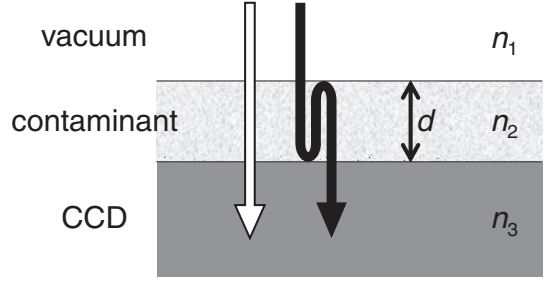}}
\vspace{3.0mm}
\caption{
Anti-reflection effect by the contaminant accumulated on CCD.
}
\label{fig:contam CCD}
\end{figure}

Our interpretation of the oscillation is that is caused by interference
within a contaminant layer (anti-reflection effect). 
Figure~\ref{fig:contam CCD} illustrates anti-reflection with a contaminant layer on the CCD.
Both the direct ray (white arrow) and the reflected rays (black arrow) enter the CCD.
Assuming three layers, vacuum, contaminant, and CCD, transmission of the visible light [$\mathcal{T}$] into the CCD is expressed as
\begin{equation}
\mathcal{T} \equiv \frac{I_\mathrm{t}}{I}
= \frac{4 n_1 n_3}{\left( n_1 + n_3 \right)^2} \left[1 - \sin^2 k_2 d \times \left\{\frac{\left(n_3^2-n_2^2\right)\left(n_2^2-n_1^2\right)}{n_2^2 \left(n_1+n_3\right)^2}\right\} \right]^{-1}
\label{eq:T_G}
\end{equation}
(see Appendix~\ref{app:transmission of rays through three layers} for the derivation of this equation),
where $I$ is the intensity in vacuum, $I_t$ transmitted intensity into CCD, $n$ refractive index of each layer,
$k_2$ wavenumber of G-band in the contaminant layer, and $d$ thickness of the contaminant layer.
The enhancement $E$ of the G-band intensity is then given as
\begin{equation}
E \equiv \frac{\mathcal{T} - \mathcal{T}_{d=0}}{\mathcal{T}_{d=0}}
= \left[1 - \sin^2k_2d \times \left\{\frac{\left(n_3^2-n_2^2\right)\left(n_2^2-n_1^2\right)}{n_2^2 \left(n_1+n_3\right)^2}\right\} \right]^{-1} - 1.
\label{eq:anti-refraction}
\end{equation}
This equation clearly shows that the increasing thickness [$d$] of contamination causes periodic intensity oscillation.
Thickness of contaminant for a local minimum in the intensity enhancement is given by
\begin{equation}
d_\mathrm{min} = \frac{m \pi}{k_2} ,
\label{eq:osc2}
\end{equation}
where $m$ is a positive integer ($m$ = 1, 2, 3, $\cdots$).
The period of the intensity oscillation can be related to the contamination thickness as
\begin{equation}
\Delta d_\mathrm{min} = \frac{\pi}{k_2} = \frac{\lambda_2}{2} = \frac{\lambda_0}{2 n_2} ,
\label{eq:osc}
\end{equation}
where $n_2$ is a refractive index of the contaminant and $\lambda_0$ is the wavelength of the incident visible light in
vacuum (4300~{\AA} in G-band).
This means that if we know the refractive index [$n_2$] of contaminant material, then we can measure the thickness of contaminant
using the measured period of G-band intensity oscillation.

In the above, we considered only contaminants on the CCD. But it is possible that the contaminants also accumulated on
the G-band glass filter. (Note this is indeed the case for the metal filters; see Appendix~\ref{subsubsec:X-ray analysis in Phase 3}.)
Thus, the possible effect of contaminants accumulated on the G-band glass filter should be considered.
As will be detailed in Appendix~\ref{subsubsec:X-ray analysis in Phase 3},
we found that the contaminant is a long-chain organic compound which consists of carbon, hydrogen, and oxygen, and that its refractive index is close to 1.5.
With this information, let us next consider the anti-reflection effect with the contaminant accumulated on the G-band filter.
Figure~\ref{fig:anti-reflection} shows expected amplitude (maximum value) of intensity enhancement
caused by the anti-reflection effect with Equation~(\ref{eq:anti-refraction}).
The material (refractive index) of G-band filter, CCD and contaminant are glass ($n_3~\approx~1.6$), silicon ($n_3~\approx~4.9$) and contaminant ($n_2~\approx~1.5$), respectively.
As seen in the green area of Figure~\ref{fig:anti-reflection}, intensity enhancement caused by (possible) contaminant
on the G-band filter is close to zero due to the proximity of $n_2~\approx~1.5$ and $n_3~\approx~1.6$.
Meanwhile, the contaminant on the CCD can enhance the G-band intensity enough (more than 0.4) for the observed enhancement shown in Figure~\ref{fig:G-band 2nd bake}.
Hence, even if the thickness of contaminant accumulated on the G-band filter changes as a function of time,
we can ignore the G-band intensity enhancement caused by the contaminant on the G-band filter.

By assuming that the refractive index of the contaminant is 1.5 and from Equation~(\ref{eq:osc}), one period of intensity oscillation corresponds to
1433~{\AA} thickness of contaminant as an average over 512~$\times$~512 pixels at the center of CCD.
The accumulation rate of the contaminant there is estimated to be 9800~{\AA}~{year$^{-1}$},
because the period of the G-band intensity oscillation at the CCD center is about 53.3~days.

\subsubsection{Model}
\label{subsubsec:measurement of contam thickness on CCD}

\begin{figure}
\centerline{\includegraphics[width=8.0cm,clip=]{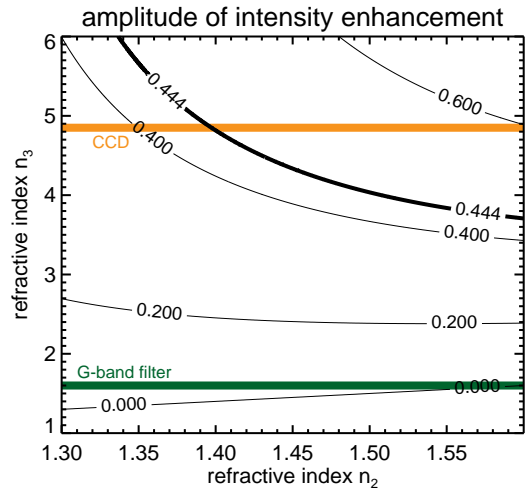}}
\caption{
Expected amplitude (maximum value) of the intensity enhancement by the anti-reflection effect ($a$ in Equation~(\ref{eq:empirical model})) is shown by contours.
The observed amplitude of the intensity enhancement (0.444; see Figure~\ref{fig:G-band 2nd bake}) is indicated by a thick contour.
The refractive indexes of CCD ($n_3~\approx~4.9$) and G-band filter ($n_3~\approx~1.6$) are shown in orange and green, respectively.
}
\label{fig:anti-reflection}
\end{figure}

In order to avoid excess accumulation of contaminant on the CCD, we started to perform regular CCD bakeout
every three\,--\,four weeks since the third bakeout.
This means that, for the period after then, there is not much time to see the intensity oscillation
in the G-band profile between two adjacent bakeouts. Hence, a quantitative estimate on the contaminant thickness directly
from the G-band intensity turned out to be not possible. Instead, for the period after the third bakeout,
we developed an empirical model of the G-band intensity oscillation using the data
between the second and third CCD bakeouts.
On the basis of this model, we estimated time-dependent thickness of the contaminant on the CCD
after the third CCD bakeout in addition to the period between the second and third CCD bakeouts.

\begin{figure}
\centerline{\includegraphics[width=9.0cm,clip=]{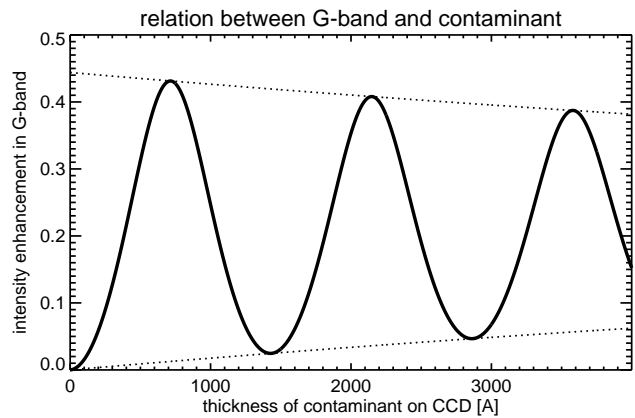}}
\caption{
Relation between intensity enhancement in G-band and thickness of contaminant on CCD.
The dotted curves show the exponentially decaying amplitude of the intensity enhancement oscillation.
}
\label{fig:G-band model}
\end{figure}

In the profile of the observed intensity enhancement (Figure~\ref{fig:G-band 2nd bake}), the periodic oscillation
and the decay of amplitude are seen. The empirical model incorporates these two properties.
Modifying Equation~(\ref{eq:anti-refraction}), the empirical model, which relates intensity enhancement in G-band ($E_\mathrm{model}$) to the thickness of contaminant [$d$],
is expressed by the following equation whose profile is shown in Figure~\ref{fig:G-band model}.
\begin{equation}
E_\mathrm{model} = \left[ \left\{ \left(1 - \sin^2k_2d \times \frac{a}{1+a} \right)^{-1} - 1 \right\} - \frac{a}{2} \right] \times \exp\left(-b \times d\right) + \frac{a}{2} ,
\label{eq:empirical model}
\end{equation}
where $a$ and $b$ are the amplitude (maximum value) and decay coefficient of intensity enhancement, respectively.
The inside of \{ \} in Equation~(\ref{eq:empirical model}) is the rewriting of Equation~(\ref{eq:anti-refraction}) with
\begin{equation}
\frac{a}{1+a} = \frac{\left(n_3^2-n_2^2\right)\left(n_2^2-n_1^2\right)}{n_2^2 \left(n_1+n_3\right)^2} .
\label{eq:amplitude}
\end{equation}
The best-fit values to the observed G-band intensity enhancement (see Figure~\ref{fig:G-band 2nd bake}) are $a = 0.444$ and $b = 8.23 \times 10^{-5}$.

In this empirical model, there are two notes:
\begin{itemize}
\item[\textit{i})] With the observed amplitude (maximum value) of intensity enhancement ($a~\approx~0.444$) and the refractive index of CCD ($n_3~\approx~4.9$),
the refractive index of contaminant is calculated as $n_2~\approx~1.4$ (see Figure~\ref{fig:anti-reflection}).
This value is certainly different from our assumption that the refractive index of contaminant is 1.5.
\item[\textit{ii})] In the observed intensity profile (Figure~\ref{fig:G-band 2nd bake}),
there is  slight decrease in the amplitude, which is incorporated as an exponential
decay in the model ($b$ in Equation~(\ref{eq:empirical model})) and as indicated by dotted lines in Figure~\ref{fig:G-band model}.
\end{itemize}
As will be discussed in Appendix~\ref{subsec:discussion and summary 1},
we conclude that they do not significantly affect the results of XRT calibration.

\begin{figure}
\centerline{\includegraphics[width=10.0cm,clip=]{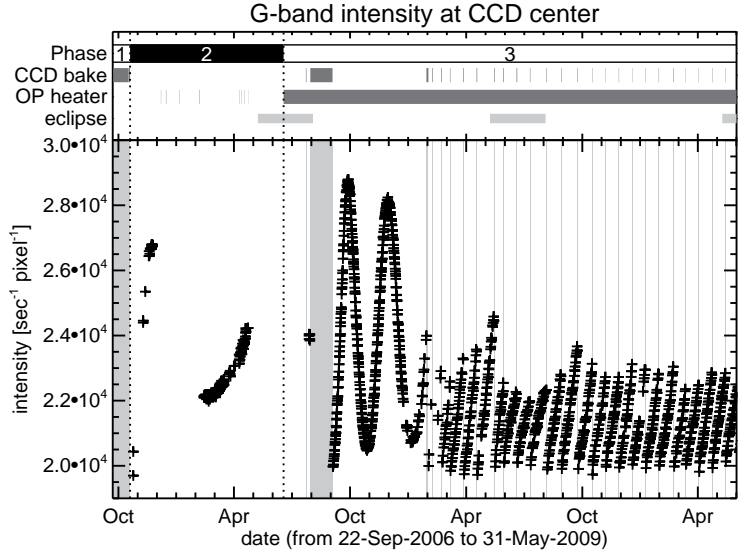}}
\caption{
Temporal evolution of G-band intensity monitored with 512~$\times$~512 pixels at CCD center.
}
\label{fig:G-band}
\end{figure}

\begin{figure}
\centerline{\includegraphics[width=6.0cm,clip=]{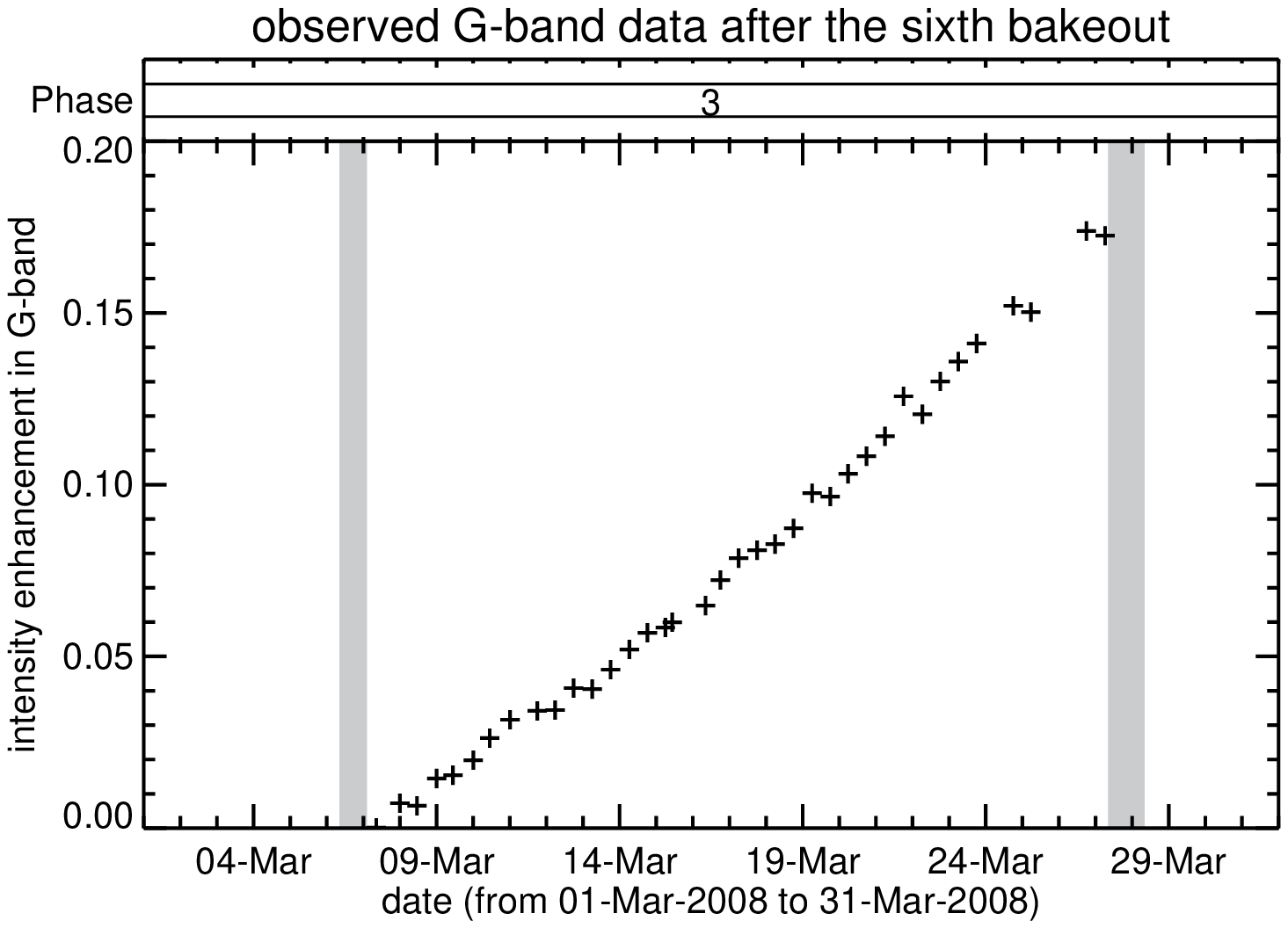}
            \hspace*{0.0cm}
            \includegraphics[width=6.0cm,clip=]{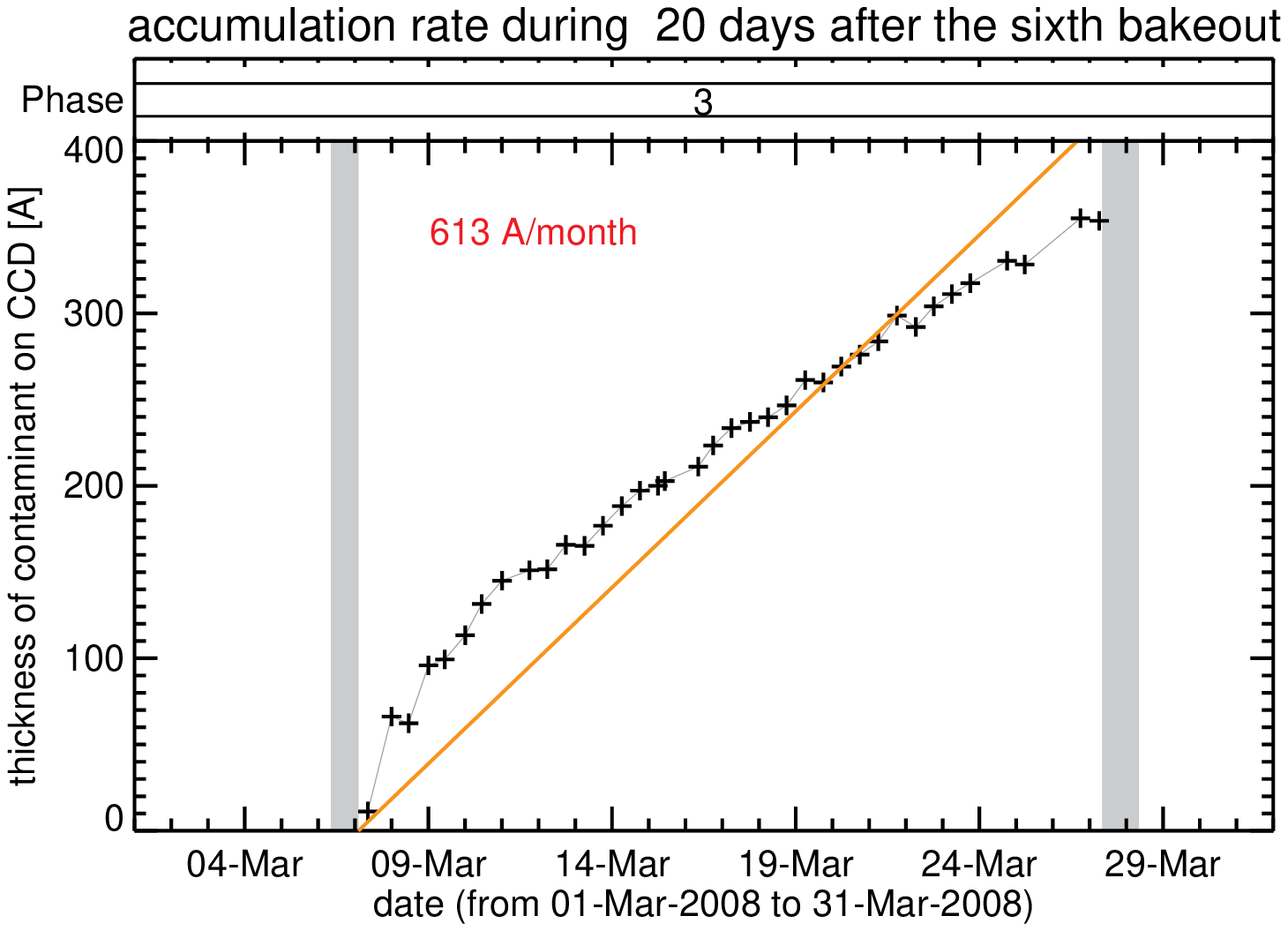}
           }
\caption{
G-band data (left panel) and estimated thickness of contaminant on the CCD (right panel) after the sixth bakeout.
}
\label{fig:6th bakeout}
\end{figure}

We continuously monitor the temporal evolution of G-band intensity (Figure~\ref{fig:G-band}).
As seen in Figure~\ref{fig:G-band}, only a portion of an entire cycle of the intensity enhancement,
namely initial rise of the intensity, is available for G-band data for each bakeout after 1 February 2008 (after the third bakeout).
We apply the empirical model given by Equation~(\ref{eq:empirical model}) to that period.
The left panel of Figure~\ref{fig:6th bakeout} is the observed G-band intensity enhancement
after the sixth bakeout, and the right panel shows the thickness of contaminant on the CCD derived from the model.
The orange line in the right panel shows the linear least-square fit to the thickness,
which yields an accumulation rate of contaminant on the CCD of about 613~{\AA}~month$^{-1}$,
where 30~days are taken as one month unless otherwise noted.
In this article, we assume a constant accumulation rate for the contaminant between two successive bakeouts.

We note that the X-ray intensity ratio and G-band intensity are
recovered to almost the same level by each CCD bakeout as seen in Phase~3 in the bottom panel of Figures~\ref{fig:DN QS} and \ref{fig:G-band}, respectively.
This suggests that after each bakeout the contaminant reduces to the same thickness.
Furthermore, the recovered G-band intensity in Phase~3 is consistent with the G-band intensity taken
just after the end of Phase~1 (until which we expect no contaminant on the CCD due to continuous CCD bakeout since launch). 
Hence, we conclude that this ``same thickness" is actually zero.

Table~\ref{tbl:bakeout} in Appendix~\ref{sec:CCD bakeout} summarizes the thus-derived accumulation rate of the contaminant between each bakeout,
and Figure~\ref{fig:contam on CCD} indicates temporal evolution of contaminant thickness on the CCD.
The similar plots to Figure~\ref{fig:6th bakeout} after all bakeouts are distributed
with \textsf{Solar Software (SSW)}
as \url{SSW_DIR/hinode/xrt/idl/response/contam/xrt_contam_on_ccd.pdf}, where \textsf{SSW\_DIR} indicates SSW directory
in your environment. This file and database of contaminant thickness will be regularly updated as the CCD is baked out repeatedly.

\subsubsection{Spatial distribution of the contaminant across the CCD}
\label{subsubsec:spatial distribution}

\begin{figure}
\centerline{\includegraphics[width=5.5cm,clip=]{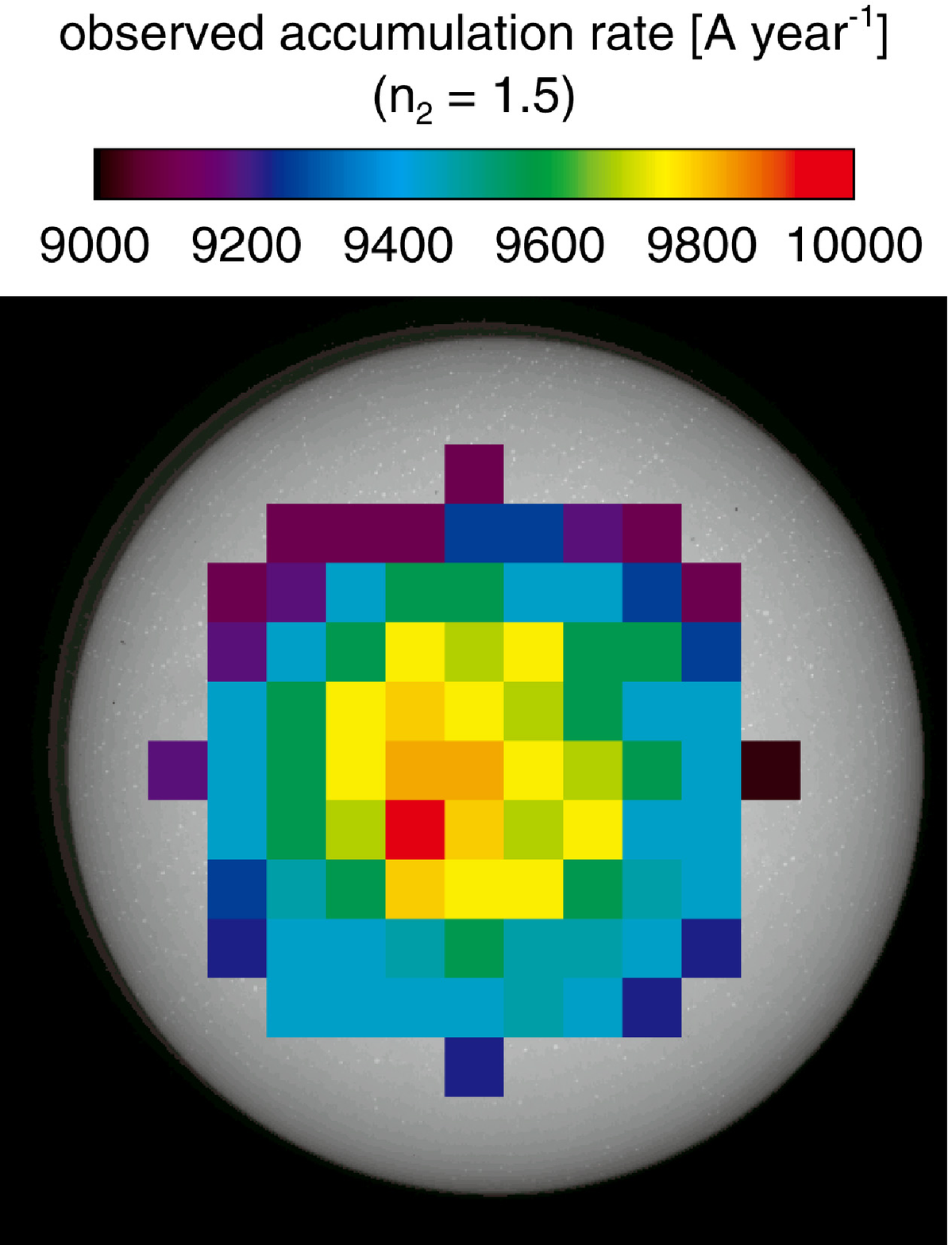}
            \hspace*{0.2cm}
            \includegraphics[width=5.5cm,clip=]{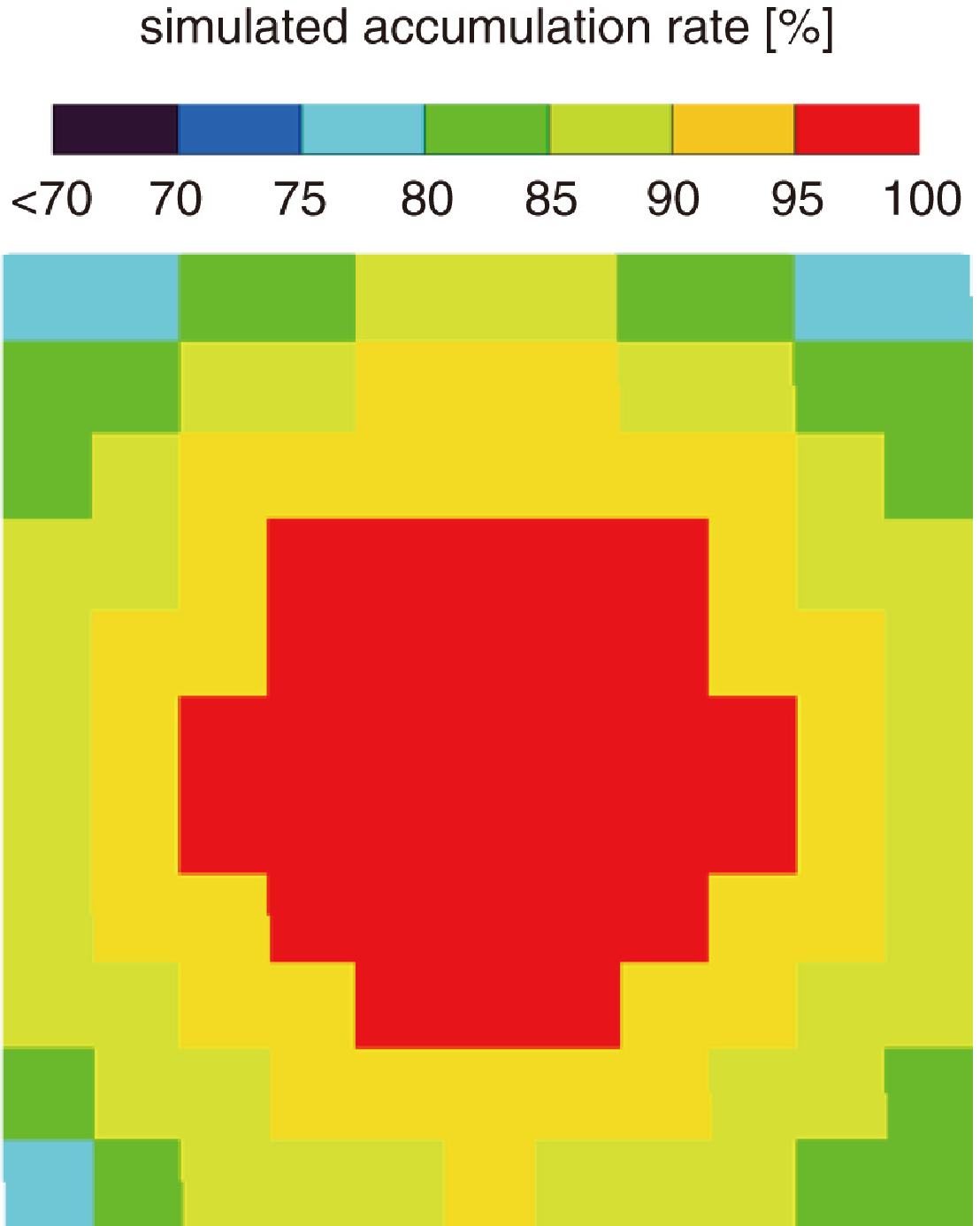}
           }
\vspace{3.0mm}
\caption{
Observed and simulated spatial distribution of the contaminant across the CCD.
The left panel is the observed accumulation rate map with G-band intensity oscillation.
The background is a full-Sun (synoptic) image in G-band.
The image size is the full CCD area ($2048 \times 2048$ pixels).
Each colored box corresponds to $128 \times 128$ pixels on the CCD.
The right panel shows the simulated accumulation-rate map under the condition where
the sources of contaminant located in the forward direction from the CCD.
The simulated area in the right panel is also full CCD area.
}
\label{fig:contam rate map}
\end{figure}

To monitor the accumulation rate of contaminant on the CCD with
G-band data, we have used the 512~$\times$~512 pixels at the center of the CCD.
On the other hand, when looking into fine structure of contaminant accumulation for the entire imaging area of the CCD,
we note that there is certain spatial distribution of accumulation rate of contaminant across the CCD
(see the left panel of Figure~\ref{fig:contam rate map}).
The accumulation rate decreases towards the edges of the CCD from its center.
The difference of accumulation rate between the CCD center and outermost area
in the left panel of Figure~\ref{fig:contam rate map} is about 10\%.

By building a numerical contamination model in which geometry and temperature of structures
around the CCD have been incorporated, \inlinecite{ura08} studied the expected spatial distribution of contaminants
accumulated on the CCD (see the right panel of Figure~\ref{fig:contam rate map}).
The resultant distribution was quite similar to the observed one (see Figure~\ref{fig:contam rate map})
when the contaminants come from
the forward direction of the CCD, while not when the contaminants come from the side
of the CCD where the pre-amplifier unit for the CCD is located.
The similarity is seen not only in the distribution of the contaminant
but also in the relative difference in the accumulation rate of about 10\%
between the CCD center and the outermost area in the observed data (left panel of Figure~\ref{fig:contam rate map}).
Since the edge of the CCD is located outside the solar disk (see the left panel of Figure~\ref{fig:contam rate map})
with synoptic G-band exposure,
we can not measure the accumulation rate at the edges of the CCD.
However, on the basis of the comparison between the results of observation and numerical simulation, we conclude
that the difference of accumulation rate between the CCD center and its edges is about 20\%.

\subsection{Analysis of Contaminant with X-Ray Data}
\label{subsec:contam analysis with X-ray}

Next, we examine whether the decrease in quiet-Sun X-ray intensity shown in Figure~\ref{fig:DN QS}
is consistent with the effect of the contaminant accumulated on the CCD
whose thickness derived from G-band data.
For this analysis, we used data sets in the period between the second and third bakeouts (in Phase~3),
because of availability of data sets both in G-band and in X-rays.

\begin{figure}
\centerline{\includegraphics[width=10.0cm,clip=]{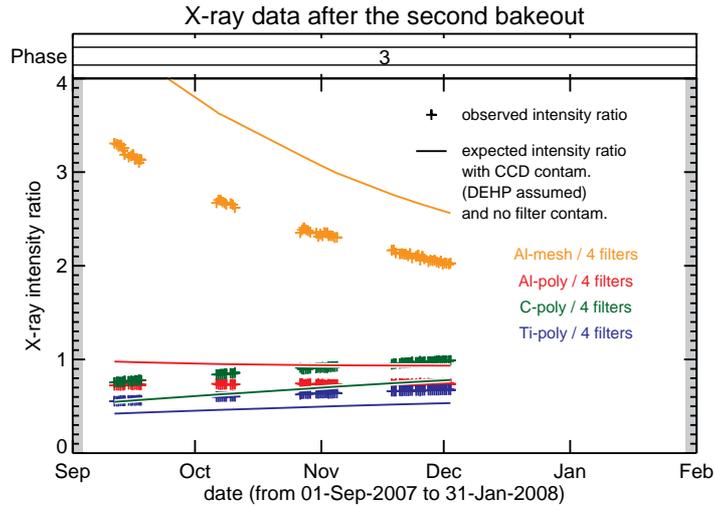}}
\caption{
Observed and expected X-ray intensity ratio after the second CCD bakeout in case of no
contaminant on focal-plane analysis filters.
}
\label{fig:X-ray no filter contam}
\end{figure}

Plus signs in Figure~\ref{fig:X-ray no filter contam} indicate temporal evolution of quiet-Sun intensity ratios
after the second CCD bakeout made from sets of simultaneously taken four filter images: Al-mesh, Al-poly, C-poly, and Ti-poly.
In order to remove possible variation of emission measure in different quiet-Sun regions, each intensity was normalized by the geometric mean of
intensities observed with the four filters,
\begin{equation}
\left(I_{\mathrm{Al}\mbox{\scriptsize -}\mathrm{mesh}} \times I_{\mathrm{Al}\mbox{\scriptsize -}\mathrm{poly}}
\times I_{\mathrm{C}\mbox{\scriptsize -}\mathrm{poly}} \times I_{\mathrm{Ti}\mbox{\scriptsize -}\mathrm{poly}} \right)^{1/4} ,
\label{eq:geometric mean}
\end{equation}
where $I_\mathrm{filter}$ is the X-ray intensity with each filter.
Data points in this plot were obtained from full-Sun images showing no active region in order to
remove the effect of scattered light from active regions.
Data gaps in this plot are due to the appearance of active regions.
In the plot, a significant decrease in intensity ratio with the Al-mesh filter
and a slight increase in those with the other three filters are clearly seen.
We consider that these changes are caused by the effect of accumulating contamination, which absorbs X-rays.
Note that the increase in intensity ratio with Al-poly, C-poly, and Ti-poly filters
is due to the large decrease in intensity with Al-mesh filter (see the top panel of Figure~\ref{fig:DN QS}).

Next, we calculate the expected intensity of the quiet Sun observed with the contaminant-accumulated CCD
using the differential emission measure (DEM) at the quiet Sun.
From Equation~(\ref{eq:A_eff}), the effective area of XRT $A_\mathrm{eff}^\mathrm{contam}$, including the effect of contamination, is given by
\begin{equation}
A_\mathrm{eff}^\mathrm{contam} = A_\mathrm{eff} \times \mathcal{T}_\mathrm{contam}
\label{eq:A_eff contam}
\end{equation}
where $\mathcal{T}_\mathrm{contam}$ is the transmission of the contaminant.
For the DEM of the quiet Sun, we adopted the DEM profile derived by \inlinecite{bro06}
with an assumption that the profile of the quiet-Sun DEM is almost constant in time.
The procedure for deriving the expected intensity from the thus-assumed DEM is detailed in Appendix~\ref{subsec:when XRT observes the solar spectra}.
We will discuss the effect of the difference in the DEM model on our analysis in Appendix~\ref{subsec:discussion and summary 3}.

By adopting the thickness of contaminant on the CCD (from the G-band data) for deriving $\mathcal{T}_\mathrm{contam}$,
the combination of $A_\mathrm{eff}^\mathrm{contam}$ and the DEM profile gives the expected profiles of the intensity ratio,
which are overlaid in Figure~\ref{fig:X-ray no filter contam}.
(The chemical composition of DEHP (Diethylhexyl phthalate: C$_{24}$H$_{38}$O$_4$) is adopted as the contaminant on the CCD for making the profiles in the figure;
see Appendix~\ref{subsubsec:X-ray analysis in Phase 3} for details.)
The figure clearly shows that the observed and expected X-ray intensity ratios are inconsistent with each other.
Especially, in addition to the large deviation in Al-mesh filter, although the expected intensity ratio with the Al-poly filter is larger than that for C-poly filter,
the observed intensity ratios with these two filters are contrary to expectation.
In the following, we investigate the cause of this inconsistency.

\subsubsection{X-ray Analysis for Phase~3
}
\label{subsubsec:X-ray analysis in Phase 3}

In order to solve the intensity controversy between Al-poly and C-poly filter,
in this section, we characterize the contaminant accumulated on each FPAF.
We note that contamination on each FPAF is the only possible candidate to account this controversy,
because the other optical elements are common for the data taken with any of the X-ray filters.
As discussed in Appendix~\ref{sec:thermal distribution inside XRT}, it is expected that there was no additional
accumulation of contaminant on the FPAFs in Phase~3.
As the period shown in Figure~\ref{fig:X-ray no filter contam} belongs to Phase~3,
the data shown in the figure is suitable for identifying thickness of contaminant
on each FPAF.
Moreover, using these data, we can also identify the properties of contaminant material, \textit{i.e.} the chemical composition, density and refractive index of contaminant material,
because the transmission of the contaminant [$\mathcal{T}_\mathrm{contam}$] in Equation~(\ref{eq:A_eff contam})
depends on the above properties as shown by Equation~(\ref{eq:T}),
where the thickness [$d$] of contaminant is estimated with refractive index of contaminant material (see Equation~(\ref{eq:osc})),
and the attenuation length [$l_\mathrm{att}\left(\lambda\right)$] is determined by the chemical composition and density.

\begin{table}
\caption{Candidates of contaminant material}
\label{tbl:contaminant}
\begin{tabular}{cccc}
\hline
\# & material                                    & density            & refractive index   \\
\hline
1  & diethylhexyl phthalate$^{(a)}$              & 0.986 g cm$^{-3}$  & $n = 1.5^{(b)}$    \\
   & (DEHP) C$_{24}$H$_{38}$O$_{4}$              &                    &                    \\
\\
2  & tetramethyl tetraphenyl trisiloxane$^{(c)}$ & 1.07  g cm$^{-3}$  & $n = 1.6^{(d)}$    \\
   & (MPS)  C$_{28}$H$_{32}$O$_{2}$Si$_{3}$      &                    &                    \\
\\
3  & polydimethylsiloxane$^{(c)}$                & 0.971 g cm$^{-3}$  & $n = 1.4^{(e)}$    \\
   & (PDMS) (C$_{2}$H$_{6}$OSi)$_{n}$            &                    &                    \\
\hline
\end{tabular}
\begin{description}

\item[$^{(a)}$] DEHP is a plasticizer.
\item[$^{(b)}$] catalog of chemical reagent by Wako Pure Chemical Industries, Ltd
\item[$^{(c)}$] MPS and PDMS are classified Siloxane.
\item[$^{(d)}$] \inlinecite{osa82}
\item[$^{(e)}$] catalog of TORAY Silicone, Ltd
\end{description}
\end{table}

Before proceeding to the above calibration,
we first try to identify the possible material of the contamination.
We picked three candidate materials, which are listed in Table~\ref{tbl:contaminant}.
These three candidates are widely used for satellites, and are well known as possible source of contamination:
DEHP (Diethylhexyl phthalate: C$_{24}$H$_{38}$O$_4$) is the representative of a long-chain organic compound which consists of carbon, hydrogen, and oxygen.
MPS (tetramethyl tetraphenyl trisiloxane: C$_{28}$H$_{32}$O$_{2}$Si$_{3}$) and PDMS (polydimethylsiloxane: (C$_{2}$H$_{6}$OSi)$_{n}$) are long-chain organic compounds
including silicon.
Although these candidate of contaminant classified into two types, namely a long-chain organic compound without and with silicon,
their density and refractive index are almost the same: 1~g~cm$^{-3}$ and 1.5, respectively.
We note that, though DEHP is not used in the XRT, if the actual contaminant material is a long-chain organic compound without silicon,
the chemical composition, density, and refractive index would be similar to DEHP.
Hence, in our analysis, we refer to DEHP as the representative of a long-chain organic compound without silicon.

\begin{table}
\caption{Fitting parameter to observed X-ray data}
\label{tbl:parameter}
\begin{tabular}{cccc}
\hline
parameter                     &   examined range      & expected value       & best fit value \\
\hline
contaminant                   &   DEHP, MPS or PDMS   & --                   & DEHP           \\
contam. on Al-mesh filter     &    0\,--\,4000~{\AA}  & --                   & 1200~{\AA}     \\
contam. on Al-poly filter     &    0\,--\,4000~{\AA}  & --                   & 2900~{\AA}     \\
contam. on C-poly filter      &    0\,--\,4000~{\AA}  & --                   &  500~{\AA}     \\
contam. on Ti-poly filter     &    0\,--\,4000~{\AA}  & --                   &  400~{\AA}     \\
metal thick. of pre-filter    & 1000\,--\,2000~{\AA}  & 1538~{\AA}$^{(a)}$   & 1550~{\AA}$^{(b)}$     \\
oxide thickness$^{(c)}$       &    0\,--\,75~{\AA}    & $<$ 75~{\AA}$^{(d)}$ & 75~{\AA} / 0~{\AA}$^{(e)}$      \\
\hline
\end{tabular}
\begin{description}
\item[$^{(a)}$] This value is based on the certification sheet (see Figure~\ref{fig:PF} and Table~\ref{tbl:FPAF}).
\item[$^{(b)}$] The expected metal thickness when it was fabricated. At fabrication, the metals had not oxidized at all.
                The on-orbit pure and oxidized-metal thicknesses are shown in Table~\ref{tbl:FPAF}.
\item[$^{(c)}$] The oxide thickness of metal filters. We examine it for open side and for polyimide side.
\item[$^{(d)}$] The oxide thickness of well-oxidized Aluminum filter is 150~{\AA} \cite{pow90}.
                Hence, the expected oxide thickness for one side (open side) is $<$ 75~{\AA}.
\item[$^{(e)}$] Best fit value of ``oxide thickness for open side" / ``oxide thickness for polyimide side".
\end{description}
\end{table}

We obtained filter parameters that best fit to the observed intensity ratio.
The parameters and their examined range for the fitting are summarized in Table~\ref{tbl:parameter}.
The examined range of contamination thicknesses accumulated on the four thin FPAFs is 0\,--\,4000~{\AA},
since contaminant thickness larger than 4000~{\AA} gave results that are inconsistent with the observed data.
In this calibration, we cannot investigate the contamination thickness on thicker FPAFs, because
the thicker FPAFs cannot observe the quiet Sun enough.
The metal thickness of the pre-filter was examined in the range of 1000\,--\,2000~{\AA}, \textit{i.e.} approximately
covering the range of $\pm$~500{\AA} from the thicknesses provided by the manufacturer.
According to \inlinecite{pow90}, the oxide thickness of a well-oxidized stand-alone aluminum filter is 150~{\AA} in total for both sides,
\textit{i.e.} 75~{\AA} for each side. Hence, we examine oxide thickness for open side of a filter in a range of 0\,--\,75~{\AA}
under the assumption that any metal filters were oxidized with the same thickness.
Some of the FPAFs have a support made of polyimide.
We also investigated oxide thickness for polyimide side in the range of 0\,--\,75~{\AA} under the above assumption.

\begin{figure}
\centerline{\includegraphics[width=10.0cm,clip=]{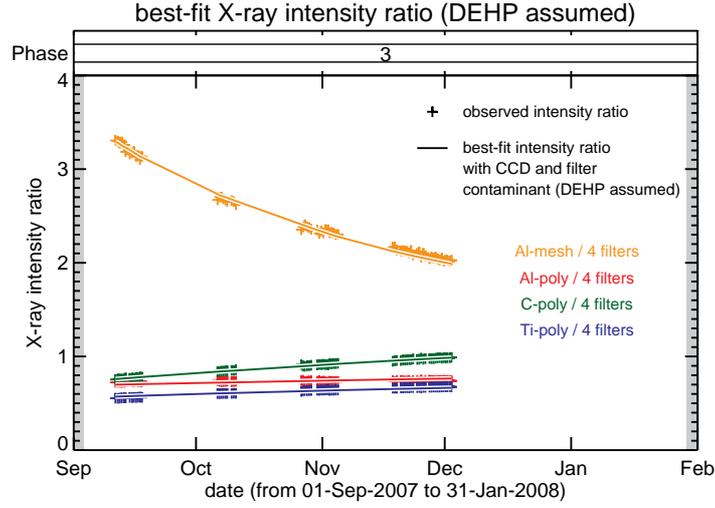}}
\caption{
Best-fit X-ray intensity ratio after the second CCD bakeout, where for this analysis we have taken the material to be similar to DEHP.
}
\label{fig:X-ray DEHP filter contam}
\end{figure}

\begin{figure}
\centerline{\includegraphics[width=6.0cm,clip=]{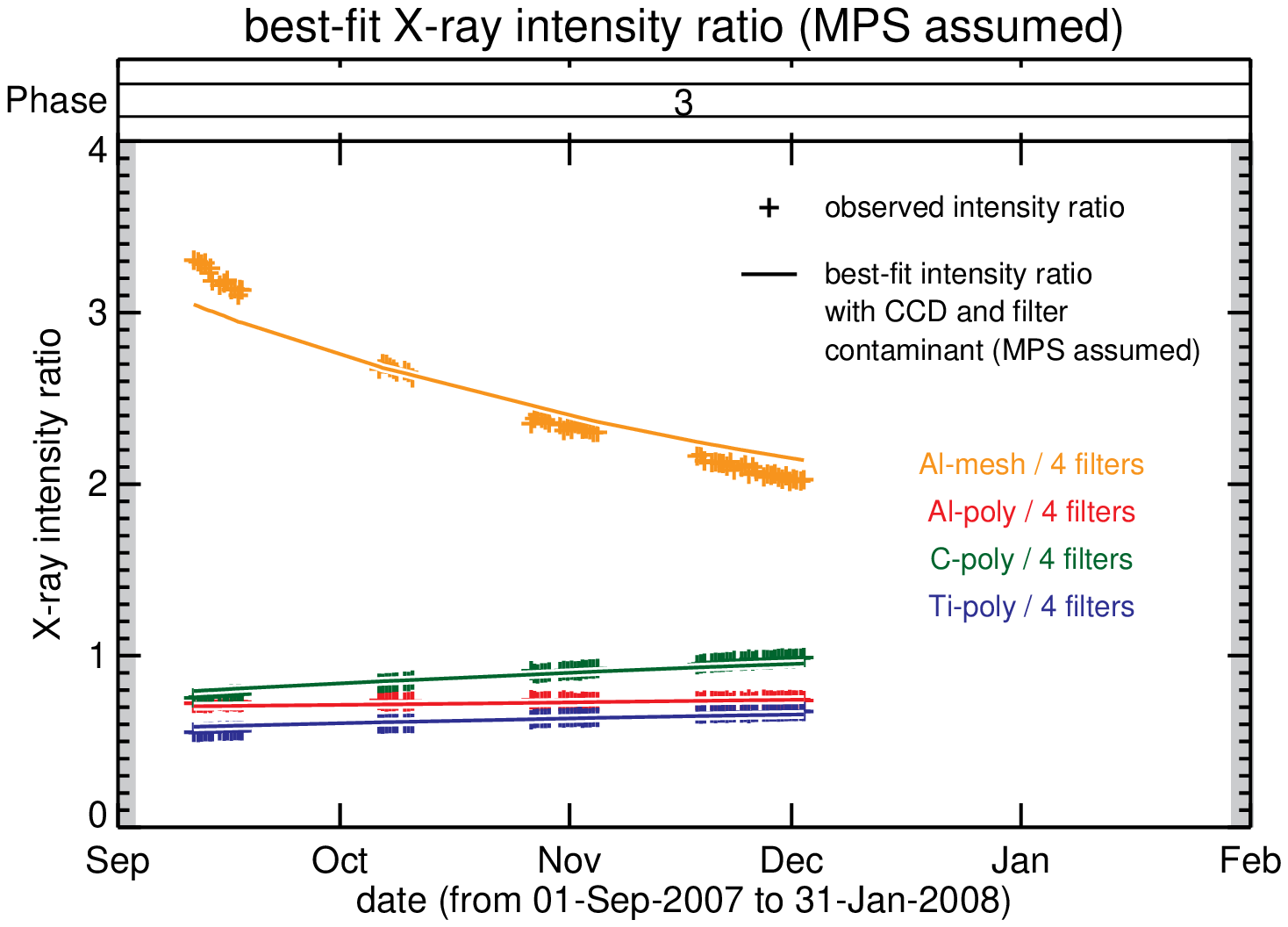}
            \hspace*{0.0cm}
            \includegraphics[width=6.0cm,clip=]{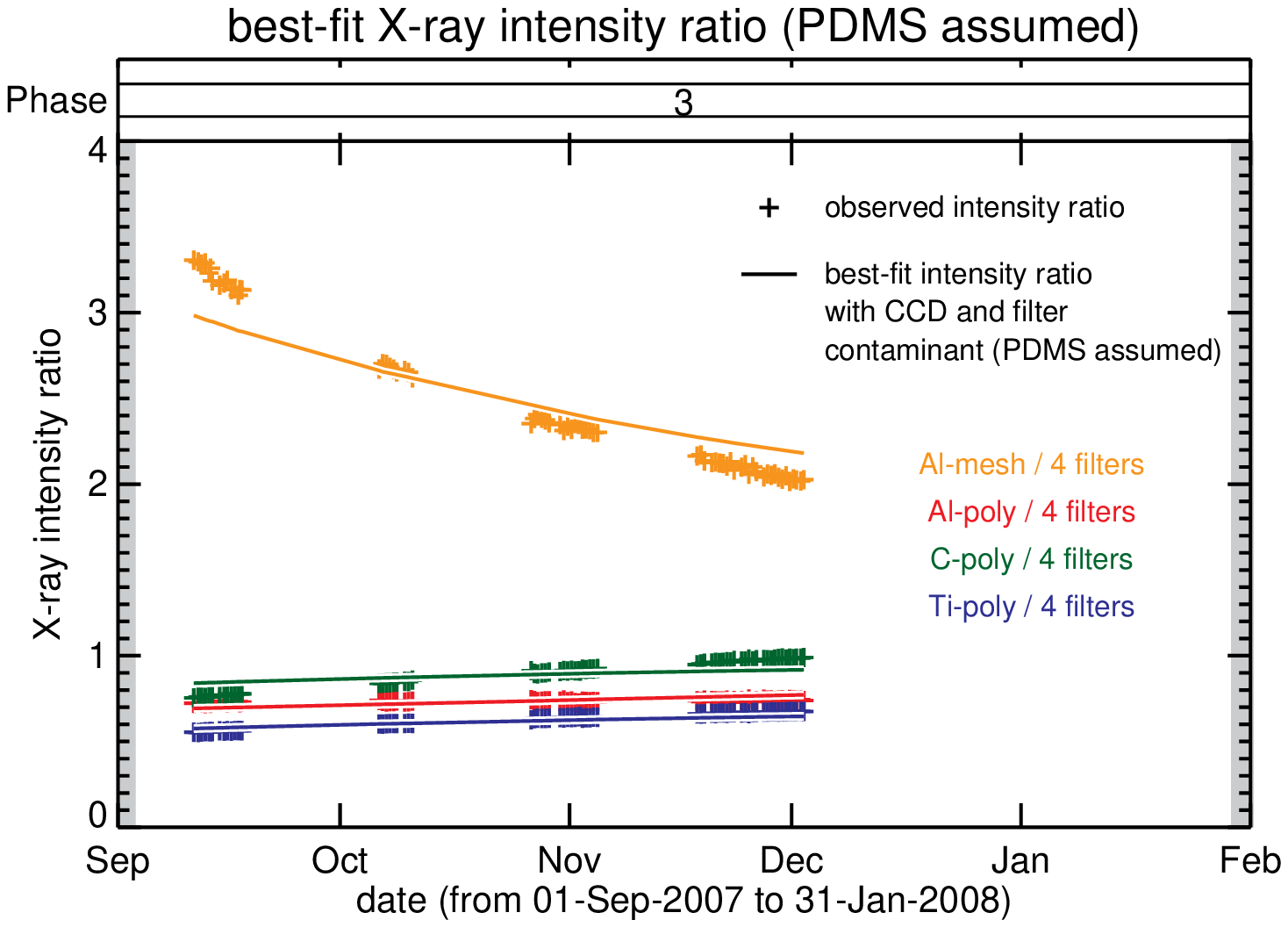}
           }
\caption{
Best-fit X-ray intensity ratio after the second CCD bakeout, where the materials of contaminant are assumed to be MPS (left panel) and PDMS (right panel).
}
\label{fig:X-ray Si filter contam}
\end{figure}

Figure~\ref{fig:X-ray DEHP filter contam} shows the estimated X-ray intensity ratio from the best-fit parameters
summarized in Table~\ref{tbl:parameter},
including material for the contaminant, which was identified to be consistent with DEHP, namely the representative material of a long-chain organic compound without silicon.
The area-weighted average thickness of the pre-filter was calibrated to be 1550~{\AA},
which is the expected metal thickness when it was fabricated. At fabrication, the metals had not oxidized at all.
This value is very close to the value of 1538~{\AA} derived from the manufacturer-supplied information on thickness for each piece of the pre-filter (see Figure~\ref{fig:PF}).
Hence, considering the oxidization (using Equation~(\ref{eq:oxide}) of Appendix~\ref{sec:oxidization of metal}),
the on-orbit pure and oxidized-metal thicknesses of pre-filter are calibrated to be 1492~{\AA} and 75~{\AA}, respectively, as summarized in Table~\ref{tbl:FPAF}.
The calibrated thickness of contaminant on each FPAF turned out to be different.
The thickness of contaminant on the Al-poly filter was identified to be 2900~{\AA} which was the thickest.
We will see in Appendix~\ref{subsubsec:contam on FPAF in Phase 2}
that this difference in thicknesses is consistent with the frequency of filter usage.

Figure~\ref{fig:X-ray Si filter contam} indicates difference in the fit among different candidates
for the contaminant material. The left panel shows the best-fit result with
MPS, while right panel PDMS.
Clearly, these materials are not able to account for the measured profiles, especially that for
the Al-mesh filter.
Difference in chemical compositions among candidate materials for contamination
most affect transmission at longer wavelength X-rays,
hence intensity with the thinnest Al-mesh filter.
This results in the situation that the Al-mesh profiles show the largest sensitivity against to the different candidates.
Furthermore, for MPS and PDMS, the metal thickness of the pre-filter
which gives the best fit to the measurements turned out to be 1000~{\AA} and 1050~{\AA}, respectively.
These thickness are unlikely because of significant deviation from the averaged pre-filter thickness
measured by the manufacturer (1538~{\AA}).
Hence, we can conclude that the contaminant material is
a long-chain organic compound whose characteristics, namely chemical composition, density, and refractive index, are similar to DEHP.
We will discuss this identified contaminant material in Appendix~\ref{subsec:discussion and summary 0}.

\subsubsection{X-Ray Analysis for Phase~2 -- Contamination on the CCD}
\label{subsubsec:contam on CCD in Phase 2}

In Appendix~\ref{subsec:contam analysis with G-band}, we established the method to monitor
the contaminant on the CCD with G-band data. Using this method, we can measure the thickness of contaminant
on the CCD for the Phase~3 period. However, for Phase~2, we cannot apply this method because there is not enough G-band data.
Instead, we rely on the X-ray data for investigating accumulation of contaminant on the CCD.

\begin{figure}
\centerline{\includegraphics[width=6.0cm,clip=]{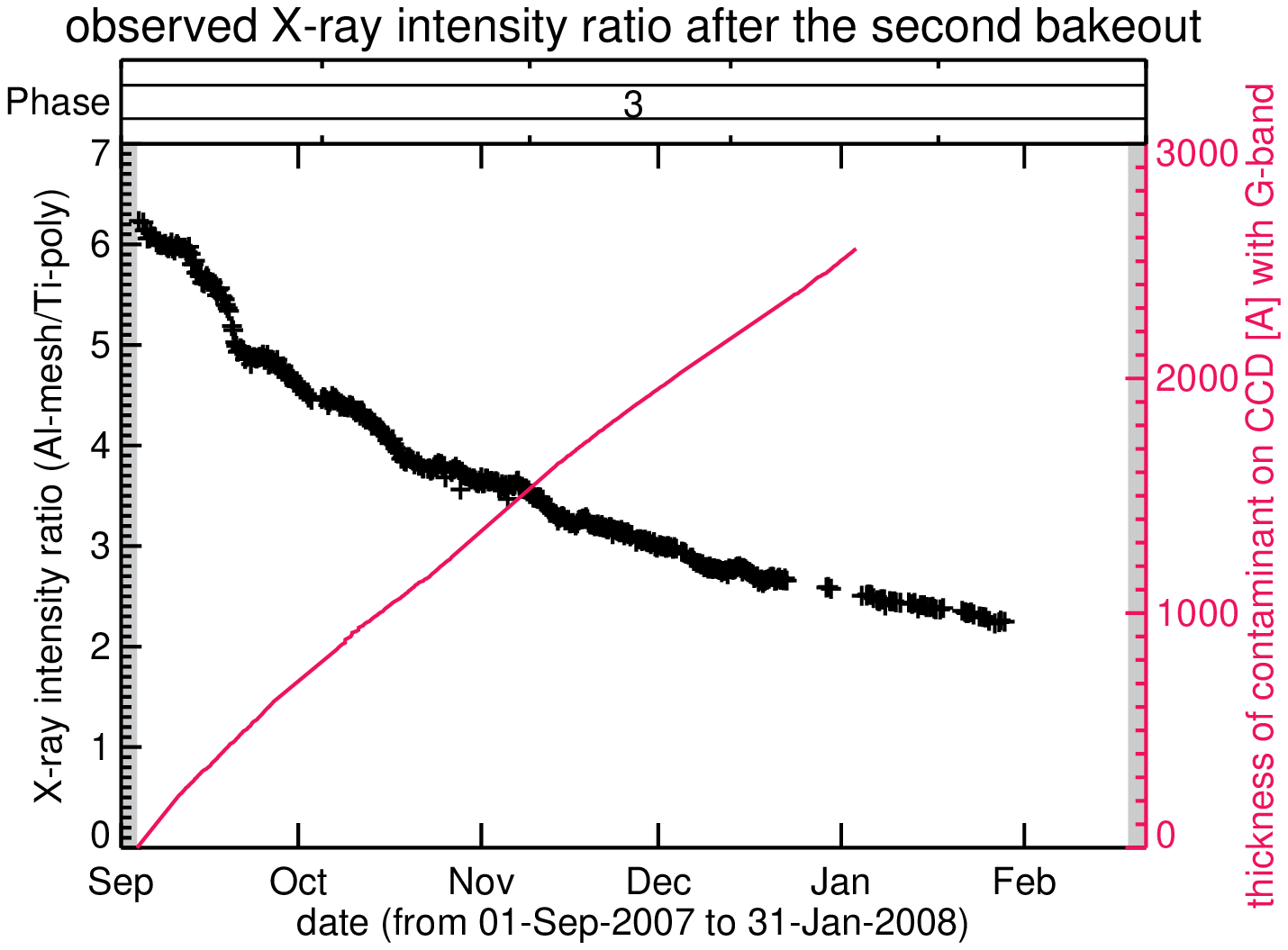}
            \hspace*{0.0cm}
            \includegraphics[width=6.0cm,clip=]{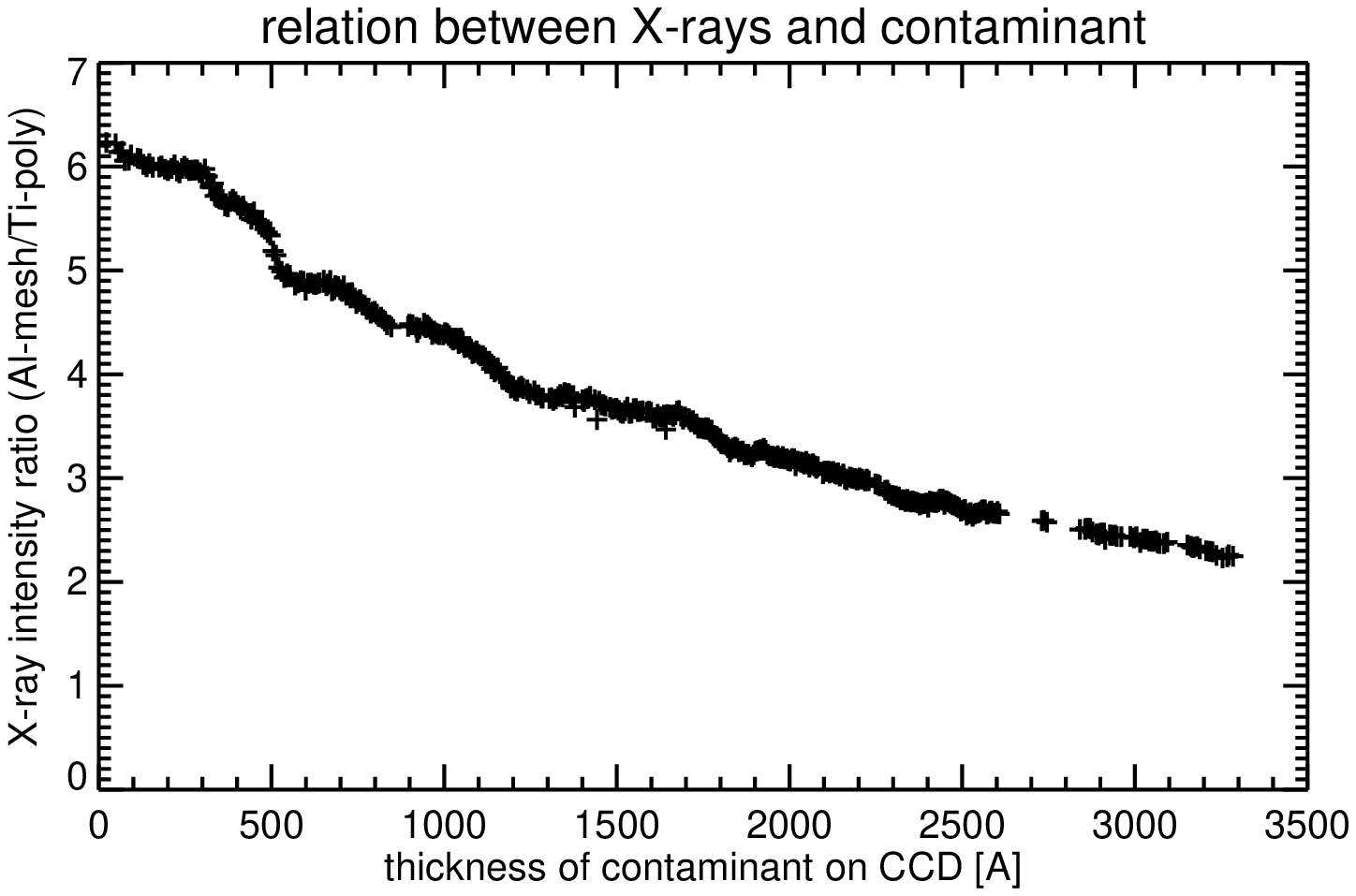}
           }
\caption{
left panel: Observed X-ray intensity ratio (black $+$) and
measured thickness of contaminant on the CCD with G-band (magenta line) after the second bakeout.
right panel: Relation between X-ray intensity ratio and thickness of contaminant on the CCD after the second bakeout.
}
\label{fig:X-ray model}
\end{figure}

For this purpose, let us first derive the relationship between G-band and X-ray data for the
period after the second bakeout (which took place in Phase~3) where both data are available.
The measured thickness of contaminant on the CCD with G-band data is shown in magenta in the left panel of
Figure~\ref{fig:X-ray model}. In this panel, the observed X-ray intensity ratio between
Al-mesh and Ti-poly filters is also shown in black color.
Using these data, we obtain the relation to convert the X-ray intensity ratio
to the thickness of contaminant on the CCD which is shown in the right panel.
As demonstrated in Appendix~\ref{sec:thermal distribution inside XRT},
the contaminant on FPAFs is constant in Phase~3.
Note the necessary condition for applying the right-panel relation of Figure~\ref{fig:X-ray model}
(hereafter, ``XSC relation", where XSC stands for X-ray-suggested CCD contamination)
is that the contaminant thicknesses on the FPAFs are the same as those in Phase~3.

\begin{figure}
\centerline{\includegraphics[width=10.0cm,clip=]{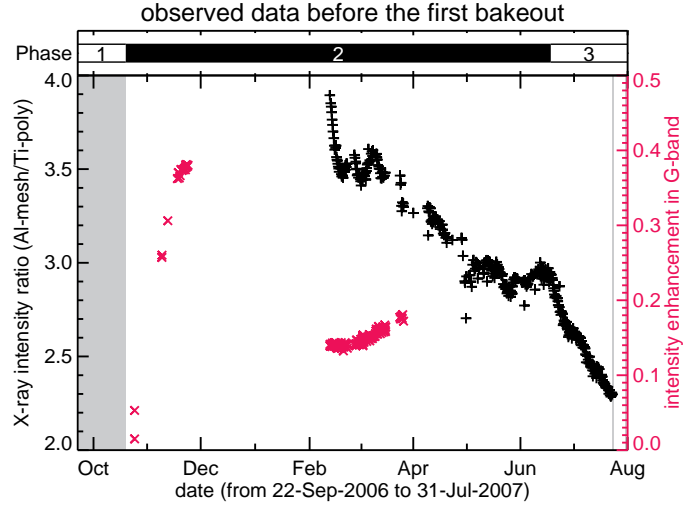}}
\caption{
Observed X-ray intensity ratio (black $+$) and intensity enhancement in G-band (magenta $\times$) before the first bakeout.
}
\label{fig:data 1st bakeout}
\end{figure}

\begin{figure}
\centerline{\includegraphics[width=10.0cm,clip=]{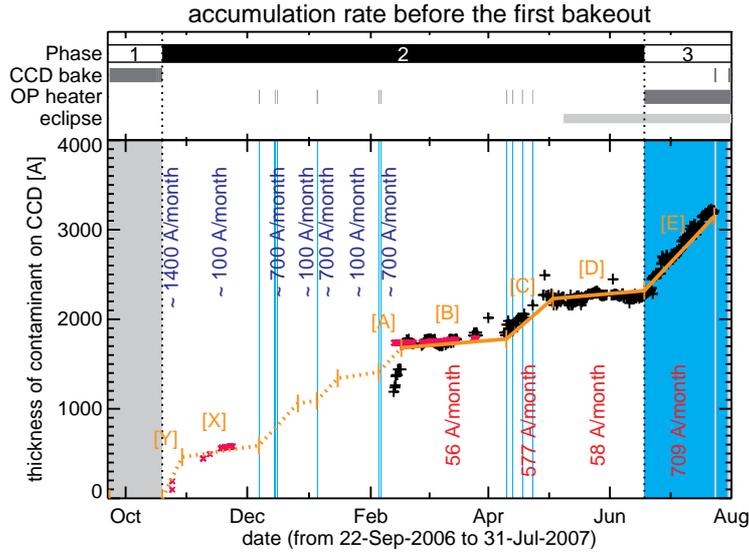}}
\caption{
Accumulation rate of contaminant on CCD before the first bakeout.
This rate is derived from two kinds of data sets:
X-ray intensity ratio between Al-poly and Ti-poly filters (black $+$),
and G-band intensity (magenta $\times$).
The orange solid and dotted lines show the measured and expected thickness of contaminant on the CCD, respectively.
The measured and expected accumulation rates are also shown with red and blue characters.
The [A]\,--\,[E], [X], and [Y] show the classified periods from the profile of contaminant thickness (orange lines).
}
\label{fig:1st bakeout}
\end{figure}

Now, let us check whether the above necessary condition is satisfied in Phase~2 too.
Figure~\ref{fig:data 1st bakeout} shows the temporal evolution of X-ray intensity ratio (black $+$) and
intensity enhancement in G-band (magenta $\times$) observed before the first CCD bakeout.
Note the entire period of Phase~2 is contained in the plot.
By applying the XSC relation to the observed X-ray intensity ratio,
we estimated the thickness of contaminant on the CCD as shown by the black pluses in Figure~\ref{fig:1st bakeout}.
Note the thickness was estimated by assuming the existence of contaminant on the FPAFs whose thicknesses are identical
to those in Phase~3. Thickness of the contaminant on the CCD derived with the G-band data is also plotted
in Figure~\ref{fig:1st bakeout} in magenta crosses.
This profile has some trends. We classify the profile into A\,--\,E as shown in Figure~\ref{fig:1st bakeout}.
In periods~A, C, and E, the thickness of contaminant increases at a certain rate,
but in B and D, the thickness is almost constant.
As discussed earlier, the variation of G-band data is solely affected by the contaminant on the CCD,
not by the contaminant on the G-band filter.
Hence, using the G-band data, we can monitor the contaminant on the CCD independently.
The magenta crosses in Figure~\ref{fig:1st bakeout} are the estimated thickness of contaminant on the CCD from G-band data.
As clearly seen, in period~B the thickness of contaminant on the CCD estimated
from X-ray intensity ratio (black $+$) is consistent with that derived with G-band data (magenta $\times$).
This suggests that the above necessary condition of the XSC relation is valid.
Now that we know that the thicknesses of contaminant on the FPAFs in period~B are the same as those in Phase~3
and that it is not likely the thickness would decrease in any certain period (note they did not decrease even with
FPAF temperature $\approx~+20^{\circ}$C in Phase~3),
we can conclude that the thicknesses of contaminant on the FPAFs are unchanged after (including) period~B
(after (including) the ``late period" of Phase~2 in Figure~\ref{fig:calibration}).
This means that, after period~B, the thickness of contaminant on the CCD can be measured from the XSC relation
as shown by the black $+$.
The solid orange lines after period~B show a linear fit to the measured thickness (black $+$).

For the interval after period~B, there are episodes of rapid accumulation of contaminant on the CCD (periods~C and E),
while the rest do not show much accumulation (periods~B and D).
We note that the behavior of contaminant accumulation on the CCD is closely related to
the status of the operational heater as follows:
\begin{itemize}
\item From turning on the operational heater to about ten days after turning off,
which we call ``contamination period (CP)",
the contaminant rapidly accumulated on the CCD at a rate of $< 700$~{\AA}~month$^{-1}$ (periods~C and E
but see comments below).
The periods when the operational heater was on are indicated by blue areas in Figure~\ref{fig:1st bakeout}.
\item From about ten days after turning off to the next turning on,
which we call ``small-contamination period (SCP)",
the accumulation rate is small, namely $< 100$~{\AA}~month$^{-1}$ (periods~B and D).
\end{itemize}
Hereafter, we call this relationship the ``ODC relation" where ODC stands for operational heater driven contamination.
This relationship suggests that the contamination is triggered by the operation of the operational heater.

So far, we have seen that the XSC relation can be applied for the period from period~B
(``late period" of Phase~2 in Figure~\ref{fig:calibration}).
Here let us check if this relation still holds for the rest of Phase~2, namely period before (including) period~A on Phase~2 (``early period" of Phase~2).
From Figure~\ref{fig:XRT status}, there are the following trends in the temperatures:
\begin{itemize}
\item[\textit{i})] Until March 2007,
when the operational heater was on, the temperature around the FPAFs was $\approx~+20^{\circ}$C and at the CCD $\approx~-65^{\circ}$C.
When the operational heater was off, the temperature around the FPAFs was $\approx~+5^{\circ}$C and at the CCD $\approx~-70^{\circ}$C.
\item[\textit{ii})] From March 2007 to May 2007, the temperatures around the FPAFs and at the CCD increased a little due to the increased Earth albedo
towards the beginning of the eclipse season.
\item[\textit{iii})] In the eclipse season (from May 2007), the temperatures around the FPAFs
decreased due to the decreased time of the solar illumination, while the temperature at the CCD increased due to the increased Earth albedo.
\end{itemize}
These changes in the temperatures were negligible for the accumulation rate of the contaminant on the CCD,
since the measured accumulation rates were almost constant during periods~B and D, which correspond to
the above trends~\textit{i})\,--\,\textit{ii}) and trend~\textit{iii}), respectively.
Therefore, we expect the ODC relation is also applicable to the period before (including) period~A.

There are not sufficient X-ray nor G-band data for the interval between periods~X and A 
(``early period" of Phase~2 in Figure~\ref{fig:calibration}) to measure contaminant thickness on the CCD.
Therefore, we rely on the ODC relation for this interval and check if the relation consistently connects the periods from A back to X.
To apply the ODC relation, we assumed that the contaminant accumulated on the CCD at a rate of $700$~{\AA}~month$^{-1}$ for CP while $100$~{\AA}~month$^{-1}$ for SCP,
as shown by dotted orange lines in Figure~\ref{fig:1st bakeout}.
In performing the extrapolation with the ODC relation, we started from period~B and backwards toward period~X.
Note that the applied rates are somewhat larger than the measured rates in periods~C ($580$~{\AA}~month$^{-1}$) and periods~A and D ($60$~{\AA}~month$^{-1}$).
This is because we expect that the accumulation rate in early period in Phase~2 would be higher than the late period in Phase~2.
Note here, that the difference in contaminant thicknesses at the beginning of period~X is small ($\approx$~200~{\AA})
even if we adopt the $580$~{\AA}~month$^{-1}$ and $60$~{\AA}~month$^{-1}$ pair.
For period~Y, which is the interval between the end of Phase~1 (until which we expect no contaminant on the CCD due to continuous CCD bakeout since launch)
and the beginning of period~X, a constant accumulation rate of contaminant on the CCD was assumed.
We note that the duration of period~Y is determined to be ten days as shown in Figure~\ref{fig:1st bakeout},
since we confirmed that the X-ray intensity in the quiet Sun observed with Al-poly filter
was almost constant
for the period when more than ten days passed after the beginning of Phase~2.
Note that there is no data available in the first ten days.
The estimated accumulation rate in period~Y is $\approx$~1400~{\AA}~month$^{-1}$.
This high rate might be caused by the hot temperature of the telescope in Phase~1.
The temperature of the front-end portion of the telescope turned out to be significantly hotter than the temperature predicted before launch.
We guess that in Phase~1 a great deal of contaminant was created and filled the inside of the telescope.
This contaminant rapidly accumulated on the CCD after the CCD bake heater was
turned off (in the beginning of Phase~2, \textit{i.e.} that of period~Y).

Finally, let us check whether the expected contaminant thickness on the CCD is reliable.
In periods~X and Y, the expected thickness (dotted orange line) is consistent with
the contaminant thickness measured with G-band data (magenta $\times$).
This result strongly supports our expectation.
As discussed above, now we understand the accumulation of contaminant on CCD in all periods (see Figure~\ref{fig:contam on CCD}).

We note that in period~A there is a discrepancy between the contaminant thickness on the CCD
derived with the ODC extrapolation (dotted orange line)
and the thickness derived with the XSC relation (black pluses $+$).
This discrepancy will be discussed in the next section.

\subsubsection{X-Ray Analysis for Phase~2 -- Contamination on the FPAFs}
\label{subsubsec:contam on FPAF in Phase 2}

From discussion in Appendix~\ref{sec:thermal distribution inside XRT} and
Appendix~\ref{subsubsec:contam on CCD in Phase 2},
we know the following for the contamination on the FPAFs:
\begin{itemize}
\item The contaminant did not accumulate in Phase~3.
\item In Phase~2, the contaminant did not accumulate after the beginning of period~B.
\end{itemize}
Hence we conclude that the FPAFs contaminant must have accumulated by the end of period~A.

\begin{figure}
\centerline{\includegraphics[width=10.0cm,clip=]{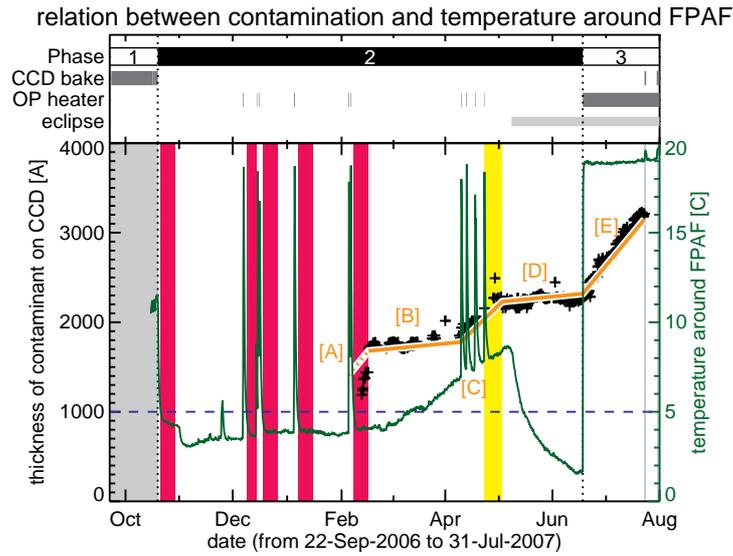}}
\caption{
Relation between the temperature around the focal-plane analysis filter (green line) and
estimated thickness of contaminant on the CCD with X-ray intensity ratio (orange line).
The contaminant would accumulate on FPAFs in the ``cool contamination periods (CCPs)" shown in magenta.
While, in the ``warm contamination period (WCP)" shown in yellow, the contaminant would not accumulated.
The blue dashed line indicates the boundary temperature between CCP and WCP ($5^{\circ}$C).
}
\label{fig:relation between contam and FPAF temperature}
\end{figure}

Figure~\ref{fig:relation between contam and FPAF temperature} shows temporal profiles of
the temperature around the FPAFs and estimated thickness of contaminant on the CCD.
In period~A, there is a discrepancy between the thickness of contaminant on the CCD derived with the ODC relation (dotted orange line)
and the thickness derived with the XSC relation (black pluses $+$).
Given the successful ODC extrapolation described in the previous section,
this discrepancy most likely implies that the thickness of contaminant on the FPAFs that we assumed (being the same as those in Phase~3)
for applying XSC relation in period~A is not correct.
In other words, the contaminant accumulated on the FPAFs at least in period~A.

Next, we investigate why the FPAFs contamination took place in period~A while not in period~C.
Both periods~A and C are ``contamination periods (CPs)"
defined in Appendix~\ref{subsubsec:contam on CCD in Phase 2}.
As we mentioned, in these CPs, the contaminant accumulated on the CCD.
The difference between periods~A and C is the temperature around the FPAFs as shown in Figure~\ref{fig:relation between contam and FPAF temperature}.
The temperature around the FPAF in period~A (about $4^{\circ}$C) was cooler than in period~C (about $8^{\circ}$C).
We classify the CPs into two according to the temperature around the FPAFs:
For the periods when the temperature is lower than $\approx$~5\,--\,8$^{\circ}$C,
we call ``cool contamination period (CCP)",
while ``warm contamination period (WCP)" for the rest of the CP.
The difference in temperature leads to the hypothesis that the contaminant generated by turning on the operational heater accumulated on the cool FPAFs in CCPs.
On the other hand, the high temperature around the FPAFs in the WCPs prevents the contaminant from accumulating.
If this hypothesis is correct, the contaminant accumulated on the FPAFs in the five CCPs shown in magenta
in Figure~\ref{fig:relation between contam and FPAF temperature}.
On the other hand, in the WCP shown in yellow, the contaminant did not accumulate on the FPAFs.
Although the difference of temperature around the FPAFs is only an indirect evidence,
there is not any reasonable timing for the FPAFs contamination other than the CCPs.
On the basis of this result, we expect that the contaminant did not accumulate on the FPAFs in Phase~1,
since the temperature around the FPAFs was higher than $10^{\circ}$C in that phase.

\begin{figure}
\centerline{\includegraphics[width=10.0cm,clip=]{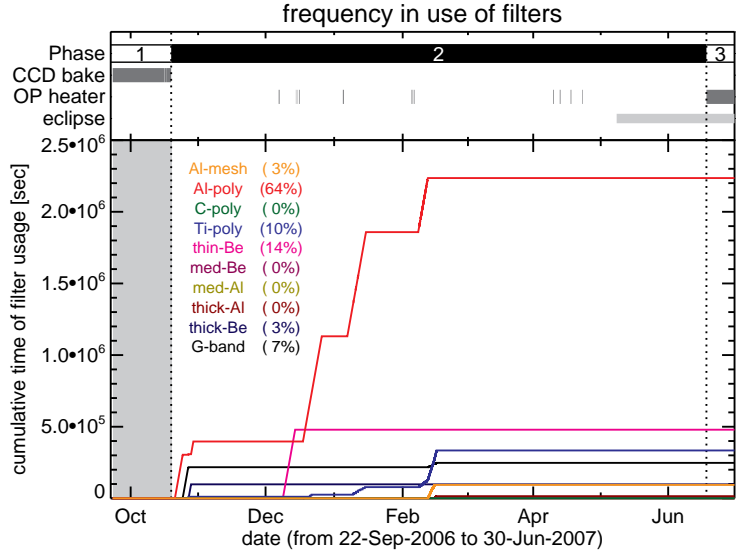}}
\caption{
Cumulative time of filter usage
during the ``cool contamination periods" colored by magenta in Figure~\ref{fig:relation between contam and FPAF temperature}.
The final percentage of cumulative time for each filter is described in a parenthesis, where the Al-poly filter has the largest cumulative time.
}
\label{fig:filter usage}
\end{figure}

\begin{figure}
\centerline{\includegraphics[width=4.0cm,clip=]{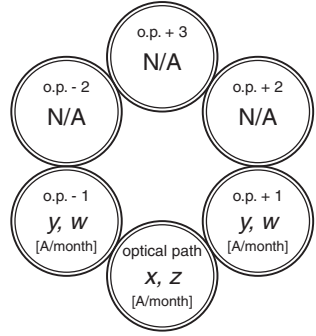}}
\vspace{3.0mm}
\caption{
Modeled accumulation rate of contaminant on each focal-plane analysis filter.
$x$ and $y$ are the rates of filters mounted on filter wheel~1,
and $z$ and $w$ are rates of filters on filter wheel~2.
}
\label{fig:contam filter model}
\end{figure}

As we identified that the CCPs are the periods when the FPAFs suffered from the contamination,
next we examine the amount of the contaminant accumulated on each FPAF as a function of time for each CCP.
Figure~\ref{fig:filter usage} gives the frequency in use of FPAFs during the CCPs.
The vertical axis shows the cumulative time for each filter located on the optical path.
We expect that this cumulative time should have a positive, linear correlation with the thickness of contaminant for each FPAF.
Hereafter, we call this relationship the ``USF relation" where USF stands for FPAF-usage-suggested FPAF contamination.
To estimate the thickness, we apply the following simple model:
We set the accumulation rates of contaminant on FPAFs to be $x$, $y$, $z$, and $w$ as shown in Figure~\ref{fig:contam filter model}.
Here, $x$ and $z$ are the accumulation rates on the filters located on the optical path (o.p.),
while $y$ and $w$ are the rates on the filters located just next to the optical path (o.p. $\pm$ 1, in Figure~\ref{fig:contam filter model}).
We assume that no contaminant accumulated on the filters far from the optical path
(o.p. $\pm$ 2 and o.p. $+$ 3).
Using these accumulation rates, we can express the thickness of contaminant accumulated
on each filter $d$ as follows:
\begin{equation}
d_\mathrm{FPAF1} = x \times t_\mathrm{FPAF1}^\mathrm{o.p.} + y \times \left( t_\mathrm{FPAF1}^\mathrm{o.p.+1} + t_\mathrm{FPAF1}^\mathrm{o.p.-1} \right) ,
\label{eq:d1}
\end{equation}
\begin{equation}
d_\mathrm{FPAF2} = z \times t_\mathrm{FPAF2}^\mathrm{o.p.} + w \times \left( t_\mathrm{FPAF2}^\mathrm{o.p.+1} + t_\mathrm{FPAF2}^\mathrm{o.p.-1} \right) ,
\label{eq:d2}
\end{equation}
where $t^\mathrm{o.p.}$ and $t^\mathrm{o.p.\pm1}$ are the cumulative time when
the filter located on and next to the optical path, respectively, and the subscripts of FPAF1 and FPAF2
mean the FPAFs mounted on FW1 and FW2, respectively.
On the basis of the estimated thicknesses of contaminant on the four thin filters (Al-mesh, Al-poly, C-poly, and Ti-poly) in Phase~3
(see Table~\ref{tbl:parameter}) and the cumulative time $t$ of filter usage
shown in Figure~\ref{fig:filter usage}, we derive the accumulation rate of contaminant on FPAFs as follows:
$x = 3220$~{\AA}~month$^{-1}$, $y = 480$~{\AA}~month$^{-1}$, $z=2080$~{\AA}~month$^{-1}$ and $w=1000$~{\AA}~month$^{-1}$.
The results of $x > y$ and $z > w$ is likely, because the filter wheels are contained in a closed structure
which has holes for the optical path and the contaminant may have come to the filters through such holes.

\begin{figure}
\centerline{\includegraphics[width=10.0cm,clip=]{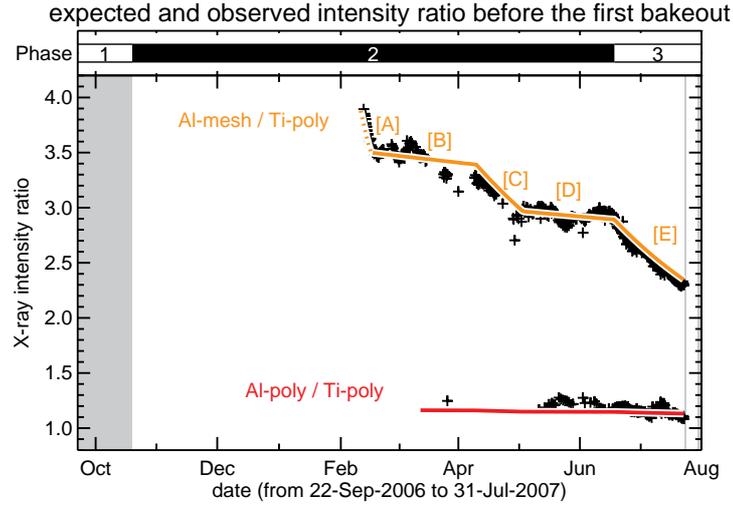}}
\caption{
Comparison between the expected (colored lines) and observed (black $+$) X-ray intensity ratio.
The expected intensity ratio is calculated with the calibrated thicknesses of the filters and contaminant.
}
\label{fig:comparison}
\end{figure}

So far, we derived coefficients of proportionality $x$, $y$, $z$, and $w$ in Equations~(\ref{eq:d1}) and (\ref{eq:d2})
using the final thicknesses of the contaminant for the four thin filters at the end of Phase~2 (beginning of Phase~3).
If the thus-derived coefficients are correct, Equations~(\ref{eq:d1}) and (\ref{eq:d2}) should give correct
contaminant thickness for each FPAF at any instant in Phase~2.
Let us now check if this is the case.
Figure~\ref{fig:contam on FPAF} indicates the temporal evolution of contaminant thickness on the FPAFs
from Equations~(\ref{eq:d1}) and (\ref{eq:d2}), while that for the CCD is shown in Figure~\ref{fig:contam on CCD}.
Figure~\ref{fig:comparison} shows comparison between observed intensity ratios in X-rays
and those expected using the thus-derived contaminant thickness.
Note the contaminant accumulated on both FPAFs and CCD in period~A while only on the CCD in periods~B\,--\,E.
The close match between observed and expected intensity ratios strongly suggests that
the estimate of contaminant thicknesses on the FPAFs with Equations~(\ref{eq:d1}) and (\ref{eq:d2}) are reasonable ones.
We now have a reasonable estimate on the thickness of contaminants
not only on the CCD but also on the FPAFs throughout the entire period since launch.

We note that, since not enough X-ray signal is acquired from the quiet Sun with the thicker filters,
the contaminant thicknesses on the thicker filters have not been directly measured in contradiction to the thinner filters.
Hence, the estimate of the thickness of contaminant on the thicker filters with Equations~(\ref{eq:d1}) and (\ref{eq:d2})
may be less accurate than the thinner filters in the sense that verification with X-ray data is not possible.
Nevertheless, we do not see any reason for Equations~(\ref{eq:d1}) and (\ref{eq:d2}) not to be applicable to the thicker filters.
Furthermore, the effect of contamination on the thicker filters is much smaller than on the thinner filters
because only short-wavelength X-rays can transmit thicker filters and the effect of contaminant is small or
negligible for such short wavelength X-rays.
Hence, our results can be applied to quantitative analyses with XRT data.

\section{CCD Bakeout}
\label{sec:CCD bakeout}

\begin{table}
\caption{Bakeout history of XRT.}
\label{tbl:bakeout}
\begin{tabular}{cccccc}
\hline
No.$^{(a)}$ &  bake heater on      & heater off      & interval       & contam. rate          & spot    \\
            &                      &                 & [days]$^{(b)}$ & [{\AA} month$^{-1}$]$^{(c)}$ & [{\%}]$^{(d)}$  \\
\hline
    $-$3    &  22 Sep 2006, 21:39  &  16 Oct, 07:53  &    0.1         & --                    & 0.00 \\
    $-$2    &  16 Oct 2006, 09:35  &  17 Oct, 08:28  &    0.1         & --                    & 0.00 \\
    $-$1    &  17 Oct 2006, 10:13  &  19 Oct, 08:12  &  277.0         & --                    & 0.00 \\
\hline
       1    &  23 Jul 2007, 09:09  &  24 Jul, 08:10  &    6.0         & --                    & --   \\
       2    &  30 Jul 2007, 08:41  &   3 Sep, 09:12  &  148.0         & 730                   & 2.61 \\
       3    &  29 Jan 2008, 08:42  &   1 Feb, 10:01  &    6.0         & 1866                  & 5.48 \\
       4    &   7 Feb 2008, 08:50  &   8 Feb, 07:52  &   13.0         & 716                   & --   \\
       5    &  21 Feb 2008, 08:19  &  22 Feb, 08:22  &   13.0         & 747                   & 5.25 \\
       6    &   6 Mar 2008, 08:18  &   7 Mar, 02:20  &   20.2         & 613                   & --   \\
       7    &  27 Mar 2008, 08:15  &  28 Mar, 08:12  &   20.0         & 656                   & --   \\
       8    &  17 Apr 2008, 09:10  &  18 Apr, 10:00  &   27.0         & 569                   & 5.26 \\
       9    &  15 May 2008, 09:18  &  16 May, 09:22  &   13.0         & 801                   & 5.25 \\
      10    &  29 May 2008, 09:50  &  30 May, 10:00  &   19.9         & 506                   & 5.27 \\
      11    &  19 Jun 2008, 08:14  &  20 Jun, 10:00  &   20.0         & 481                   & 5.28 \\
      12    &  10 Jul 2008, 09:47  &  11 Jul, 04:00  &   25.2         & 432                   & 5.24 \\
      13    &   5 Aug 2008, 08:33  &   6 Aug, 04:00  &   22.3         & 510                   & 5.23 \\
      14    &  28 Aug 2008, 10:05  &  29 Aug, 04:00  &   25.2         & 514                   & --   \\
      15    &  23 Sep 2008, 09:53  &  24 Sep, 04:00  &   22.3         & 522                   & 5.24 \\
      16    &  16 Oct 2008, 11:06  &  16 Oct, 23:07  &   20.4         & 567                   & 5.24 \\
      17    &   6 Nov 2008, 09:29  &   6 Nov, 21:29  &   20.5         & 507                   & 5.23 \\
      18    &  27 Nov 2008, 10:11  &  27 Nov, 22:17  &   20.4         & 546                   & 5.23 \\
      19    &  18 Dec 2008, 08:33  &  18 Dec, 20:39  &   20.5         & 544                   & 5.23 \\
      20    &   8 Jan 2009, 09:24  &   8 Jan, 21:23  &   20.5         & 547                   & 5.23 \\
      21    &  29 Jan 2009, 08:29  &  29 Jan, 21:23  &   22.5         & 487                   & 5.22 \\
      22    &  21 Feb 2009, 08:24  &  22 Feb, 02:26  &   18.3         & 527                   & 5.22 \\
      23    &  12 Mar 2009, 09:19  &  12 Mar, 21:20  &   20.5         & 499                   & 5.21 \\
      24    &   2 Apr 2009, 09:16  &   2 Apr, 21:25  &   20.5         & 577                   & 5.23 \\
      25    &  23 Apr 2009, 09:13  &  23 Apr, 21:14  &   --           & --                    & 5.18 \\
\hline
\end{tabular}
\begin{description}
\item[$^{(a)}$] Bakeouts from $-$3 to $-$1 were preformed before first light (23 October 2006).
We count the bakeout number up with positive number after the first light.
\item[$^{(b)}$] Interval from this bakeout to next bakeout.
\item[$^{(c)}$] Accumulation rate of contaminant on the CCD from this bakeout to the next bakeout.
This rate is measured from G-band intensity oscillation.
The details are described in Appendix~\ref{subsubsec:measurement of contam thickness on CCD}.
\item[$^{(d)}$] The ratio of contamination spot area to the full CCD area. The details of the spot are described
in Appendix~\ref{sec:Spot of contamination on CCD}.
\end{description}
\end{table}

The XRT team decided that the 800~{\AA} of contaminant thickness should be the maximum acceptable contamination.
The XRT team regularly performs CCD bakeout every three to four weeks to remove the contaminant from the CCD.
Table~\ref{tbl:bakeout} is the summary of CCD bakeout from the launch of \textit{Hinode} (22 September 2006) to the end of April 2009.

\section{Estimate of Data Number and Intensity Detected by XRT}
\label{sec:estimate of intensity detected by XRT}

We explain the method to estimate the expected data numbers (DNs) when XRT observes the X-ray spectra from an X-ray generator and from the Sun,
in order to compare them with the actually detected DNs at XRCF and on orbit, respectively.
First of all, we define the meaning of DN and intensity in this article.
The DN is the dimensionless value which is proportional to the total energy
of the incident X-rays into XRT.
Meanwhile, the intensity is the DN normalized by observation time and number of pixels [DN sec$^{-1}$ pixel$^{-1}$].

\subsection{XRCF}
\label{subsec:when XRT observes the XRCF spectra}

When incident X-rays into XRT have a photon-number spectrum of $P\left(\lambda\right)$ [cm$^{-2}$ sec$^{-1}$ {\AA}$^{-1}$],
the energy spectrum observed by XRT ($E_\mathrm{XRT}\left(\lambda\right)$ [eV sec$^{-1}$ {\AA}$^{-1}$]) is written as
\begin{equation}
E_\mathrm{XRT}\left(\lambda\right) = P\left(\lambda\right) \times A_\mathrm{eff}\left(\lambda\right) \times \frac{h c}{\lambda} \times \frac{1}{e} ,
\label{eq:E_XRT}
\end{equation}
where $h$, $c$, and $e$ are Planck's constant, speed of light, and elementary electric charge, respectively.
This incident energy generates electron hole pairs on the CCD.
One electron hole pair is generated by 3.65 eV.
The data number $DN$ detected by the XRT CCD is expressed as
\begin{equation}
DN = t \times \int \left( E_\mathrm{XRT}\left(\lambda\right) \times \frac{1}{3.65} \times \frac{1}{G} \right) \mathrm{d}\lambda ,
\label{eq:DN2}
\end{equation}
where $t$ is the exposure time in the unit of [sec], and $G$ is the system gain of the CCD camera, which is 57.5 [e DN$^{-1}$] in the XRT case \cite{kan08}.
Substituting Equation~(\ref{eq:E_XRT}) to (\ref{eq:DN2}),
we obtain the equation to derive the $DN$ from a photon-number spectrum $P\left(\lambda\right)$,
\begin{equation}
DN = t \times \int \left( P\left(\lambda\right) \times A_\mathrm{eff}\left(\lambda\right) \times \frac{h c}{\lambda} \times \frac{1}{e \times 3.65 \times G} \right) \mathrm{d}\lambda .
\label{eq:DN1}
\end{equation}

\subsection{On Orbit}
\label{subsec:when XRT observes the solar spectra}

Using the effective area (see Figure~\ref{fig:eff area}), we can estimate the intensity [$I$] detected by XRT, \textit{i.e.}
how much data number [$DN$] is detected by one pixel of XRT CCD in one second,
when XRT observes the Sun whose photon-number spectrum is $P_{\odot}\left(\lambda\right)$ [cm$^{-2}$ sec$^{-1}$ sr$^{-1}$ {\AA}$^{-1}$].
The estimated $I$ [DN sec$^{-1}$ pixel$^{-1}$] is given by
\begin{equation}
I = \int \left( P_{\odot}\left(\lambda\right) \times S \times \frac{A_\mathrm{eff}\left(\lambda\right)}{R^2} \times \frac{h c}{\lambda} \times \frac{1}{e \times 3.65 \times G} \right) \mathrm{d}\lambda ,
\label{eq:DN0}
\end{equation}
where $S$ [cm$^2$] is the solar area detected in one pixel of CCD.
$A_\mathrm{eff}$ and $R$ are the effective area of XRT and distance between the Sun and XRT ($\approx$~1~AU), respectively.
Then ${A_\mathrm{eff}}/{R^2}$ is the solid angle of XRT effective area in units of sr.
The meaning of the other terms are the same as Equation~(\ref{eq:DN1}).
Considering the geometry of the telescope, the relation of
\begin{equation}
\frac{S}{R^2}  = \frac{s}{f^2}
\label{eq:geometry}
\end{equation}
is given, where $s$ and $f$ are the area of CCD 1 pixel ($13.5^2$~$\mu$m$^2$, see \inlinecite{kan08}) and focal length of XRT (2708~mm, see \inlinecite{gol07}), respectively.
Using this relation, Equation~(\ref{eq:DN0}) can be rewritten as
\begin{equation}
I = \int \left( P_{\odot}\left(\lambda\right) \times s \times \frac{A_\mathrm{eff}\left(\lambda\right)}{f^2} \times \frac{h c}{\lambda} \times \frac{1}{e \times 3.65 \times G} \right) \mathrm{d}\lambda .
\label{eq:I}
\end{equation}

Using Equation~(\ref{eq:I}), we can derive the intensity observed by XRT.
The CHIANTI atomic database \cite{der97} gives us a photon-number spectrum
$\tilde{P}_\mathrm{iso\odot}\left(\lambda, T\right)$ [cm$^{-2}$ sec$^{-1}$ sr$^{-1}$ {\AA}$^{-1}$]
emitted from an isothermal plasma at a given temperature of $T$ and an unit column emission measure (CEM) of 1~cm$^{-5}$.
The definition of CEM is
\begin{equation}
CEM \equiv \int n_\mathrm{e} \times n_\mathrm{H} ~\mathrm{d}l  ,
\label{eq:CEM}
\end{equation}
where $n_\mathrm{e}$, $n_\mathrm{H}$, and $dl$ are the electron number density [cm$^{-3}$], hydrogen number density [cm$^{-3}$],
and unit length along the line-of-sight [cm], respectively.
We calculate $\tilde{P}_\mathrm{iso\odot}\left(\lambda, T\right)$ based on CHIANTI version 6.0.1 (\opencite{der97}; \citeyear{der09}: which is the latest version, when we analyzed)
with ionization equilibrium of \textsf{chianti.ioneq} \cite{der07} and abundance of \textsf{sun\_coronal\_ext.abund} \cite{fel92, lan02}.
Replacing $P_{\odot}\left(\lambda\right)$ by $\tilde{P}_\mathrm{iso\odot}\left(\lambda, T\right)$ in Equation~(\ref{eq:I}),
we can calculate the so-called ``temperature response" $F\left(T\right)$ [DN sec$^{-1}$ pixel$^{-1}$] of XRT,
which is the intensity observed by XRT
from a plasma at a given temperature of $T$ and a unit column emission measure of 1~cm$^{-5}$ as
\begin{equation}
F\left(T\right) \equiv \int \left( \tilde{P}_\mathrm{iso\odot}\left(\lambda, T\right) \times s \times \frac{A_\mathrm{eff}\left(\lambda\right)}{f^2} \times \frac{h c}{\lambda} \times \frac{1}{e \times 3.65 \times G} \right) \mathrm{d}\lambda .
\label{eq:F}
\end{equation}

Figure~\ref{fig:response} shows $F\left(T\right)$ for each X-ray analysis filter
at the launch of \textit{Hinode} and on-orbit.
According to this figure, we can appreciate that the XRT has a superior ability to
observe coronal plasmas over a wide temperature range from less than 1~MK to more than 10~MK.

Using this temperature response $F(T)$ of XRT,
we can estimate the X-ray intensity $I$ [DN sec$^{-1}$ pixel$^{-1}$] observed with XRT as follows:
\begin{equation}
I =  F(T) \times CEM
\label{eq:I iso}
\end{equation}
for an isothermal corona at a temperature of $T$,
and
\begin{equation}
I =  \int F(T) \times DEM\left(T\right) ~\mathrm{d}T
\label{eq:I multi}
\end{equation}
for a multi-temperature corona with the differential emission measure $DEM$ [cm$^{-5}$ K$^{-1}$], which is defined as
\begin{equation}
DEM \equiv \frac{\int n_\mathrm{e}\left(T\right) n_\mathrm{H}\left(T\right) \mathrm{d}l}{\mathrm{d}T}  .
\label{eq:DEM}
\end{equation}

The data number [$DN$] detected by XRT is given as
\begin{equation}
DN = I \times t \times p  ,
\label{eq:DN with I}
\end{equation}
where $t$ and $p$ are the observation time and the number of pixels used for obtaining the DN, respectively.
Using the volume emission measure [$VEM$: cm$^{-3}$] defined as
\begin{equation}
VEM \equiv \int n_\mathrm{e} \times n_\mathrm{H} ~\mathrm{d}l ~\mathrm{d}S = CEM \times ( S \times p )  ,
\label{eq:VEM}
\end{equation}
where $\mathrm{d}S$ is an unit area of observed solar region,
$DN$ is expressed as
Equation~(\ref{eq:DN with VEM})
for an isothermal corona at a temperature of $T$.

\section{X-Ray Spectrum at XRCF}
\label{sec:XRCF spectrum}

\begin{figure}
\centerline{\includegraphics[width=11.0cm,clip=]{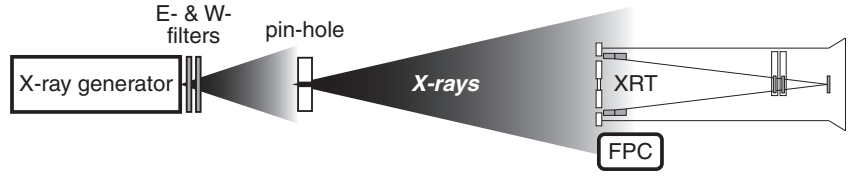}}
\vspace{3.0mm}
\caption{Configuration of the end-to-end test at XRCF.
}
\label{fig:XRCF}
\end{figure}

\begin{table}
\caption{Characteristic X-ray lines used in the end-to-end test at XRCF.}
\label{tbl:XRCF source}
\begin{tabular}{ccccccc}
\hline
line  &  energy      & wavelength    & source   &  voltage  & E-filter      & W-filter             \\
\hline
Mo-L  &   2.29~keV   & 5.41~{\AA}    & Mo mono  &  7.0~kV  & Lexan$^{(a)}$  & Mo  3.0$\mu$m        \\
Al-K  &   1.49~keV   & 8.34~{\AA}    & Al mono  &  7.0~kV  & Lexan          & Al 18.7$\mu$m        \\
Cu-L  &  0.930~keV   & 13.3~{\AA}    & Cu mono  & 1.45~kV  & Cu 2.8$\mu$m   & Lexan                \\
O-K   &  0.525~keV   & 23.6~{\AA}    & O on Al  &  4.0~kV  & Cr 0.7$\mu$m   & Lexan                \\
C-K   &  0.277~keV   & 44.7~{\AA}    & C mono   &  5.0~kV  & Lexan          & Parylene 10.0$\mu$m  \\
\hline
\end{tabular}
\begin{description}
\item[$^{(a)}$] This Lexan is the product of Luxel R/N 6370.
\end{description}
\end{table}

We estimate spectra of X-ray beams at the XRCF.
Figure~\ref{fig:XRCF} shows the configuration of the end-to-end test performed at the XRCF.
The XRCF employs a target-impact type X-ray generator.
Five X-ray lines that we used in this test are summarized in Table~\ref{tbl:XRCF source}
with target sources, applied acceleration voltage for electrons, and blocking filters.
The blocking filters, called E- and W-filters, are introduced to suppress longer and
shorter wavelengths than the selected characteristic X-ray line,
because lower levels of characteristic X-rays and continuum bremsstrahlung X-rays are also emitted.
The X-rays illuminate XRT in an almost parallel beam, because the small-sized X-ray source is located far from XRT, namely $\approx$~500~m.
A flow proportional counter (FPC) was located beside the XRT
to monitor intensity and spectral shape of the X-ray beam.

\begin{figure}
\centerline{\includegraphics[width=10.0cm,clip=]{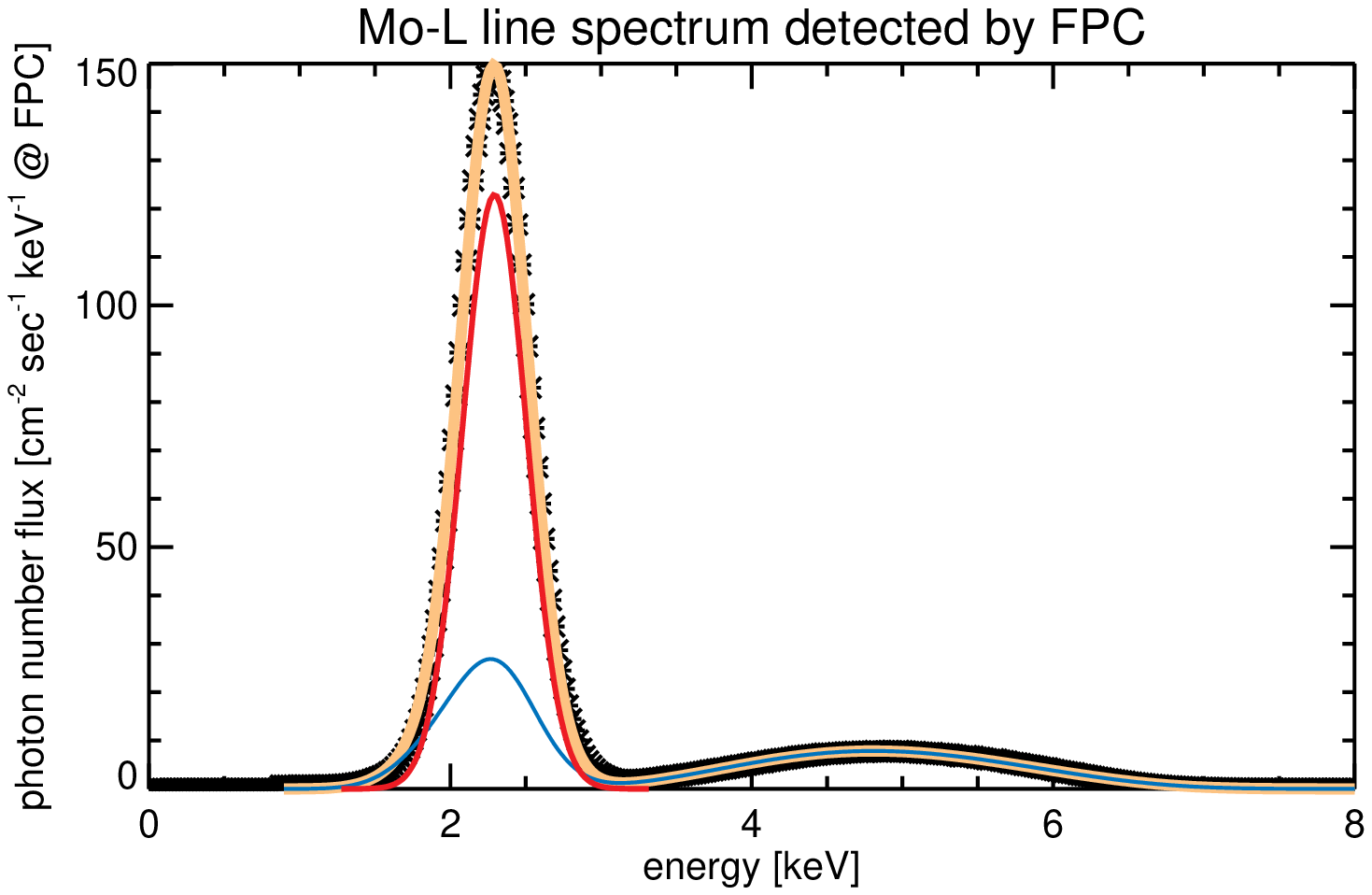}}
\vspace{0.5cm}
\centerline{\includegraphics[width=6.0cm,clip=]{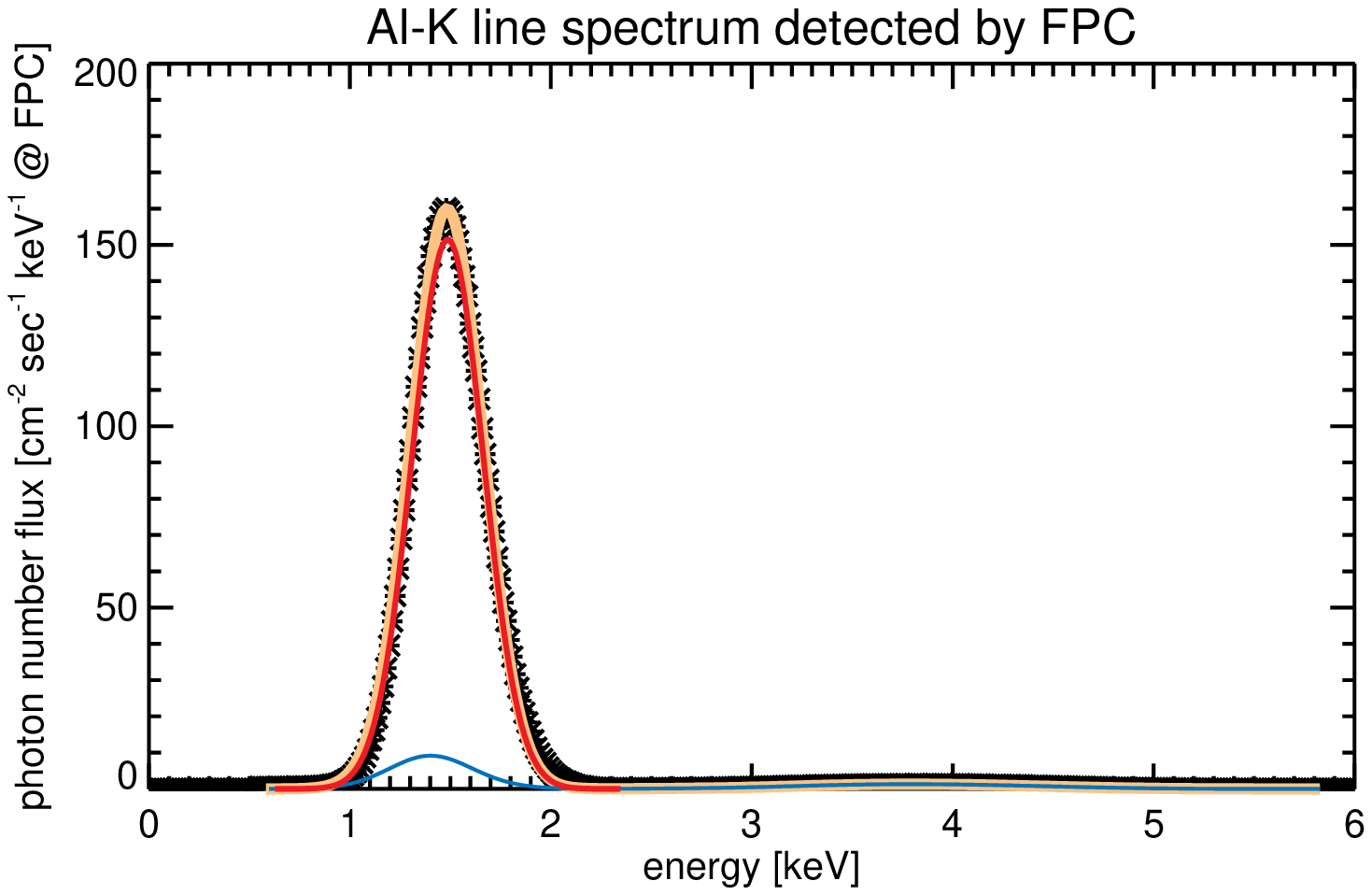}
            \hspace*{0.0cm}
            \includegraphics[width=6.0cm,clip=]{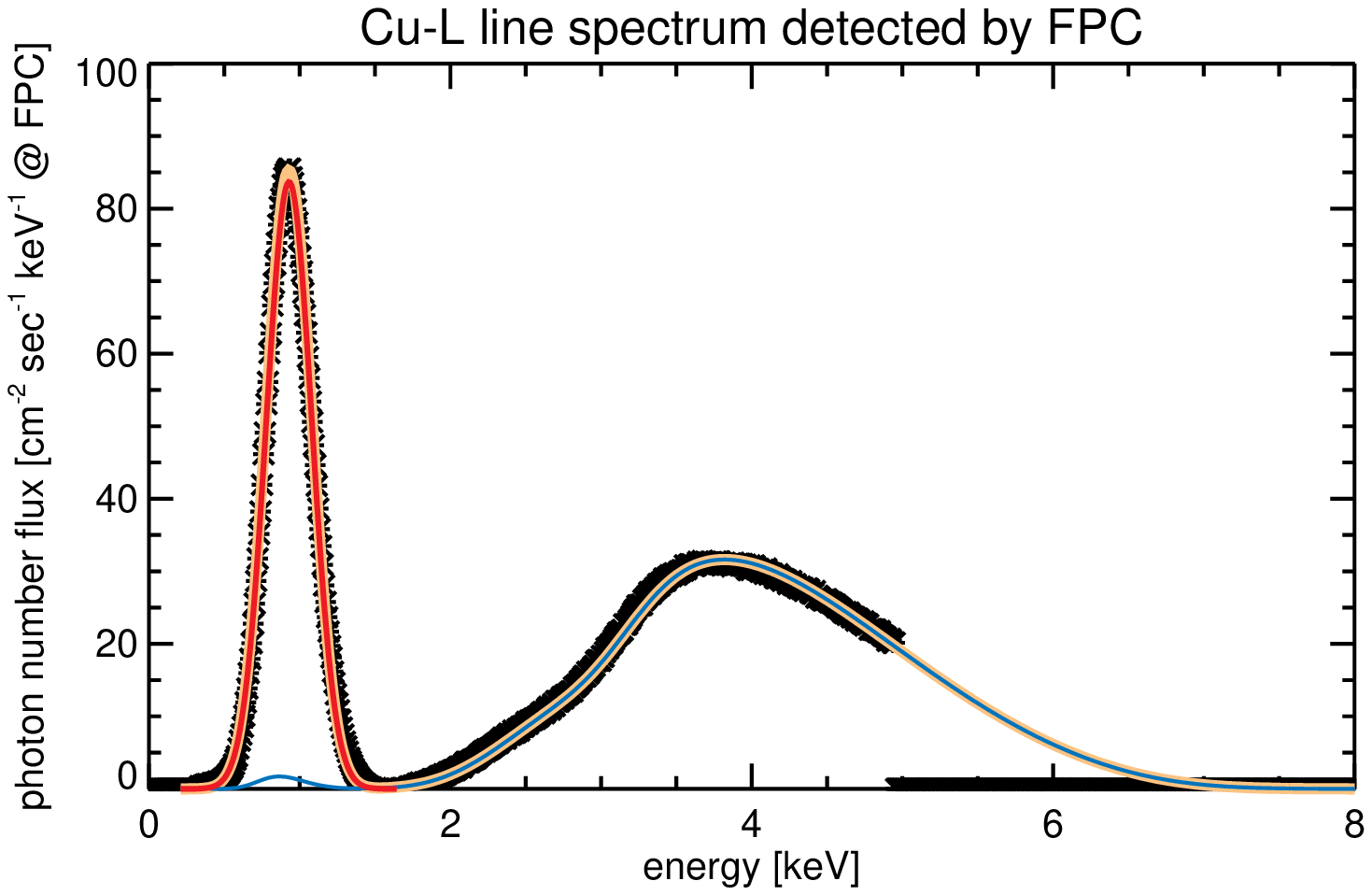}
           }
\vspace{0.5cm}
\centerline{\includegraphics[width=6.0cm,clip=]{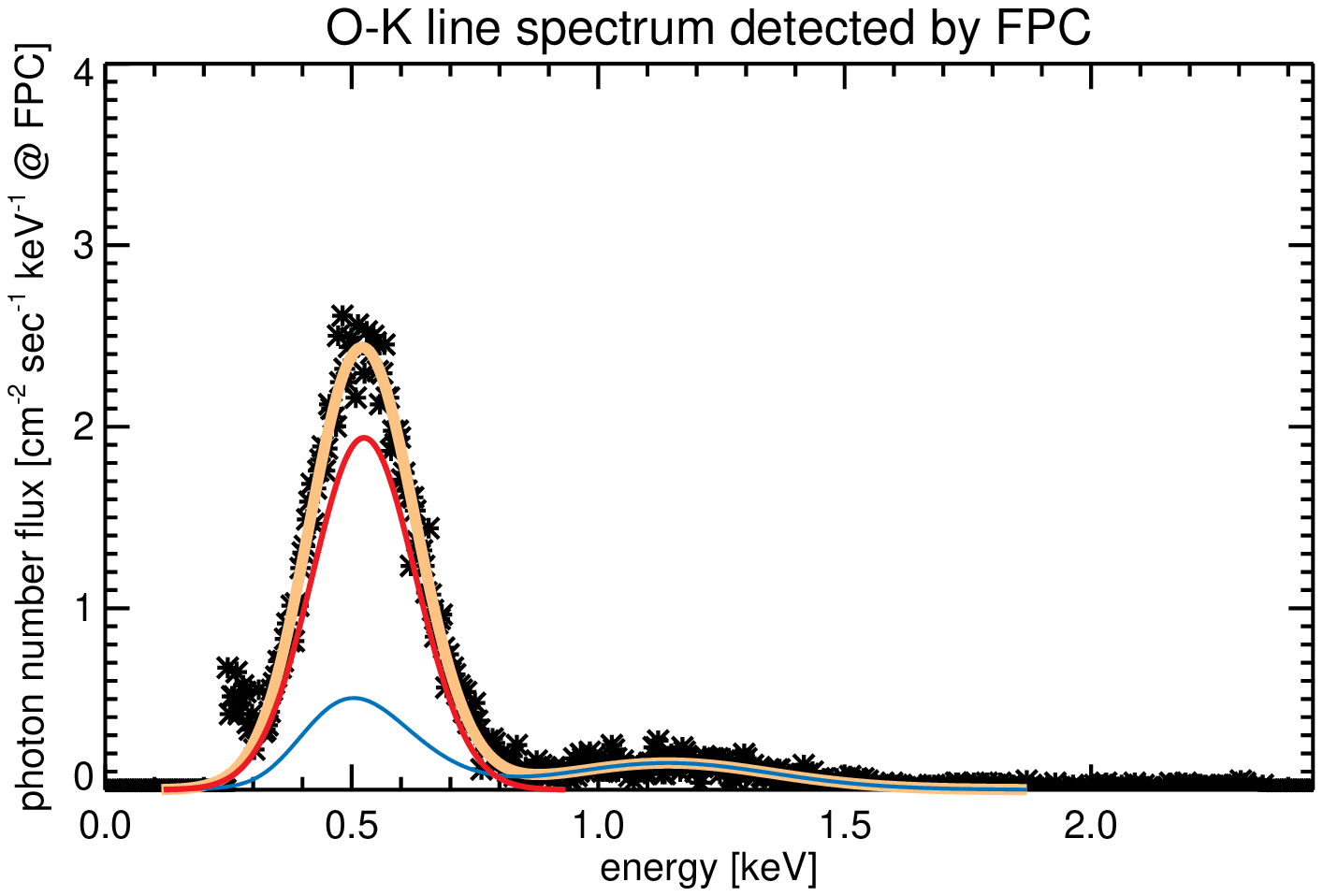}
            \hspace*{0.0cm}
            \includegraphics[width=6.0cm,clip=]{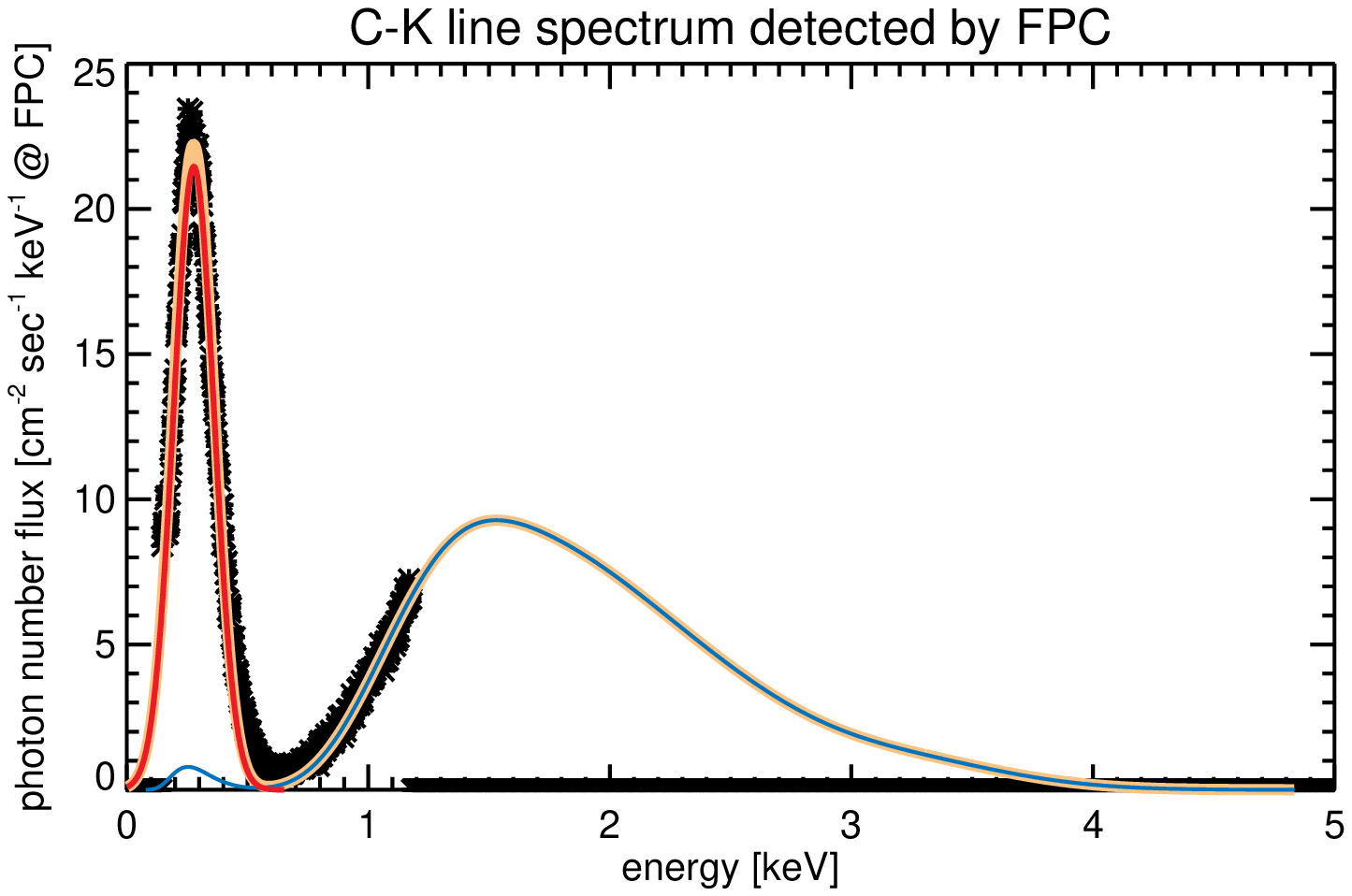}
           }
\vspace{3.0mm}
\caption{
X-ray spectra detected by the flow proportional counter (FPC).
The black marks (asterisks) are the data points detected by FPC.
The orange line shows the estimated spectrum, which consists of characteristic X-ray line
(red line) and continuum bremsstrahlung (blue line).
}
\label{fig:FPC data}
\end{figure}

\begin{figure}
\centerline{\includegraphics[width=10.0cm,clip=]{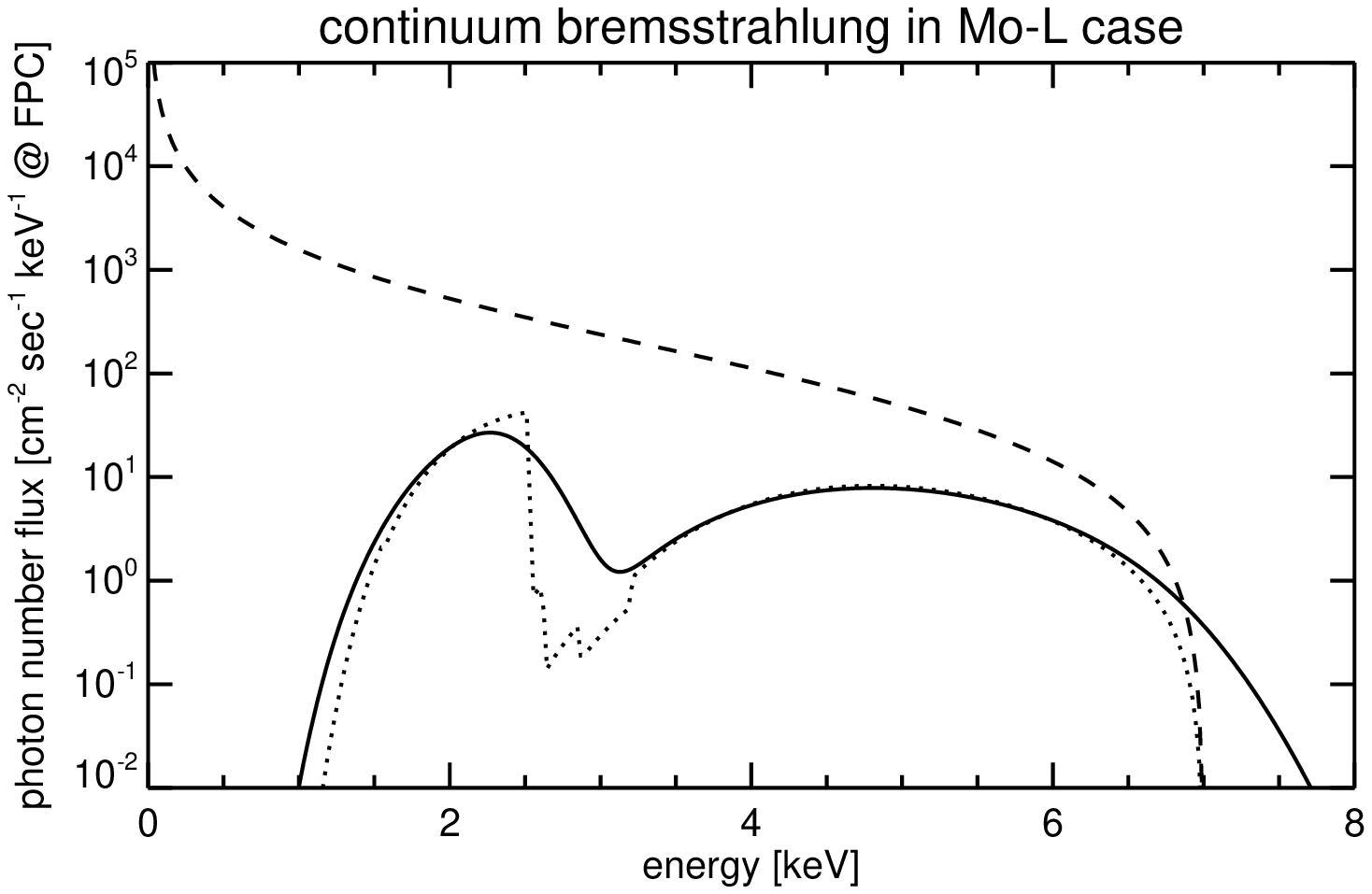}}
\caption{Continuum bremsstrahlung spectrum.
The dashed line is the continuum spectrum emitted by 7 keV electrons.
The dotted line indicates the incident continuum spectrum into FPC.
The solid line shows the continuum spectrum detected by FPC with its energy resolution.
}
\label{fig:FPC Brems}
\end{figure}

Here, using the FPC data, we derive the incident X-ray spectrum into XRT.
The FPC detected two humps in each characteristic X-ray line data (see Figure~\ref{fig:FPC data}).
Even with the blocking filters, the continuum bremsstrahlung emission component is not completely removed.
The main hump comes from the characteristic X-ray line and the other from the continuum bremsstrahlung.

To understand the incident X-ray spectrum into the XRT, we need to estimate the spectrum of continuum bremsstrahlung also.
On the basis of the thick target emission model, the continuum bremsstrahlung profile can be described as
\begin{equation}
I_\mathrm{thick}(\varepsilon) = \frac{1}{4 \pi R^2 K} \int_{\varepsilon}^{\infty} F(E_0) \int_{\varepsilon}^{E_0} E \sigma_\mathrm{B}(\varepsilon, E) ~\mathrm{d}E ~\mathrm{d}E_0 ,
\label{eq:Brems}
\end{equation}
\begin{equation}
\sigma_\mathrm{B}(\varepsilon, E) = 1.58 \times 10^{-24} \frac{1}{\varepsilon} \ln \left[ \left(\frac{E}{\varepsilon}\right)^\frac{1}{2} + \left(\frac{E}{\varepsilon} - 1\right)^\frac{1}{2} \right]
\label{eq:BH cross section}
\end{equation}
\cite{sak94}, where $\varepsilon$ is the photon energy, $E$ the electron energy, $E_0$ the initial electron energy,
$F$ the energy distribution function of electrons injected to the target
[electrons sec$^{-1}$ keV$^{-1}$],
$\sigma_\mathrm{B}$ the Bethe-Heitler cross section \cite{jac62}, $K \equiv 2 \pi e^4 \ln \Lambda$,
$\ln \Lambda$ the Coulomb logarithm \cite{spi62}, and $R$ the distance between the X-ray source and the FPC.
In our case, because single-energy electrons at an energy of $E_{00}$ were injected,
the distribution function can be expressed as:
\begin{equation}
F(E_0) = F_{00} \delta(E_0 - E_{00}) ,
\label{eq:F(E0)}
\end{equation}
where $F_{00}$ is the number of injected electrons per second.
The dashed line in Figure~\ref{fig:FPC Brems} is the estimated continuum bremsstrahlung spectrum emitted at the acceleration voltage of 7~kV
with Equations~(\ref{eq:Brems})\,--\,(\ref{eq:F(E0)}).
Using the transmission of the E- and W-filters and the efficiency of FPC,
the incident continuum spectrum into FPC is derived as shown by the dotted line in Figure~\ref{fig:FPC Brems}.
Considering the energy resolution of FPC, which can be written as
\begin{equation}
\left( \Delta E \right)_\mathrm{FWHM} = a E^{1/2}
\label{eq:FPC resolution}
\end{equation}
\cite{cha68}, where $a$ is a constant of proportionality,
the continuum spectrum detected by the FPC is simulated as shown by a solid curve in Figure~\ref{fig:FPC Brems}.

The actual spectrum detected by the FPC is fit by this simulated FPC continuum spectrum and the characteristic X-ray line (Figure~\ref{fig:FPC data}).
The fitting parameters are the constant $a$ in Equation~(\ref{eq:FPC resolution}), the number of injected electrons per second $F_{00}$,
and the strength of the characteristic X-ray line represented by a Gaussian shape.

\begin{table}
\caption{Best-fit photon number fluxes of characteristic X-ray lines and continua.}
\label{tbl:XRCF comp}
\begin{tabular}{ccccc}
\hline
line  &  energy  & voltage  & characteristic line     & continuum               \\
      &  [keV]   & [kV]     & [cm$^{-2}$ sec$^{-1}$]  & [cm$^{-2}$ sec$^{-1}$]  \\
\hline
Mo-L  &   2.29   &  7.0     & 80.7 (63.6\%) $^{(a)}$ & 46.1 (35.4\%)          \\
Al-K  &   1.49   &  7.0     & 77.4 (90.8\%)          &  7.8  (9.2\%)          \\
Cu-L  &  0.930   & 1.45     & 50.8 (36.1\%)          & 89.9 (63.9\%)          \\
O-K   &  0.525   &  4.0     &  2.2 (68.7\%)          &  1.0 (31.3\%)          \\
C-K   &  0.277   &  5.0     &  7.8 (30.9\%)          & 17.4 (69.1\%)          \\
\hline
\end{tabular}
\begin{description}
\item[$^{(a)}$] The percentage in parenthesis shows the ratio of each component to the total photon flux.
\end{description}
\end{table}

Using the best-fit results in Figure~\ref{fig:FPC data},
we obtained the incident spectrum into XRT (see Figure~\ref{fig:XRCF spectrum}).
The errors shown by gray areas in Figure~\ref{fig:XRCF spectrum} are estimated from the photon noise and the energy range of FPC data points.
For example, the error bars in O-K and C-K line are larger than for the other lines,
because the total photon number is small in the O-K line case, and because the energy range of the C-K line does not completely cover the continuum.
The photon-number fluxes of characteristic X-ray lines and continua are summarized in Table~\ref{tbl:XRCF comp}.
The contribution of continuum is not negligible even for the Al-K line.

\section{Variation in Annulus Transmission}
\label{variation in annulus transmission}

\begin{figure}
\centerline{\includegraphics[width=6.0cm,clip=]{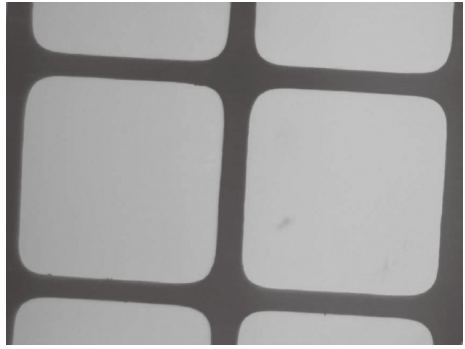}}
\vspace{3.0mm}
\caption{
Stainless mesh of Al-mesh filter taken by the filter manufacturer with a high precision digital microscope.
This mesh supports the thin Al film of Al-mesh filter.
}
\label{fig:mesh}
\end{figure}

\begin{figure}
\centerline{\includegraphics[width=10.0cm,clip=]{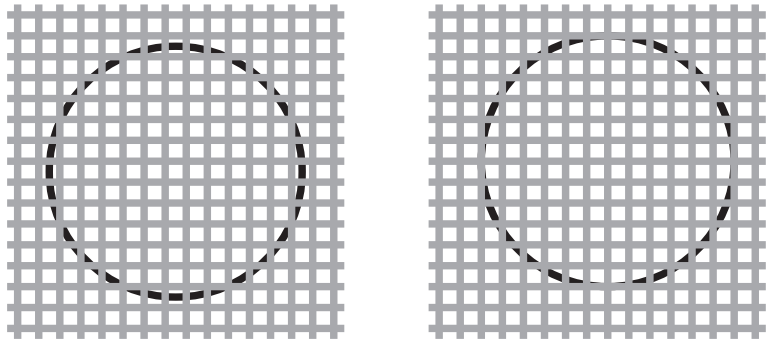}}
\vspace{3.0mm}
\caption{
Concept images of variation in ``annulus transmission". The gray area indicates the mesh of Al-mesh filter.
The black annulus shows the shape of X-rays imaged on the filter.
The left and right panels shows the difference of annulus transmission caused by the different
relative position between mesh pattern and annular X-rays.
The aspect ratio between the black annulus and gray mesh in this figure
is different from the actual aspect ratio between the annular X-rays image and stainless mesh
on Al-mesh filter.
}
\label{fig:mesh trans image}
\end{figure}

\begin{figure}
\centerline{\includegraphics[width=10.0cm,clip=]{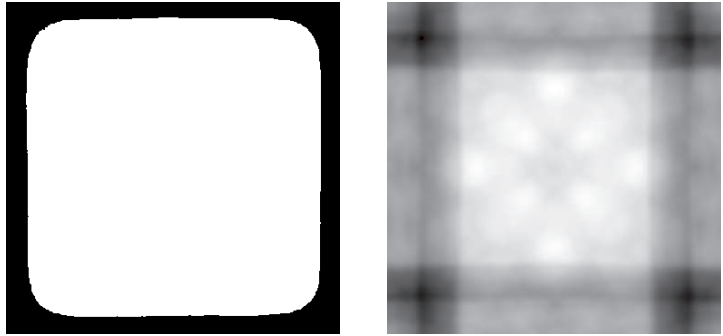}}
\caption{
Variation in ``annulus transmission".
The left panel is the mesh cell. The black and white regions indicate the mesh wire and open area, respectively.
The right panel shows the annulus transmission where the center of annular X-rays imaged is located
at the corresponding left panel. The white and black mean high and low transmission, respectively.
}
\label{fig:mesh trans}
\end{figure}

\begin{figure}
\centerline{\includegraphics[width=10.0cm,clip=]{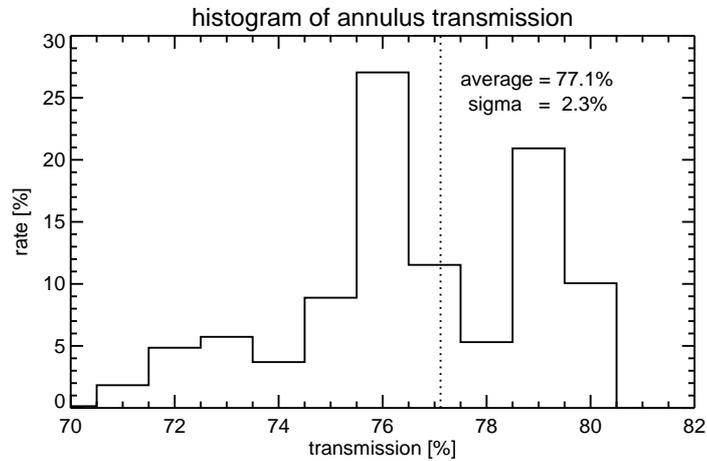}}
\caption{
Histogram of annulus transmission based on the right panel of Figure~\ref{fig:mesh trans}.
The dotted line indicates the average of annulus transmission: 77.1\%.
}
\label{fig:mesh trans histogram}
\end{figure}

The support of the Al-mesh filter is a stainless steel mesh.
Figure~\ref{fig:mesh} is a picture of the mesh taken by the filter manufacturer with a high-precision digital microscope.
According to the manufacturer, the width of the mesh wire and the open area are 38~$\mu$m and 330~$\mu$m, respectively.
On the basis of this measurement, the open area occupies 77\% of whole area.
We adopt this measured value of 77\% for the geometrical open area.

Here we define ``annulus transmission" as the fraction of
the X-rays passed though the open area of the mesh to the incident X-rays into the mesh.
Note that the X-rays pass through the open area without any loss of intensity,
while at the wires, the X-rays are completely blocked.
Hence, the average of annulus transmission should be consistent with the geometrical open area of the mesh.

The annulus transmission is not uniform and depends on the relative position of the shape of X-rays on the filter with respect to the mesh pattern.
However, we do not know this relative position for the flight XRT.
Instead, we derive distribution of the annulus transmission
for all possible relative positions, take the average of the distribution as the annulus transmission of the Al-mesh filter,
and employ the deviation of the distribution as that for the annulus transmission.
The procedure is detailed in the following.

The shape of X-rays imaged on the FPAFs is an annulus, because the shape of the XRT aperture is an annulus (see Figure~\ref{fig:PF}).
The radius and width of annular X-ray image are about 4.1~mm and 7.6~$\mu$m on the FPAFs, respectively,
because the radius and width of XRT annular aperture, focal length, and distance between focus position and FPAF are
about 170~mm, 0.32~mm, 2700~mm, and 65~mm, respectively.
Figure~\ref{fig:mesh trans image} shows some schematic examples of the relationship between
the annular X-ray image and the mesh.
The gray area indicates the mesh of Al-mesh filter and the black ring shows the X-ray annulus on the filter.
The variation in annulus transmission is shown in Figure~\ref{fig:mesh trans}.
The left panel shows a cell (unit opening) of the mesh. The width and height of this cell are both 368~$\mu$m on the basis of manufacturer's measurement.
The black and white regions indicate mesh wire and open area, respectively.
The right panel shows the annulus transmission in gray scale at the position where the center of the annular X-ray image is located
at the corresponding left panel.
Figure~\ref{fig:mesh trans histogram} shows distribution of the annulus transmission
over the varying center position of the X-ray annulus with respect to the mesh pattern
(This is the histogram of calculated annulus transmission in the right panel of Figure~\ref{fig:mesh trans}).
The average of annulus transmission is consistent with the geometrical open area of the mesh.
However, the annulus transmission has a scatter with the standard deviation $\sigma$ of 2.3\%.
This $\sigma$ is also considered in the calibration of Al-mesh filter thickness.

The uncertainty in the annulus transmission originated from the known relative position of
the annulus center with respect to the mesh pattern can thus be evaluated in this way.
In this evaluation, we assumed that the shape of the X-ray image on the filter is a perfect annulus.
But, in reality, the shape corresponds to the colored area in Figure~\ref{fig:PF} and is not symmetrical.
Hence, we should consider not only the relative position but also the relative angle
between the patterns along the image annulus and the direction of the mesh wires, which is also not known.
In order to assess the amount of additional uncertainty caused by this,
we calculated the scatter of annulus transmission due to the uncertainty of the relative position and angle.
The standard deviation $\sigma'$ of this scatter (imperfect annulus case) turned out to be
almost the same as $\sigma = 2.3\%$  (perfect annulus case).
Hence, we claim our treating the image shape as a perfect annulus is sufficient for
deriving the average fraction and $\sigma$ of the annulus transmission for the X-ray image on the Al-mesh filter.

\section{Spot-Shaped Contaminants on the CCD}
\label{sec:Spot of contamination on CCD}

\begin{figure}
\centerline{\includegraphics[width=12.0cm,clip=]{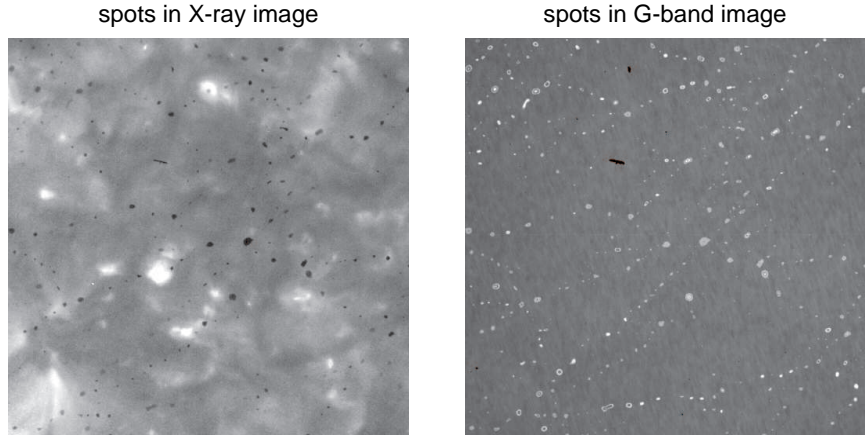}}
\caption{
Spot-shaped contaminants observed in X-rays and G-band.
The spots in the X-ray data are dark, because the spots absorb the X-rays.
Meanwhile, the spots in the G-band data are bright, since the spots work as anti-reflection coatings.
}
\label{fig:spot}
\end{figure}

\begin{figure}
\centerline{\includegraphics[width=10.0cm,clip=]{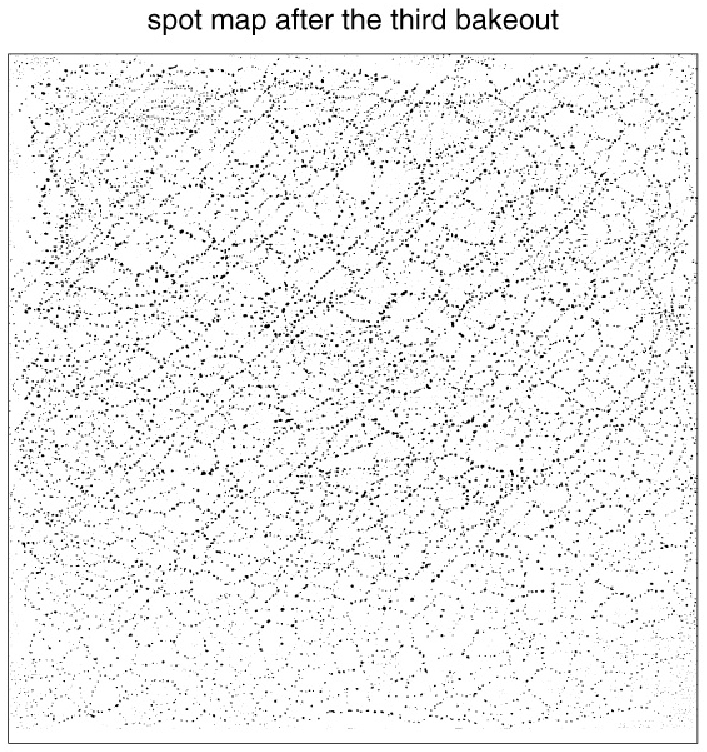}}
\caption{
Spot map after the third bakeout. The ratio of spot area to full CCD area is about 5\%.
}
\label{fig:spot map}
\end{figure}

After the first bakeout, spot-like patterns appeared in both X-ray and G-band data as shown in Figure~\ref{fig:spot}.
Subsequent analyses indicated that the estimated thickness of contaminant on the CCD was
more than 3000~{\AA} before the first bakeout (see Figure~\ref{fig:contam on CCD}).
The patterns most likely originate from the condensation of contaminant and are triggered by a bakeout
with the thickness of the contaminants accumulated on the CCD exceeding a certain threshold value
(the value itself is not known).
These spots were not removed at all even with the second bakeout that was performed after the first bakeout and
lasted for a month.
The ratio of spot area to full CCD area was 2.6\% at this point (see Table~\ref{tbl:bakeout}).
As it took five months before we agreed, as the instrument team, to proceed with performing the next (third) bakeout,
the resultant thickness of the contaminants again exceeded 3000~{\AA}, reaching even beyond 3500~{\AA} (see Figure~\ref{fig:contam on CCD}).
This resulted in the creation of additional contamination spots across the CCD, with an increased spot area ratio of 5.2\%. 
Note, however, this additional increase was rather an intentional one;
We took the option of keeping a low-temperature diagnostic capability by removing the accumulated contaminants
while accepting an increase in the spot area.
The resultant spot distribution is indicated by the black areas in Figure~\ref{fig:spot map}.
After the third CCD bakeout, we regularly perform the bakeout every three to four weeks.
As we guessed, no more spots were created after the third bakeout as shown in Table~\ref{tbl:bakeout}.
The ratio of spot area to full-CCD area has kept to be 5.2\%.

We investigate the relation between 
the thickness of some spots measured with their Newton-ring patterns in G-band and the absorbed X-ray intensities by such spots,
and conclude that the material of spot contaminant is most likely the same material as the laminar contaminant.
This result will be presented as a separate paper with more detailed analysis.

\section{Transmission of Rays Through Three Layers}
\label{app:transmission of rays through three layers}

In Appendix~\ref{subsubsec:contam monitor method},
in order to measure the contaminant thickness on the CCD,
we used the transmission of visible light into the CCD through contamination layers, Equation~(\ref{eq:T_G}).
Here, we derive this equation.

The reflected and transmitted amplitudes [$A_\mathrm{R}$ and  $A_\mathrm{T}$] of rays at
a boundary of two layers whose refractive indexes are $n_1$ and $n_2$
are written as
\begin{equation}
A_\mathrm{R} = A \left( \frac{n_1-n_2}{n_1+n_2} \right) ,
\label{eq:A_R}
\end{equation}
\begin{equation}
A_\mathrm{T} = A \left( \frac{2 n_1}{n_1+n_2} \right) ,
\label{eq:A_T}
\end{equation}
where $A$ is the amplitude of incident rays from the $n_1$ layer to the boundary.

We consider transmission of three layers as shown in Figure~\ref{fig:contam CCD}.
Here, we define $A$ as the amplitude of incident rays in vacuum ($n_1$ layer), and
$A_k$ the complex amplitude of rays which are reflected $2k$ times at the borders of layers
($n_1$\,--\,$n_2$ border and $n_2$\,--\,$n_3$ border) and transmitted into the CCD ($n_3$ layer).
Note the amplitudes of rays shown by white and black arrows in Figure~\ref{fig:contam CCD} are $A_0$ and $A_1$.
$A_k$ is expressed as
\begin{equation}
A_k = A e^{k \mathrm{i} \Delta\varphi } \left( \frac{2 n_1}{n_1 + n_2} \right) \left( \frac{2 n_2}{n_2 + n_3} \right) \left\{ \left( \frac{n_2-n_3}{n_2+n_3} \right) \left( \frac{n_2-n_1}{n_2+n_1} \right) \right\}^k  ,
\label{eq:A_k}
\end{equation}
where $\Delta\varphi$ ($= k_2 \times 2d$) gives the phase difference between $A_k$ and $A_{k+1}$.
Hence, the total amplitude [$A_\mathrm{t}$] of incident rays into the CCD is derived as
\begin{eqnarray}   
A_\mathrm{t} & \equiv & \sum^{\infty}_{k=0} A_k \nonumber \\
& = & A \left( \frac{2 n_1}{n_1 + n_2} \right) \left( \frac{2 n_2}{n_2 + n_3} \right) \left[ 1 - e^{\mathrm{i} \Delta\varphi } \left\{ \left( \frac{n_2-n_3}{n_2+n_3} \right) \left( \frac{n_2-n_1}{n_2+n_1} \right) \right\}  \right]^{-1} .
\label{eq:A_t}
\end{eqnarray}   
Using the relation between amplitude ${\cal A}$ and intensity ${\cal I}$ in a layer of refractive index $n$ given as
\begin{equation}
{\cal I} = n {\cal A} {\cal A}^\ast  ,
\label{eq:IA}
\end{equation}
where $^\ast$ denotes complex conjugate,
the intensity $I$ in vacuum ($n_1$ layer in Figure~\ref{fig:contam CCD}) is
\begin{equation}
I = n_1 A A^\ast  ,
\label{eq:I_n1}
\end{equation}
and the total intensity $I_\mathrm{t}$ detected by the CCD is
\begin{equation}
I_\mathrm{t} = n_3 A_\mathrm{t} A_\mathrm{t}^\ast  .
\label{eq:I_t}
\end{equation}
Equation~(\ref{eq:T_G}) is derived from Equations~(\ref{eq:A_t}), (\ref{eq:I_n1}), and (\ref{eq:I_t}).

\section{Oxidization of Metal}
\label{sec:oxidization of metal} 

When the molecular formula of pure metal and oxidized metal are X and X$_\mathrm{A}$O$_\mathrm{B}$,
and the molecular weight of X and O are $M_\mathrm{X}$ and $M_\mathrm{O} (= 16.00)$, respectively,
the reduced thickness [$\Delta d_\mathrm{pure}$] of pure metal by the creation of 
oxidized metal with a thickness of $d_\mathrm{ox}$ is derived to be
\begin{equation}
\Delta d_\mathrm{pure} = d_\mathrm{ox} \times \frac{M_\mathrm{X} \times A}{M_\mathrm{X} \times A + M_\mathrm{O} \times B} ,
\label{eq:oxide}
\end{equation}
based on the conservation of the number of metal atoms.
The information about materials used in our calibration is summarized in Table~\ref{tbl:material}.

\section{Meaning of Photon Noise Calculated with Filter-Ratio Temperature}
\label{sec:meaning of filter-ratio noise}

\begin{figure}
\centerline{\includegraphics[width=8.0cm,clip=]{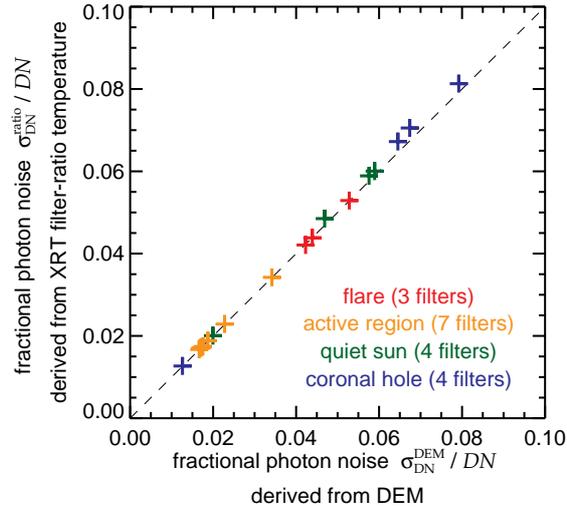}}
\caption{
Correlation between the fractional photon noise derived with DEM [$\sigma_\mathrm{DN}^\mathrm{DEM} / DN$] and with the filter-ratio temperature [$\sigma_\mathrm{DN}^\mathrm{ratio} / DN$]
for four DEMs shown in Table~\ref{tbl:coronal structure}.
This plot is the case when 1 month passed after the CCD bakeout, \textit{i.e.} 800~{\AA} of contaminant on the CCD.
The dashed line indicates the positions where $\sigma_\mathrm{DN}^\mathrm{ratio} / DN$ is equal to $\sigma_\mathrm{DN}^\mathrm{DEM} / DN$.
The number of analyzed filters is shown in parenthesis.
}
\label{fig:filter ratio noise}
\end{figure}

The photon noise can be derived with Equation~(\ref{eq:sigma_I}).
For the Condition~(B) in Section~\ref{subsec:suitable filter pair},
we investigated how much photon noise is expected in the observed data with the single temperature of the filter-ratio temperature.
However, the actual solar corona may have a multi-temperature structures.
Hence, here we discuss the meaning of photon noise estimate with the filter-ratio temperature.
Figure~\ref{fig:filter ratio temperature} shows the correlation between
the fractional photon noise derived with DEM [$\sigma_\mathrm{DN}^\mathrm{DEM} / DN$] and with the filter-ratio temperature [$\sigma_\mathrm{DN}^\mathrm{ratio} / DN$]
for four DEMs (regions) summarized in Table~\ref{tbl:coronal structure}.
$\sigma_\mathrm{DN}^\mathrm{ratio}$ is the photon noise calculated with the filter-ratio temperature [$T_\mathrm{ratio}$] and Equation~(\ref{eq:sigma_I}).
$\sigma_\mathrm{DN}^\mathrm{DEM}$ is the photon noise derived with the DEM as following steps:
\textit{i}) We calculate the photon noise [$\sigma_\mathrm{DN}$] for each temperature bin of DEM with Equation~(\ref{eq:sigma_I}).
\textit{ii}) We derive the net photon noise of step~\textit{i}) as $\sigma_\mathrm{DN}^\mathrm{DEM} = \sqrt{\sum{\sigma_\mathrm{DN}^2}}$.
The $\sigma_\mathrm{DN}^\mathrm{ratio}$ and $\sigma_\mathrm{DN}^\mathrm{DEM}$ is calculated for all filters
which are included in the suitable filter pairs in Figure~\ref{fig:suitable filter pair 800}
with the typical values of coronal structures in Table~\ref{tbl:coronal structure}.
On the basis of this figure, we can say that $\sigma_\mathrm{DN}^\mathrm{ratio}$ well matches $\sigma_\mathrm{DN}^\mathrm{DEM}$ which is the actual photon noise.

This equality can be understood as following:
Since the amount of plasma at the DEM peak is dominant, its photon noise mainly affects $\sigma_\mathrm{DN}^\mathrm{DEM}$.
In the temperature range where the DEM peak is located,
the conversion factor [$K^{(2)}$] from observed DN to $\sqrt{\sigma_{DN}}$ (see Equation~\ref{eq:sigma_I}) is almost constant
for the filters which are included in the suitable filter pairs (see Figures~\ref{fig:suitable filter pair 0} and \ref{fig:suitable filter pair 800}).
For example, in a temperature range of 3\,--\,4~MK where the active region has the DEM peak (see Figure~\ref{fig:filter pair2}),
$K^{(2)}$ for Al-poly, Ti-poly, and med-Be filters varies within factors of 1.07, 1.11, and 1.02, respectively (see the bottom panel in Figure~\ref{fig:K}).
Hence, $\sigma_\mathrm{DN}^\mathrm{DEM}$ well matches $\sigma_\mathrm{DN}^\mathrm{ratio}$,
since $\sigma_\mathrm{DN}^\mathrm{ratio}$ is derived from $K^{(2)}$ at $T_\mathrm{ratio}$,
and $T_\mathrm{ratio}$ is close to the DEM peak temperature (see Section~\ref{subsec:filter ratio temperature}).
We note that, for temperature ranges other than the DEM peak, $K^{(2)}$ varies with a factor of typically 2\,--\,4.
However, since \textit{i}) the photon noise is a function of $\sqrt{K^{(2)}}$ (see Equation~\ref{eq:sigma_I}) and furthermore,
\textit{ii}) in this temperature range, the contribution of photon noise to $\sigma_\mathrm{DN}^\mathrm{DEM}$ is smaller than at the DEM peak,
the variation of $K^{(2)}$ in these temperature ranges does not significantly affect $\sigma_\mathrm{DN}^\mathrm{DEM}$.

\section{Notes}
\label{app:notes}

\subsection{Contaminant Material}
\label{subsec:discussion and summary 0}

In Appendix~\ref{subsubsec:X-ray analysis in Phase 3}, we identified the contaminant material as
a long-chain organic compound whose chemical composition, density, and refractive index are
similar to those of DEHP, although DEHP is not used in the XRT.
This means that the actual contaminant material is narrowed down to
the material whose refractive index and density are close to 1.5 and 1~g~cm$^{-3}$, respectively.

Next let us discuss the chemical composition of the actual contaminant material.
Since the atomic number of silicon is about two times larger than carbon and oxygen,
\textit{i.e.} the cross-section of silicon to X-rays is certainly larger, the existence of silicon affects on the X-ray transmission certainly.
In fact, when we assume that the contaminant contains some silicon atoms,
the discrepancy between the observed and expected X-ray intensity ratios rises as shown in Figure~\ref{fig:X-ray Si filter contam}.
Hence, we conclude that there are no silicon atoms in the contaminant material.

On the other hand, since the atomic numbers of carbon and oxygen (also nitrogen) are comparable
and that of hydrogen is much smaller than carbon and oxygen, \textit{i.e.} hydrogen is much more transparent to X-rays, 
some slight differences in numbers of such atoms among a long-chain organic compound do not make any significant difference in the transmission of X-rays.
Hence, although we cannot exactly identify the chemical composition of the contaminant material,
it is acceptable for the calibration of the XRT that we tentatively employed DEHP as the contaminant material.
We note that the identified property of the contaminant, where chemical composition is a long-chain organic compound without silicon, refractive index is $\approx$~1.5, and density is $\approx$~1~g~cm$^{-3}$,
is a common property of materials
which are widely used for satellites, and which are well known as possible source of contamination.

\subsection{G-Band Method for CCD Contamination Analysis}
\label{subsec:discussion and summary 1}

In Appendix~\ref{subsubsec:contam monitor method}, we explained the method to
measure the thickness of contaminant on the CCD using G-band data.
The G-band intensity oscillation shown in Figure~\ref{fig:G-band 2nd bake} is essentially expressed by Equation~(\ref{eq:anti-refraction}).
However, there are two differences between the calculated G-band intensity enhancement with the anti-reflection effect of Equation~(\ref{eq:anti-refraction})
and the observed one that is characterized by the empirical model given by Equation~(\ref{eq:empirical model}).

The first one is the refractive indexes of the contaminant.
In this article, we obtained the refractive index of contaminant with following two methods:
\begin{itemize}
\item[\textit{i})]
In Appendix~\ref{subsubsec:measurement of contam thickness on CCD},
from the observed amplitude of intensity enhancement in G-band, the refractive index of contaminant is estimated as $\approx$~1.4.
\item[\textit{ii})]
In Appendix~\ref{subsubsec:X-ray analysis in Phase 3}, the decrease in X-ray intensities caused by the accumulation of contaminant
can be well explained by adopting a well-known material whose refractive index and density of contaminant are 1.5 and 0.986~g~cm$^{-3}$, respectively, as the contaminant material.
\end{itemize}
This difference may be explained with the possible hypothesis where the density of actually-accumulated contaminant is lower than the contaminant in its usual state (as, \textit{e.g.}, products),
because the contaminant material is an organic compound and its accumulation was formed
under vacuum deposition process but with much slower rate than standard ones in laboratories,
and the accumulated contaminant most likely consists of sparsely-structured molecules (rather than the dense crystalline structure). Hereafter, we call this state as sparse state.
The same trend as this hypothesis is seen in the case of C-poly filter as described in the last paragraph of Appendix~\ref{subsec:calibration of FPAF}.
Generally, the refractive index [$n_\mathrm{sparse}$] and density [$\rho_\mathrm{sparse}$] of material in sparse state are smaller than those ($n_\mathrm{usual}$ and $\rho_\mathrm{usual}$) in the usual state.
If the above hypothesis holds, the method~\textit{i}) gives the refractive index of contaminant in sparse state.
On the other hand, in method~\textit{ii}), we identified the refractive index and density in usual state
from among the possible candidates of contaminant (see Table~\ref{tbl:contaminant}).
Hence, the difference in the refractive indexes in methods~\textit{i}) and \textit{ii}) is caused by the different state of the contaminant.

Next, we consider the effect of this difference on the calibration results.
On the basis of Equation~(\ref{eq:osc}), the smaller refractive index gives the thicker contaminant.
Meanwhile, the smaller density gives the longer attenuation length [$l_\mathrm{att}$] in Equation~(\ref{eq:T}).
Hence, the X-ray transmission calculated with $n_\mathrm{sparse}$ and $\rho_\mathrm{sparse}$ would be close to the transmission with $n_\mathrm{usual}$ and $\rho_\mathrm{usual}$.
We note that even if the actual state of contaminant is not sparse, \textit{i.e.} even if the density of actual contaminant is the same as the usual state,
the difference in refractive indexes (1.4 and 1.5) is only 7\%, and then the difference in the estimated thickness of contaminant on the CCD is also 7\%.
Since the maximum thickness of contaminant is about 3600~{\AA} (see Figure~\ref{fig:contam on CCD}), the maximum error in the estimate is about 250~{\AA}.
This error is smaller than the accumulated thickness of contaminant on the thinner FPAFs (see Figure~\ref{fig:contam on FPAF}),
and is negligibly small against the metal thicknesses of the thicker filters.
Hence, we conclude that, whether the hypothesis holds or not, the difference in refractive indexes does not significantly affect the results of XRT calibration,
and adopt $n_\mathrm{usual} = 1.5$ and $\rho_\mathrm{usual} = 0.986$~g~cm$^{-3}$ for the calibration.

The second one is the decay of oscillation amplitude.
We consider that this is caused by the inhomogeneous accumulation rate of the contaminant.
In Appendix~\ref{subsubsec:spatial distribution}, we demonstrated
the accumulation rate of contaminant at the CCD center is larger than at the edges of the CCD.
This effect might appear in the 512~$\times$~512 pixels data which is used to make the G-band intensity plot (Figure~\ref{fig:G-band 2nd bake}).
The difference of accumulation rate will result in an asynchronous phase in the intensity oscillation across the 512~$\times$~512 pixels.
The G-band intensity plot is the average of such different oscillations which are sine curves with different phases.
Because the phase difference becomes larger and larger as time passes, the oscillation amplitude of the G-band intensity enhancement decays.

On the basis of Equation~(\ref{eq:osc}), the thickness of contaminant on the CCD is derived
from only the refractive index of the contaminant and the period of the intensity oscillation.
Hence, we expect that the refractive index of $\approx$~1.5 and the oscillation period in Figure~\ref{fig:G-band 2nd bake}
give the average thickness of contaminant on the CCD area of 512~$\times$~512 pixels, which is reliable enough for the XRT calibration.

\subsection{Accumulation Profile of Contaminant on the CCD}
\label{subsec:discussion and summary 2}

When we estimated the thickness of the contaminant accumulated on the CCD with the visible light intensity profile
(Appendix~\ref{subsubsec:measurement of contam thickness on CCD}),
we assumed a constant accumulation rate for the contaminant between two successive bakeouts.
However, actually, we see rapid accumulation of contaminant right after the CCD bakeout
while for the rest of each period the rate is almost constant (see right panel of Figure~\ref{fig:6th bakeout}).
Nevertheless, the assumption of constant accumulation for the entire period between two successive bakeouts
is a valid one for the calculation of the XRT effective area,
because the contaminants accumulated on the FPAFs are much thicker than
the difference between the actual and estimated thickness with the above assumption,
\textit{i.e.} such difference is negligible.

\subsection{DEM Model Used for the Analysis}
\label{subsec:discussion and summary 3}

In Appendix~\ref{subsec:contam analysis with X-ray}, we calibrated the contaminant thickness
on the FPAFs with the observed X-ray data and DEM in the quiet Sun.
When we performed the calibrations described in this article,
the DEM model in the quiet Sun derived by \inlinecite{bro06} was the latest result available.
Hence, we adopted their DEM model for our analysis.
\inlinecite{bro06} derived the DEM of quiet Sun from a data set taken
with \textit{SOHO}/EIT, \textit{SOHO}/CDS, and \textit{TRACE} on 1 May 1998,
while more recently \inlinecite{bro09} analyzed the DEM with 45 data sets
observed with \textit{Hinode}/EIS in the period from January to April 2007.
The profiles of DEMs studied in these two papers are
very similar up to at least $\log T~\approx~6.2$~K in all cases.
We confirmed that the calibrated contaminant thicknesses on the FPAFs with
the DEMs from \inlinecite{bro06} and \inlinecite{bro09} are consistent with each other.

\subsection{Source Location of the Contaminant}
\label{subsec:discussion and summary 4}

We found that there is remarkable similarity between the observed spatial distribution of the contaminant across the CCD
and the simulated result by \inlinecite{ura08} for the case
where the contaminant was assumed to come from the direction in front of the CCD (see Figure~\ref{fig:contam rate map}).
Meanwhile, we found that the contamination is triggered by the operational heater which warms the rear end of telescope tube
up to about $20^{\circ}$C.
On the basis of these circumstantial evidences,
we suspect that the contaminant most likely originates from somewhere in front of the CCD, in the telescope tube.

\subsection{Future plan for the calibration}
\label{subsec:discussion and summary 5}

As we mentioned in Appendix~\ref{subsec:calibration of FPAF},
we consider that the calibration of thick filters (med-Be, thick-Al and thick-Be filters)
with ground-based test data is not ideal,
though the difference between calibrated and actual X-ray transmission of them should be small.
For further calibration of thick filters, we need to take enough data sets
where XRT observes intense X-ray sources, \textit{e.g.}, active regions and flares, with thick filters.
Calibration with on-orbit observation data is our future work.

Also, in order to supplement our calibration, we plan to perform the following cross-calibrations:
\textit{i}) between \textit{Hinode}/XRT and \textit{Hinode}/EIS, and
\textit{ii}) between XRT and \textit{GOES13}/SXI, which is the grazing incidence X-ray telescope like XRT.

\begin{acks}
The authors thank members of the XRT team for useful discussions and comments.
We acknowledge D. Brooks and H. Warren for variable information on quiet Sun and active region DEMs.
\textit{Hinode} is a Japanese mission developed and launched by ISAS/JAXA,
collaborating with NAOJ as a domestic partner, NASA and STFC (UK)
as international partners.
Scientific operation of the \textit{Hinode} mission is conducted
by the \textit{Hinode} science team organized at ISAS/JAXA.
This team mainly consists of scientists from institutes
in the partner countries.
Support for the post-launch operation is provided by
JAXA and NAOJ (Japan), STFC (UK), NASA, ESA, and NSC (Norway).
CHIANTI is a collaborative project involving the NRL (USA), the Universities of
Florence (Italy) and Cambridge (UK), and George Mason University (USA). 
We wish to express our sincere gratitude to the late Takeo Kosugi,
former project manager of \textit{Hinode} (\textit{SOLAR-B}) at ISAS, who passed away
in November 2006. Without his leadership in the development of
\textit{Hinode}, this mission would have never been realized.
We express our sincere gratitude to those who participated in, or supported, 
XRT end-to-end calibration measurement at MSFC XRCF that was carried out for 
two weeks in May-June 2005. 
First of all, we are very much indebted to an extremely-talented 
team of people at XRCF led by C. Reily and J. McCracken, including R. Siler, 
E. Wright, J. Carpenter, J. Keegley and G. Zirnstein, and also 
M. Baker, H. Haight, B. Hale, T. Hill, B. Hoghe, D. Javins, J. Norwood, 
H. Rutledge, G. St. John, J. Tucker, and D. Watson. Without their continuous 
and enormous support including preparation of LN2-cooled cold plate facing 
XRT radiators in the cryogenic chamber of XRCF, from early preparation phase 
of the XRCF experiment and throughout the experiment period, the filter 
calibration at XRCF would have been totally impossible. 
We are also grateful to people from SAO who supported 
the XRCF measurement; G. Austin, W. Podgorski, E. Dennis, J. Chappel, 
D. Caldwell, W. Martell, M. Harris, M. Cosmo, D. Weaver, S. Park and T. Kent 
who participated in the experiment, and P. Cheimets, J. Bookbinder, 
J. Boczenowski. Also A. Sabbag of Naval Research Laboratory is appreciated 
for his support during pre-shipment instrument check performed at SAO. 
K. Kumagai and M. Tamura of NAOJ, K. Yaji of Rikkyo U., and K. Kobayashi of 
MSFC are greatly appreciated for preparing for, and participating in, the 
XRCF experiment. 
Also, people from MSFC, L. Hill, J. Owens, B. Cobb, D. Coleman, R. Jayroe, 
T. Perrin, D. Schultz, A. Sterling and C. Talley are sincerely acknowledged 
for their various support throughout our stay in Huntsville and also for our 
transportation to and from XRCF.
Finally, the authors grateful to the referees and editor for their detailed and careful reviews.
\end{acks}


\end{article} 

\end{document}